\documentclass[a4paper,onecolumn,11pt,accepted=2026-06-26]{quantumarticle}
\pdfoutput=1
\usepackage[utf8]{inputenc}
\usepackage[english]{babel}
\usepackage[T1]{fontenc}
\usepackage{csquotes}
\usepackage{microtype}
\usepackage[bookmarksnumbered]{hyperref}
\usepackage[datamodel=quantum,bibstyle=quantum,maxbibnames=11]{biblatex}
\usepackage{amsmath,amsfonts}
\usepackage{tikz}
\usetikzlibrary{shapes}
\usetikzlibrary{decorations.pathreplacing}
\usepackage{changepage}
\usepackage{braket}
\usepackage{physics}
\usepackage{mathpartir}
\usepackage{xcolor}
\usepackage{multirow}
\usepackage{tabu}
\usepackage{xspace}
\usepackage{hyperref}
\usepackage[capitalize,noabbrev]{cleveref} %
\usepackage{paralist}
\usepackage[normalem]{ulem}
\usepackage{soul}

\crefformat{section}{\S#2#1#3}
\crefformat{subsection}{\S#2#1#3}
\crefformat{subsubsection}{\S#2#1#3}
\crefrangeformat{section}{\S\S#3#1#4 to~#5#2#6}
\crefmultiformat{section}{\S\S#2#1#3}{ and~#2#1#3}{, #2#1#3}{ and~#2#1#3}
\creflabelformat{equation}{#2\textup{#1}#3}

\usepackage{xfrac}
\usepackage{stmaryrd,bbold}
\usepackage{multicol}

\usepackage{alltt}
\usepackage{listings,lstcoq}
\definecolor{ltblue}{rgb}{0,0.4,0.4}
\definecolor{dkblue}{rgb}{0,0.1,0.6}
\definecolor{dkgreen}{rgb}{0,0.35,0}
\definecolor{dkviolet}{rgb}{0.3,0,0.5}
\definecolor{dkred}{rgb}{0.5,0,0}
\newcommand{\ctrl}[2]{\draw[fill=black] (#1,#2) circle [radius=0.12];}
\newcommand{\qnot}[2]{\draw (#1,#2) circle [radius=0.25]; \draw (#1,#2-0.25) -- (#1,#2+0.25);}
\newcommand{\cnot}[3]{\qnot{#1}{#3}\ctrl{#1}{#2}\draw (#1,#2) -- (#1,#3);}
\newcommand{\cz}[3]{\ctrl{#1}{#3}\ctrl{#1}{#2}\draw (#1,#2) -- (#1,#3);}

\newcommand{\unitary}[3]{\draw[fill=white] (#2-0.4,#3-0.4) rectangle node {\texttt{#1}} (#2+0.4,#3+0.4);}

\newcommand{\meas}[2]{\draw[fill=white] (#1-0.8,#2-0.4) rectangle node {\MEAS} (#1+0.8,#2+0.4);}

\newcommand{\oracle}[5]{\draw[fill=white] (#2-0.4,#3-0.4) rectangle node {\texttt{#1}} (#4+0.4,#5+0.4);}

\renewcommand{\vec}[1]{\ensuremath\mathbf{#1}}%
\newcommand{\sqt}{\tfrac{1}{\sqrt 2}}

\newcommand{\qwire}{\ensuremath{\mathcal{Q}\textsc{wire}}\xspace}

\newcommand{\type}[1]{\ensuremath{\mathbf{#1}}\xspace}
\newcommand{\pred}[1]{\ensuremath{\mathbf{#1}}\xspace}

\newcommand{\hoare}[3]{\ensuremath{\left\{\type{#1}\right\} ~\mathtt{#2}~ \left\{\type{#3}\right\} }} %
\newcommand{\hoarepre}[3]{\ensuremath{\left\{\type{#2}\right\} ~\mathtt{#1}~ \left\{\type{#3}\right\} }}

\newcommand{\axiom}[1]{\inferrule*{ }{#1}}

\newcommand{\Z}{\type{Z}}
\newcommand{\X}{\type{X}}
\newcommand{\Y}{\type{Y}}
\newcommand{\I}{\type{I}}

\newcommand{\Gt}{\type{G}}
\newcommand{\T}{\type{T}}

\newcommand{\A}{\type{A}}
\newcommand{\B}{\type{B}}

\newcommand{\Ut}{\type{U}}
\newcommand{\Ct}{\type{C}}
\newcommand{\D}{\type{D}}
\newcommand{\blank}{\boxdot}

\newcommand{\normal}{\text{norm}}

\newcommand{\complex}{\mathbb{C}}
\renewcommand{\P}{\mathcal{P}}

\newcommand{\CNOT}{\ensuremath{\mathtt{CNOT}}\xspace}
\newcommand{\CZ}{\ensuremath{\mathtt{CZ}}\xspace}

\renewcommand\H{\ensuremath{\mathtt{H}}}
\newcommand\MEAS{\ensuremath{\mathtt{MEAS}}}

\newcommand{\floor}[1]

\newcommand{\denote}[1]{\ensuremath{[\![#1]\!]}}



\usepackage[utf8]{inputenc}
\usepackage{newunicodechar}
\newunicodechar{→}{\ensuremath{\to}}
\newunicodechar{⊗}{\ensuremath{\otimes}}
\newunicodechar{×}{\ensuremath{\times}}
\newunicodechar{∩}{\ensuremath{\cap}}
\newunicodechar{⊤}{\ensuremath{\top}}
\newunicodechar{⊥}{\ensuremath{\bot}}

\usepackage{etoolbox} %
\newtoggle{comments}
\toggletrue{comments}

\iftoggle{comments}{
\newcommand{\anote}[1]{\textcolor[rgb]{0.7,0,0.3}{(Aarthi: #1)}}
\newcommand{\bnote}[1]{\textcolor{orange}{(Brad: #1)}}
\newcommand{\anew}[1]{\textcolor[rgb]{0.7,0,0.3}{#1}}
\newcommand{\rnr}[1]{\textcolor[rgb]{0.1,0.7,0.1}{(Robert: #1)}}
\newcommand{\knote}[1]{\textcolor{blue}{(Kartik: #1)}}
\newcommand{\ynote}[1]{\textcolor{blue}{(Youngchan: #1)}}
\newcommand{\todo}[1]{\textcolor{red}{#1}}
\newcommand{\done}[1]{\textcolor{green}{Done #1}}

}{
  \newcommand{\anote}[1]{}
  \newcommand{\bnote}[1]{}
  \newcommand{\anew}[1]{}
  \newcommand{\rnr}[1]{}  
  \newcommand{\knote}[1]{}
  \newcommand{\ynote}[1]{}
  \newcommand{\todo}[1]{}
  \newcommand{\done}[1]{}
  
}

\usepackage{anyfontsize}

\usepackage{amsthm,amssymb,amsfonts,thmtools}
\usepackage{algorithm}
\usepackage{algpseudocode}
\newcommand{\FullComment}[1]{%
  \Statex \(\triangleright\) #1
}
\algtext*{EndIf}
\algtext*{EndFor}
\algtext*{EndWhile}
\newtheorem{theorem}{Theorem}
\newtheorem{corollary}[theorem]{Corollary}
\newtheorem{lemma}[theorem]{Lemma}
\newtheorem{proposition}[theorem]{Proposition}
\newtheorem{example}[theorem]{Example}
\newtheorem{fact}[theorem]{Fact}
\newtheorem{definition}[theorem]{Definition}
\newtheorem{remark}[theorem]{Remark}

\Crefname{fact}{Fact}{Facts}

\usepackage{float}
\usepackage{yquant}
\begin{document}

\title{Hoare meets Heisenberg:\newline A Lightweight Logic for Quantum Programs}

\author{Aarthi Sundaram}
\affiliation{Microsoft Quantum, Redmond, WA}
\email{aarthi.sundaram@microsoft.com}
\orcid{0000-0002-4740-1886}

\author{Robert Rand}
\affiliation{University of Chicago, Chicago, IL}
\email{rand@uchicago.edu}
\orcid{0000-0001-6842-5505}

\author{Kartik Singhal}
\affiliation{University of Chicago, Chicago, IL}
\thanks{now at Quantinuum, Broomfield, CO}
\email{ks@cs.uchicago.edu}
\orcid{0000-0003-1132-269X}

\author{Youngchan Cho}
\affiliation{University of Chicago, Chicago, IL}
\email{youngchan@uchicago.edu}
\orcid{0009-0007-4523-4619}

\author{Brad Lackey}
\affiliation{Microsoft Quantum, Redmond, WA}
\email{brad.lackey@microsoft.com}
\orcid{0000-0002-3823-8757}

\maketitle

\def\titlerunning{Hoare meets Heisenberg}
\def\authorrunning{Sundaram, Rand, Singhal \& Lackey}

\begin{abstract}

    We show that \citeauthor{Gottesman1998}'s (\citeyear{Gottesman1998}) semantics for Clifford circuits based on the Heisenberg representation gives rise to a lightweight Hoare-like logic for efficiently characterizing a common subset of quantum programs. Our applications include (i) certifying whether auxiliary qubits can be safely disposed of, (ii) determining if a system is separable across a given bipartition, (iii) checking the transversality of a gate with respect to a given stabilizer code, and (iv) computing post-measurement states for computational basis measurements. Further, this logic is extended to accommodate universal quantum computing by deriving Hoare triples for the $T$-gate, multiply-controlled unitaries such as the Toffoli gate, and some gate injection circuits that use associated magic states. A number of interesting results emerge from this logic, including a lower bound on the number of $T$ gates necessary to perform a multiply-controlled $Z$ gate.

\end{abstract}

\section{Introduction}

Quantum programs are notoriously complex in both the traditional and computational sense: They are hard to write correctly and extremely expensive to simulate. The same applies to program logics for quantum programs: It can be hard to come up with an assertion about a quantum program (typically an observable or a projector), and it is doubly hard to prove that it holds upon execution of a program. Typically this will either involve complex calculations or high-level reasoning. To address this problem, we propose a logic that uses the stabilizer formalism for efficient reasoning about Clifford circuits. We extend this system to handle universal quantum gate sets, both by explicitly adding the $T$ gate and by showing how to handle arbitrary unitary gates. We also expand the system to handle measurement on stabilizer circuits and a restricted set of Clifford+$T$ circuits. This system is designed for efficient analysis: In particular, given a precondition and a Clifford circuit, we can derive the most general postcondition in linear time. Given a non-Clifford circuit, determining the postcondition will double in the worst case for each non-Clifford gate in the circuit.

The starting point to understanding our system is the Heisenberg interpretation of quantum mechanics. This interpretation treats quantum operators as functions on operators rather than on quantum states. For instance, given an arbitrary quantum state $\ket{\phi}$, the Hadamard operator $H$ satisfies
\begin{equation}
\label{eq:unitaryH}
HZ\ket{\phi} = XH\ket{\phi}.
\end{equation}
In other words, the operator $H$ can be viewed as a function that takes $Z$ to $X$ and similarly takes $X$ to $Z$.
\textcite{Gottesman1998} used this representation to present the rules for how the Clifford set ($H$, $S$ and $\mathit{CNOT}$) operates on Pauli $X$ and $Z$ matrices. Hence, we can use Hoare-style triple to describe the action of $H$ on both the $X$ and $Z$ operators.
\begin{equation}
\label{eq:typeH}
\hoare{X}{H}{Z} \qquad \hoare{Z}{H}{X}
\end{equation}
Note that it suffices to just specify how $H$ acts on $\X$ and $\Z$ as we can derive the action of $H$ on $\Y$ by treating the operator $Y$ as $iXZ$ (since $\sigma_y = i \sigma_x \sigma_z$). More specifically,

\begin{align*}
    HY \ket{\phi} &= H(iXZ) \ket{\phi} \\
    &= i(HX)Z \ket{\psi} \\
    &= i(ZH)Z \ket{\psi} \\
    &= iZ(HZ) \ket{\psi} \\
    &= iZXH \ket{\psi} \\
    &= -YH \ket{\psi}
\end{align*}

Throughout this work, we develop a logic motivated by this interpretation of Pauli matrices as predicates. Formally, the syntax of the logic, as found in \Cref{fig:types} and \Cref{fig:app-rules}, is axiomatic with the semantics described above being a sound interpretation. For example, we can represent the general form of this last deduction by the following deductive rule:
\begin{equation*}
\inferrule*[right=Y]{\hoare{X}{U}{A} \\ \hoare{Z}{U}{B}}{\hoare{Y}{U}{iAB}}
\end{equation*}
Here \pred{A} and \pred{B} are assumed to be Paulis, so the product of \pred{A} and \pred{B} is simply the third Pauli, possibly negated or multiplied by $i$. This is indicative (and a special case) of the kinds of rules we will use throughout the paper.

In Gottesman's paper, the end goal was to fully describe quantum programs and prove the \emph{Gottesman-Knill theorem}, which shows that any Clifford circuit can be classically simulated efficiently. Our goal is to take the rules in \cref{eq:typeH} and use them as a starting point to build a logical system for characterizing quantum programs (\Cref{sec:types-semantics}). Furthermore, we move beyond Clifford circuits and expand the predicates to characterize some magic states, the $T$ gate, and other gates in the Clifford hierarchy (\Cref{sec:additive-types}). A key feature of our system is that the predicates correspond to unitary Hermitian operators: when restricted to stabilizer quantum computing, these are (tensor products of) Pauli matrices, and for universal quantum computing, they are general unitary Hermitian matrices. Notationally, we use uppercase letters $U, V, \ldots$ to denote unitary gates or matrices and the boldface $\pred{U}, \pred{V}, \dots$ to denote the corresponding predicates.

\paragraph{The semantics of stabilizer predicates}
In our system, a judgment of the form $\pred{P}(\ket{\psi})$ admits a straightforward interpretation: $\ket{\psi}$ is a $+1$ eigenstate of $P$. As a result,  $\hoare{A}{U}{B}$ means that $U$ maps a $+1$ eigenstate of $A$ to a $+1$ eigenstate of $B$. This closely mirrors the stabilizer formalism used for error correcting codes~\cite{gottesman1996class}. It works well as long as we restrict to Clifford circuits and are fine with very coarse judgments in the face of measurements.
However, for more accurate judgments when measurements are performed and to work with more general gates, we will associate $\pred{P}(\ket{\psi})$ with the fact that $\ket{\psi}$ lies in the image of the projection $\Pi_P^+ := \frac{1}{2} (I + P)$ i.e., $\frac{1}{2} (I + P) \ket{\psi}= \ket{\psi}$.

We use the tensor operand $\otimes$ to represent multi-qubit predicates.
Using our first interpretation, $\ket{\psi}$ satisfies $\pred{A} \otimes \pred{B}$ if $\ket{\psi}$ is a $+1$-eigenstate of $A\otimes B$. Observe that this does not restrict $\ket{\psi}$ to be a product state $\ket{\phi_1} \otimes \ket{\phi_2}$ such $\ket{\phi_1}$ satisfies \A and $\ket{\phi_2}$ satisfies \B. For instance, $\A \otimes \B$ holds of $\ket{\psi'}$ when $\ket{\psi'} = \ket{\phi'_1} \otimes \ket{\phi'_2}$ such that $-\A(\ket{\phi'_1}) $ (i.e., $\ket{\phi'_1}$ is a $-1$-eigenstate of $A$) and $-\B(\ket{\phi'_2})$. Moreover, arbitrary superpositions of $\ket{\psi}$ and $\ket{\psi'}$ also satisfy $\A \otimes \B$.

In our predicate language, conjunction and intersection coincide: if $\ket\psi$ is a $+1$-eigen-state of both $A$ and $B$, then it satisfies $\A \cap \B$. Note that for this to hold $A$ and $B$ must commute, because Pauli operators that do not commute will instead anticommute and have no common eigenvectors. %
We use disjunctions to represent post-measurement states when the outcome is probabilistic. In this case, $(\A \uplus \B)(\ket{\psi})$ denotes that $\ket\psi$ is either a +1-eigenstate $A$ or a +1-eigenstate of $B$, without making any claims to the likelihood of which case is true. In the measurement context, it means that one outcome results in the system satisfying $\pred{A}$ and the other outcome satisfies $\B$. %

While our statements about conjunction and disjunction are focused on the stabilizer formalism, and hence refer to Pauli types, they extend to commuting additive predicates as well. For additive predicates, failing to commute does mean they must anticommute and hence in complete generality the situation can be quite complex. 

\paragraph{Applications}
Our syntax and derivation rules for Clifford circuits and stabilizer states are methodically developed in~\Cref{sec:types-semantics}. The full list of our rules are in~\Cref{fig:types,fig:app-rules}.
The most straightforward use of our system is in characterizing properties of Clifford circuits, particularly entanglement and separability. For the textbook case of Deutsch's algorithm, we are easily able to verify three key properties: (i) the first qubit is $\ket{0}$ whenever the function is constant, (ii) the first qubit is $\ket{1}$ whenever the function is balanced, and (iii) that the two qubits are never entangled, and therefore the second can be safely discarded. These are three common and broadly useful properties to check.

Ideally, our predicates would be unique: any $\A$ and $\B$ that have the same set of eigenvectors should be equal. This would allow us to prove program equivalence, given fully descriptive predicates for a program. While this uniqueness does not hold by construction, we can obtain a canonical representation for our predicates that guarantees this property. Inspired by the \emph{row echelon form} of a matrix, we describe in~\Cref{sec:normal} an efficient algorithm to generate a canonical representation for intersections of predicates. This allows us to use our logic to \emph{efficiently} track whether a given sub-system is separable from the rest of the system in \Cref{sec:separability}. In \Cref{sec:GHZ}, we generate and then disentangle a GHZ state $\frac{1}{\sqrt{2}}(\ket{000} + \ket{111})$ to show how the logic is capable of tracking both the creation and destruction of entanglement.

A crucial method used by quantum circuits to extract or output classical information is measurement (usually in the computational basis). It is challenging to tune our logic to accommodate measurement in light of the fact that it requires managing the operation on all the basis states, unlike the evolution of a single Pauli operator. However, measurement on stabilizer states is well understood, and this allows us to construct a procedure to generate a measurement outcome and post-measurement as discussed in~\Cref{sec:stab_meas}. In particular, when the measurement outcome is random, we use disjunctions to capture the fact that the system could be in one of several states, depending on the outcome. As a simple example, applying a z-basis measurement on an $\X$ qubit to get a random $0$ or $1$ outcome is represented as %
\begin{align*}
    \axiom{\hoare{X}{\textsf{Meas}}{Z \uplus -Z}}
\end{align*}

Using all the elements described above, in~\Cref{sec:qecc},
we demonstrate how our logic can be used to verify the working of a stabilizer error correcting code (the Steane code on $7$ qubits~\cite{steane1997active}). Specifically, we (i) derive a predicate for a logical qubit in the Steane code; (ii) verify that the encoding circuit constructs the appropriate logical qubit state; and (iii) show the transversality of the $H$ and $S$ gates as well as the non-transversality of the $T$ gate for the Steane code.

\paragraph{Additive predicates and their applications}
All the ideas discussed up to this point deal with the realm of stabilizer states and Clifford circuits, which are not universal for quantum computing. For instance, while we can add the axiom $\hoare{\Z}{T}{\Z}$ to our system, the stabilizer formalism is incapable of expressing the action of the $T$ gate on $\X$. We address this shortcoming by developing \emph{additive predicates} in~\Cref{sec:additive-types}, which are expressed as linear combinations of our basic predicates. This allows us to express the action of $T$ on $\X$ as $\hoare{\X}{T}{\sqt(\X + \Y)}$.

Since the $T$-gate is not the only way to achieve universal quantum computation, we produce a straightforward algorithm for adding new gates to the system (such as the Toffoli) by fully deriving their corresponding rules. A particularly nice application has to do with multiply-controlled $Z$ gates, which have very succinct postconditions when applied to \X or \Z qubits. In fact, comparing their derived rules to that of the $T$ gate, we can easily show that synthesizing an $n$-controlled $Z$-gate requires at least $(2n-2)$ $T$-gates.

In~\Cref{sec:meas-add}, we discuss how to determine the postcondition of a single-qubit computational basis measurement given one- and two-qubit additive preconditions. Putting these pieces together in~\Cref{sec:magic}, we derive rules for gate injection circuits that use associated magic states to implement non-Clifford circuits. We focus on single-qubit unitaries that correspond to a rotation about the $Z$ axis, i.e., that rotate qubits in the $\X/\Y$-plane by some angle $\theta$.

\paragraph{Applying the logic in practice.} We conclude with a discussion on the complexity of determining the most descriptive postcondition given a program and a precondition in~\Cref{sec:complexity}. Unsurprisingly, fully characterizing a circuit with high $T$-depth proves to be intractable in the general case. However, proving interesting properties of circuits with a few $T$ gates is often quite possible. Moreover, Clifford circuits can be efficiently characterized to any degree of precision, allowing us to flexibly analyze a broad range of quantum programs. Note that efficiency here means that the procedure scales linearly with the number of gates in the operation and polynomially in the size of the system.

\paragraph{} We place our work in the context of related work in~\Cref{sec:related} and discuss possible future applications and extensions to this system in \Cref{sec:future}.

\section{Our Hoare-style Logic and its Semantics}
\label{sec:types-semantics}

Here we present the internal syntax of our predicates and several semantic interpretations. We will extend this to more general predicates in \S{\ref{sec:additive-types}} below. Our atomic predicates correspond to the Pauli matrices and are denoted $\X, \Y, \Z$. We denote basic operators (or \emph{gates}) by $H$, $S$, $\mathit{CNOT}$ and later $T$.

\subsection{A Simple Quantum Language}

Our language is given by the following grammar:
\begin{align*}
    g &:= H~i \mid S~i \mid \mathit{CNOT}~i~j \mid T~i \mid \text{Meas}~i \mid g ; g
\end{align*}
where the semicolon corresponds to sequencing. $T$ and Meas are both significantly more difficult to reason about than the other operators and will have their own sections devoted to them (\Cref{adding-T} and \Cref{sec:stab_meas}). Note that we can express other common operators like $Z$, $\mathit{CZ}$, $T^\dag$, and $\mathit{CCX}$ (or \emph{Toffoli}) in terms of the operators above. We will introduce each of these in turn and derive their associated pre- and postconditions ($Z$ in \Cref{eg:ZonZ,eg:ZonX}, $\mathit{CZ}$ in \Cref{eg:CZ}, $T^\dag$ in \cref{eg:tdag}, and $\mathit{CCX}$ in \Cref{sec:toffoli}).

\subsection{Atomic Predicates}

Our core interpretation for $\X, \Y$ and $\Z$ is that each of these predicates is inhabited by a single qubit state, the $+1$-eigenstate of the Pauli operators $\sigma_x, \sigma_y$, and $\sigma_z$. %
The following judgments hold using the standard quantum computing notation:

\begin{equation*}
\X(\ket{+}), \qquad
\Y(\ket{i}), \qquad
\Z(\ket{0}).
\end{equation*}

We can also negate these operators to obtain their $-1$-eigenstates:
\begin{equation*}
-\X(\ket{-}), \qquad
-\Y(\ket{-i}), \qquad
-\Z(\ket{1}).
\end{equation*}

Note that we can equally well read $-\X(\ket{-})$ as ``$\ket{-}$ is a $+1$-eigenstate of $-X$'' or ``$\ket{-}$ is a $-1$-eigenstate of $X$''. We prefer the former since it will generalize better to multiplication by numbers other than $-1$ in subsequent sections.

Unlike \X, \Y, and \Z, every single-qubit state is a $+1$ eigenvector of \I:
\begin{equation*}
\forall \ket{\psi}, \I(\ket{\psi}).
\end{equation*}
In this sense, \I corresponds to the proposition ``True''.

Note that all of our atomic predicates correspond to unitary and Hermitian operators. This ensures that they all have $+1$-eigenstates. In fact, throughout this paper, our predicates will be unitary and hermitian and, with the exception of \I (and \I tensored with itself) of trace $0$. This guarantees that they have an equal number of $+1$ and $-1$ eigenstates.

\subsection{Basic Hoare Triples and Sequencing}
\label{sec:arrow}

To define our basic Hoare triples, we turn to the characterization of $H$, $S$ and $\mathit{CNOT}$ by \textcite{Gottesman1998}:

\begin{proposition}
\label{prop:eigenstate}
    Given a unitary $U : A \rightarrow B$ in the Heisenberg interpretation, $U$ takes every eigenstate of $A$ to an eigenstate of $B$ with the same eigenvalue.
\end{proposition}

\begin{proof}
    From eq. [1] in \textcite{Gottesman1998}, given a state $\ket{\psi}$ and an operator $U$,
    \[ U N \ket{\psi} = U N U^{\dagger} U \ket{\psi}. \]
    In the Heisenberg interpretation, this can be denoted as $U : N \rightarrow U N U^{\dagger}$. Suppose that $\ket{\psi}$ is an eigenstate of $N$ with eigenvalue $\lambda$ and let $\ket{\phi}$ denote the state after $U$ acts on $\ket{\psi}$. Then,
    \[ \lambda \ket{\phi} = U (\lambda \ket{\psi}) = U N \ket{\psi} = U N U^{\dagger} U \ket{\psi} = (U N U^{\dagger}) \ket{\phi}.
    \]

    Hence, $\ket{\phi}$ is an eigenstate of the modified operator $U N U^{\dagger}$ with eigenvalue $\lambda$.
\end{proof}

Therefore, in our logic, $\hoare{A}{U}{B}$ will mean that the unitary program corresponding to $U$ takes a $+1$-eigenstate of $A$ to a $+1$-eigenstate of $B$. In particular, for the identity operator $\hoare{A}{I}{A}$ for any predicate $\A$. %

As a result, our basic Hoare triples for $\H$ and \texttt{S} follow precisely from Gottesman (we will postpone the introduction of $\mathit{CNOT}$ until the next section):
\begin{equation}
\axiom{\hoare{X}{H}{Z}} \qquad
\axiom{\hoare{Z}{H}{X}} \qquad
\axiom{\hoare{X}{S}{Y}} \qquad
\axiom{\hoare{Z}{S}{Z}} \qquad
\label{rule:HS}
\end{equation}
So, for instance, $H$\footnote{For readability, we use $U$ as shorthand for $U$ applied to qubit $1$ when $U$ is a one qubit unitary, and $V$ as shorthand for $V~1~2$ when $V$ is a two-qubit unitary, particularly when these are applied to one- and two-qubit states, respectively. We discuss applying unitaries to larger states in \cref{sec:multi-qubit}.} takes $\ket +$ to $\ket 0$.

Note that every qubit is an $+1$-eigenstate of $I$, and similarly, every quantum state is an $+1$-eigenstate of $I^k$ (our notation for $I^{\otimes k}$ where $k$ is the number of qubits in the system) so we have the following rule for any single qubit unitary $U$:
\begin{equation}\label{rule-identity1}
\inferrule*[Right=\I]{~}{\hoare{I}{U}{I}}
\end{equation}

We adopt the standard sequencing rule from Hoare logic:
\begin{equation}\label{rule-seq}
\inferrule*[Right=seq]{\hoare{A}{p_1}{B} \and \hoare{B}{p_2}{C}}{\hoare{A}{p_1 ; p_2}{C}}
\end{equation}

\begin{example}
\label{eg:ZonZ}
For instance, here is our derivation of the postcondition for applying $Z$ = $S;S$ to a state satisfying $\Z$:
\begin{equation*}
\inferrule*[Right=seq]{\axiom{\hoare{Z}{S}{Z}} \and \axiom{\hoare{Z}{S}{Z}}}
{\hoare{Z}{S;S}{Z}}
\end{equation*}    
\end{example}

Since $S$ has $\Y$ as a postcondition given the precondition $\X$ and our basic Hoare judgments only have $\X$ and $\Z$ preconditions, we will need to introduce rules for coefficients and multiplication that generalize the \pred{Y} rule presented in the introduction:

\vspace{1em}
\begin{minipage}{0.4\linewidth}
\begin{equation}
\inferrule*[Right=scale]{\hoare{A}{p}{B}}{\hoare{cA}{p}{cB}}\label{eq:rule-scale}
\end{equation}
\end{minipage}
\begin{minipage}{0.55\linewidth}
\begin{equation}
\inferrule*[Right=mul]{\hoare{A}{p}{B} \and \hoare{C}{p}{D}}
{\hoare{AC}{p}{BD}}\label{eq:rule-mul}
\end{equation}
\end{minipage}
\vspace{1em}

In \textsc{scale}, \pred{c} can be any complex number, although in any derivation that stays within the Clifford group, non-vacuous predicates will only use $c \in \{-1,i,-i\}$. We should note that there is no matrix multiplication happening when we apply the \textsc{mul} rule: There are only $16$ possible combinations of two Paulis, each of which produces a Pauli, so we can efficiently simplify these symbolically. The same is true for $c \in \{1,-1,i,-i\}$: We discuss simplification in more detail in \Cref{sec:intersection}. We should also note that many intermediate deductions will prove vacuous: For instance, we can easily derive that \hoare{XZ}{H}{ZX} which simplifies to \hoare{-i\Y}{H}{i\Y}. In itself, this is not useful, as neither $-iY$ nor $iY$ have $+1$-eigenvectors. However, by applying the scale rule with $c=i$, we get \hoare{\Y}{H}{-\Y}, which is certainly meaningful.

\begin{example}
\label{eg:ZonX}
We can now derive a postcondition for $Z$ given the precondition $\X$:
\begin{equation*}
\inferrule*[Right=seq]{\axiom{\hoarepre{S}{X}{Y}} \and
\inferrule*[Right=mul]{\axiom{\hoarepre{S}{X}{Y}} \and \axiom{\hoarepre{S}{Z}{Z}}}
{\inferrule*[Right=scale]{\hoarepre{S}{XZ}{YZ}}{\hoarepre{S}{Y}{iYZ}}}}
{\hoarepre{S;S}{X}{-X}}
\end{equation*}    
\end{example}

In this deduction, we simplify $i\X\Z$ to $\Y$ and $i\Y\Z$ to $-\X$.

\begin{lemma}
Defining \texttt{X} as \texttt{H;Z;H} and \texttt{Y} as \texttt{S;X;Z;S}, the following triples are all derivable and hence admissible as rules in the logic:
\[
\inferrule*{~}{\hoare{X}{X}{X}} \quad \inferrule*{~}{\hoare{Z}{X}{-Z}} \quad \inferrule*{~}{\hoare{X}{Y}{-X}} \quad \inferrule*{~}{\hoare{Z}{Y}{-Z}}
\]
\end{lemma}
We will frequently make use of these rules in the text.

\subsection{Predicates over multi-qubit systems}
\label{sec:multi-qubit}

In order to do anything interesting, we are going to need to consider multi-qubit systems. We can write a predicate $\pred P_1 \otimes \pred P_2 \otimes \dots \otimes \pred P_n$ for Pauli predicates $\pred{P_i}$ to characterize an $n$ qubit system following our semantics. We use $\pred{T}[i]$ to refer to $\pred P_i$ from that tensor product and $U~i$ to apply $U$ to the $i^\text{th}$ qubit in a quantum state. We can therefore introduce the following rule for applying a single-qubit operator to a multi-qubit state:
\begin{equation}\label{rule-tensor1}
    \inferrule*[right=\ensuremath{\otimes_1}]{\pred{T}[i] = \A \and \hoare{A}{U}{B}}{\hoare{T}{U~i}{\pred{T}[i \mapsto \B]}}
\end{equation}
Here, $\T[i \mapsto \B]$ replaces the predicate for the $i^\text{th}$ qubit in the tensor product with $\B$.

We can now introduce the Hoare triples corresponding to Gottesman's rules for $\CNOT$:
\begin{align}
    & \axiom{\hoarepre{CNOT}{X \otimes I}{X \otimes X}} \quad
    & \axiom{\hoarepre{CNOT}{I \otimes X}{I \otimes X}} \nonumber\\
    & \axiom{\hoarepre{CNOT}{Z \otimes I}{Z \otimes I}} \quad
    & \axiom{\hoarepre{CNOT}{I \otimes Z}{Z \otimes Z}} \label{rule:cnot}
\end{align}

To apply $\CNOT$ to multi-qubit states, we'll need a new rule:
\begin{equation}\label{rule-tensor2}
    \inferrule*[right=\ensuremath{\otimes_2}]{\pred{T}[i] = \A \and \pred{T}[j] = \B \and \hoarepre{U}{A\otimes B}{C\otimes D}}
    {\hoarepre{U~i~j}{T}{T[i \mapsto \pred C; j \mapsto \pred D]}}
\end{equation}

Note that we'll often need to use this in conjunction with the \textsc{mul} rule, where multiplication distributes over addition. Consider this simple derivation:

\begin{equation*}
    \inferrule*[right=\ensuremath{\otimes_2}]{
    \inferrule*[right=mul]{
    (\Z\otimes \Y \otimes \X)[1] = \Z \\ (\Z\otimes \Y \otimes \X)[3] = \X \\
    \hoare{Z \otimes I}{CNOT}{Z \otimes I} \\ \hoare{I \otimes X}{CNOT}{I \otimes X}}
      {\hoare{Z\otimes X}{CNOT}{Z\otimes X}}}
    {\hoare{Z\otimes Y \otimes X}{CNOT~1~3}{Z\otimes Y \otimes X}}
\end{equation*}

Note that multiplication distributes componentwise over tensors. Note that sometimes we will obtain a coefficient of $-1$ when multiplying two terms; we move these to the outside of the tensor so that $\X \otimes -\Z$ becomes $-(\X \otimes \Z$) (the parentheses are generally not needed as a result).

The single qubit identity rule (\ref{rule-identity1}) clearly extends to any number of qubits; on this basis, it is easy to show that $\I^k$ is a universal predicate for all quantum programs, where $\I^k$ corresponds to $I^{\otimes k}$ and $k$ is greater than or equal to the number of qubits in our program:
\begin{equation}\label{rule-identity}
    \inferrule*[right={id}]{~}
    {\hoarepre{p}{\I^k}{\I^k}}
\end{equation}
In particular, for our $\CNOT$ gate:
\begin{equation*}
    \inferrule*{~}
    {\hoarepre{CNOT}{I \otimes I}{I \otimes I}}
\end{equation*}

We will often want a complete description of a given gate or circuit $C$. Instead of making $2^n$ judgements to characterize an $n$-qubit Clifford circuit $C$ it suffices to make the $2n$ judgements
\[
\hoare{I \otimes \cdots \otimes X \otimes \cdots \otimes I}{C}{P_j} \quad \text{ and } \quad \hoare{I \otimes \cdots \otimes Z \otimes \cdots \otimes I}{C}{Q_j}
\]
for each position $j$. As noted in \cite{Gottesman1998}, any Pauli operator can be generated as the tensor product of weight one Pauli operators and so by the scale and multiply rules (\Cref{eq:rule-scale,eq:rule-mul}) we can derive the triple associated to any other Pauli predicate. In particular, the \CNOT gate is fully characterized by rules (\ref{rule:cnot}).

\begin{example}\label{eg:CZ}
We can now derive the rules for the controlled-Z gate on two qubits. We know that \textnormal{\texttt{CZ 1 2 = H 2; CNOT 1 2; H 2}}. We will start with \CZ on \pred{Z \otimes I}:
\[
\inferrule*[right=seq]
{\inferrule*
    {\inferrule*[right=\ensuremath{\otimes_1}]{\hoare{I}{H}{I}}{\hoare{X \otimes I}{H~2}{X \otimes I}} \; \inferrule*[right=\ensuremath{\otimes_2}]{\hoare{X \otimes I}{CNOT}{X \otimes X}}{\hoare{X \otimes I}{CNOT~1~2}{X \otimes X}}}
    {\hoare{X \otimes I}{H~2;~CNOT~1~2}{X \otimes X}} \quad \inferrule*[right=\ensuremath{\otimes_1}]{\hoare{X}{H}{Z}}{\hoare{X \otimes X}{H~2}{X \otimes Z}}}
{\hoare{X \otimes I}{H~2;~CNOT~1~2;~H~2}{X \otimes Z}}
\]
\end{example}

In future proofs, we will generally elide the $\otimes_1$ and $\otimes_2$ rules, which are easy to derive from context. We leave the remaining rules for \CZ as exercises for the reader:
\begin{gather*}
          \hoare{I \otimes X}{CZ~1~2}{Z \otimes X}
    \qquad \hoare{Z \otimes I}{CZ~1~2}{Z \otimes I}
    \qquad \hoare{I \otimes Z}{CZ~1~2}{I \otimes Z}
\end{gather*}

The reader may further verify that \texttt{CZ 2 1} has the same behavior as \texttt{CZ 1 2} on all four predicates, that is, \CZ is symmetric in its arguments.

\subsection{Intersections and Consequence Rules}
\label{sec:intersection}

If we want to fully describe an operator's behavior, we need to add conjunctions; or, from the perspective of describing a set of eigenstates, intersections. The derivation rule for intersection is exactly what we would expect:
\begin{equation}\label{rule-intersection}
    \inferrule*[Right=\ensuremath{\cap}]{\hoare{A}{p}{B} \and \hoare{C}{p}{D}}{\hoare{A \cap C}{p}{B \cap D}}
\end{equation}
The soundness of this rule follows from Proposition 1; we will discuss this in more detail in \S{\ref{section:soundness}}.

Of course, if \hoare{A}{p}{B \cap C} is a valid triple, then \hoare{A}{p}{B} should as well -- we're simply weakening the postcondition. Similarly, we should be able to strengthen the precondition to $\pred{A} \cap \pred{E}$, even if this winds up being the empty set. This brings us to the rule of consequence:
\begin{equation}\label{rule-consequence}
    \inferrule*[Right=cons]{\pred{A'} \Rightarrow \pred{A}  \and \hoare{A}{p}{B} \and \pred{B} \Rightarrow \pred{B'}}{\hoare{A'}{p}{B'}}
\end{equation}
Here $\pred{P} \Rightarrow \pred{Q}$ means that the $+1$-eigenvectors of the operator $P$ are contained in the $+1$-eigenvectors of $Q$.

It's worth noting some implications relevant to intersection that we use throughout this paper, which follow directly from the semantics:
\begin{equation}
\label{rule-imp_base}
    \begin{aligned}
    & \A \cap \B \Rightarrow \A \\
    & \A \cap \B \Rightarrow \B \cap \A \\
    & \A \cap (\B \cap \Ct) \Leftrightarrow (\A \cap \B) \cap \Ct,
    \end{aligned}    
\end{equation}

One linear algebraic rule is important and will be used repeatedly in our normalization section (\cref{sec:normal}). Consider the following derivation for \CNOT:
\[
\inferrule*[Right=\ensuremath{\cap}]{\hoare{\Z \otimes \I}{CNOT}{\Z \otimes \I} \and \hoare{\I \otimes \Z}{CNOT}{\Z \otimes \Z}}
{\hoarepre{CNOT}{\Z \otimes \I \cap \I \otimes \Z}{\Z \otimes \I \cap \Z \otimes \Z}}
\]

This should become quite easy to interpret: $\Z \otimes \I$ characterizes states where the first qubit is $\ket{0}$ and $\I \otimes \Z$ does the same for the second qubit. Hence, their intersection describes exclusively the state $\ket{00}$. However, while we can check mathematically that $\Z \otimes \I \cap \Z \otimes \Z$ describes exactly the same state, it is rather less obvious from looking at it. Hence, we introduce the implication
\begin{equation}\label{rule-imp_mul}
\A \cap \B \Leftrightarrow \A \cap \A\B
\end{equation}
where multiplication distributes component-wise over tensors. 

Recall that multiplying Pauli predicates is trivial: Multiplying any two non-identity Pauli operators yields the third, with an $i$ and sometimes a minus sign. i.e. $\Z\X = i\Y$ and $\Y\X = -i\Z$. We spell out the simplification rules in \Cref{fig:app-rules} of the appendix, using $\Z\X \leadsto i\Y$ to represent simplification. Since these predicates are trivial to normalize, we will often treat $\Z\X$ and $i\Y$ as identical in our presentation.

We will discuss the soundness of \Cref{rule-imp_mul} in more detail in \S{\ref{section:soundness}}; for now we can continue our derivation above and we get
\[
\Z \otimes \I \cap \Z \otimes \Z \Rightarrow \Z \otimes \I \cap \Z\Z \otimes \Z\I \leadsto \Z \otimes \I \cap \I \otimes \Z
\]
Hence we can derive the desired triple
\[
\hoarepre{CNOT}{\Z \otimes \I \cap \I \otimes \Z}{\Z \otimes \I \cap \I \otimes \Z}
\]

\subsection{Example: Deutsch's Algorithm}
\label{sec:deutsch}

A complete list of our rules and grammar for Pauli predicates is given in~\Cref{fig:types,fig:app-rules}. Here, we show an example of how we can apply these rules to make non-trivial judgments about quantum programs.

Many quantum circuits introduce ancillary qubits that perform some classical computation and are then discarded in a basis state.
Several efforts have been made to verify this behavior: The Quipper~\cite{Green2013} and Q\#~\cite{Svore2018} languages allow us to \emph{assert} that ancillae are separable and can be safely discarded, while \qwire allows us to manually verify this~\cite{Rand2018}. More recently, Silq~\cite{Silq} allows us to define ``qfree'' functions that never put qubits into a superposition. We can use our logic to avoid this restriction and automatically guarantee ancilla correctness by showing that the ancillae are discarded as they satisfy the predicate $\Z$ and, more specifically, are separable from the rest of the system.

A simple example to demonstrate this ability to safely discard auxiliary qubits is Deutsch's algorithm~\cite{Deutsch85}. Given a function $f : \{0, 1\} \rightarrow \{0, 1\}$, the algorithm uses oracle access to $f$ and a single auxiliary qubit to determine if $f$ has a constant value or is balanced.

\begin{figure}[ht]
\begin{center} \begin{tikzpicture}[scale=0.7,y=-1cm]

    \draw (-0.4,2) node[left] {$x \; \ket{0}$} -- (8, 2);
    \draw [double] (8, 2) -- (10, 2) node[right] {$b$};
    \draw (-0.4,3) node[left] {$y \; \ket{0}$} -- (10, 3) node[right] {$\ket{0}$};
    \draw (10,2.75) -- (10,3.25) ;

	\unitary{H}{2}{2}
	\unitary{X}{0.5}{3}
	\unitary{H}{2}{3}

    \oracle{$U_f$}{3.5}{2}{5}{3}

	\unitary{H}{6.5}{2}
	\draw [dotted] (7.25, 1.5) -- (7.25, 3.5);
	\meas{8.5}{2}
	\unitary{H}{8.5}{3}

\end{tikzpicture} \end{center}

\caption{Deutsch's algorithm to check if $f:\{0, 1\} \rightarrow \{0, 1\}$ is constant or balanced}
\label{fig:deutsch}
\end{figure}

We want to show that the qubit $y$ is never entangled with qubit $x$ despite the application of the oracle $U_f : \ket{x}\ket{y} \rightarrow \ket{x}\ket{y \oplus f(x)}$. In this case, it would be safe to discard qubit $y$ just after the dotted line in~\Cref{fig:deutsch} (i.e., even before measurement destroys any hypothetical entanglement).

Before analyzing the circuit, consider the possible behaviors for $f$. Acting on a single bit, one can conclude that $f(x) \in \{0, 1, x, (1-x)\}$. It is easy to derive the oracle application by a case-by-case analysis:
\begin{align*}
    \mathtt{U_f \ 1 \ 2} = \begin{cases}
    \mathtt{I \ 2} & \text{if } f(x) = 0 \\
    \mathtt{X \ 2} & \text{if } f(x) = 1 \\
    \mathtt{CNOT \ 1 \ 2} & \text{if } f(x) = x \\
    \mathtt{X \ 1; \ CNOT \ 1 \ 2; \ X \ 1} & \text{if } f(x) = 1 - x.
    \end{cases}
\end{align*}

Clearly, the first two cases are not entangling gates. The last case is a 0-controlled $\CNOT$ which is equivalent to the $\CNOT$ gate for our purposes. Hence, we analyze the circuit for the case where $\mathtt{U_f \ 1 \ 2} \equiv \mathtt{CNOT \ 1 \ 2}$. The precondition for this circuit is two qubits initialized in the computational basis, or equivalently $\Z \otimes \I \cap \I \otimes \Z$.
Instead of building a full proof tree for \coqe{deutsch}, we'll use the standard approach for Hoare logic, in which we write an annotated program with the intermediate predicates in between the commands. For convenience, we'll do our derivations in parallel, rather than combining them with the $\cap$ rule at the end:
\begin{coq}
deutsch :=
  {I ⊗ Z ∩ Z ⊗ I}
  X 2;        (* y is set to 1 *)
  {-I ⊗ Z ∩ Z ⊗ I}
  H 1;
  {-I ⊗ Z ∩ X ⊗ I}
  H 2;
  {-I ⊗ X ∩ X ⊗ I}
  U_f 1 2;   (* U_f = CNOT *)
  {-I ⊗ X ∩ X ⊗ X}
  H 1;
  {-I ⊗ X ∩ Z ⊗ X} => 
  {I ⊗ -X ∩ -Z ⊗ I}
\end{coq}

The last line is obtained through the implication rule from~\Cref{rule-imp_mul} and the distributivity of negation. This is precisely what Deutsch's algorithm is supposed to produce -- two separable qubits (implied by $\A \otimes \I$, see \cref{sec:separability}), the first of which is an eigenstate of $-Z$, corresponding to a $\ket{1}$ qubit. Note that we could return the second qubit to $\Z$ by applying a Hadamard and an $X$ gate. However, as we have verified that the second qubit is unentangled with the first, we may freely discard it and optimize away the final $\mathtt{H \ 2; X \ 2}$.

\section{Normal Forms}
\label{sec:normal}

Our intersection predicates have the property that there exists a canonical form with which to describe them. This allows us to verify the equality of intersection predicates by verifying the equality of their canonical forms.
The canonical form we use is inspired by the \emph{row echelon form} of a matrix, in which every row has its first nonzero term before any subsequent row. We translate this into $\I$ being $0$ and further impose that $\X \prec \Y \prec \Z \prec \I$ so $\I \otimes \X$ precedes $\I \otimes \Z$. Further, an intersection predicate involving commuting, independent terms can be viewed as a matrix with each term corresponding to a row and each column corresponding to a qubit.
Then, in the canonical form, any column contains at most one \X or \Y, and any column without an \X or \Y has at most one \Z. The unique \X, \Y or \Z in each column will be called the \X, \Y or \Z pivot.

The implication rule for intersection predicates (\Cref{rule-imp_mul}),  $\A \cap \B \Leftrightarrow \A \cap \A\B$, will be useful to reduce the predicates to their canonical forms.
Given an $n$-qubit intersection predicate with $m$ independent terms $\pred{A}_{(1)} \cap \ldots \cap \pred{A}_{(m)}$, apply~\Cref{alg:normalize}:

\begin{algorithm}
\caption{The normalization procedure for $n$-qubit intersection predicates}
\begin{algorithmic}
    \State \textbf{Input:} $n$-qubit intersection predicate with $m$ terms $\A = \pred{A}_{(1)} \cap \ldots \cap \pred{A}_{(m)}$ 
    \State $\P \gets []$ \Comment{$\P$ is an ordered list of n-qubit Pauli predicates}
    \For{$i$ in $1..n$} \Comment{Apply the Pivot Rule to each qubit}
        \State pivot $\gets$ False %
        \For{$j$ in $1..m$} \Comment{check for X pivot}
            \If{$A_{(j)} \notin \P$ and $\A_{(j)}[i] \in \{\X, \Y\}$}
                \State pivot $\gets$ True \Comment{X pivot found}
                \State $\P \gets \P + A_{(j)}$
                \For{$k$ in $(j+1)..m$}
                    \State \algorithmicif\ $\A_{(k)}[i] \in \{\X, \Y\}$ \algorithmicthen\ $\A_{(k)} \gets \A_{(j)} \A_{(k)}$
                \EndFor
            \EndIf
            \State \algorithmicif\ pivot \algorithmicthen\ break
        \EndFor
        \State \algorithmicif\ pivot \algorithmicthen\ continue \Comment{If X pivot is found, move to the next qubit}
        \For{$j$ in $1..m$} \Comment{check for Z pivot}
            \If{$A_{(j)} \notin \P$ and $\A_{(j)}[i] \in \{\Z\}$}
                \State pivot $\gets$ True \Comment{Z pivot found}
                \State $\P \gets \P + A_{(j)}$
                \For{$k$ in $(j+1)..m$}
                    \State \algorithmicif\ $\A_{(k)}[i] =\Z$ \algorithmicthen\ $\A_{(k)} \gets \A_{(j)} \A_{(k)}$
                \EndFor
            \EndIf
            \State \algorithmicif\ pivot \algorithmicthen\ break
        \EndFor
    \EndFor
    \FullComment{Return the intersection as the predicates in $\P$ followed by the remaining predicates sorted by the lexicographic ordering $\X \prec \Y \prec \Z \prec \I$}
    \State $\A \gets$ toIntersection($\P$ + sorted($\A \setminus \P$))
\end{algorithmic}
\label{alg:normalize}
\end{algorithm}
Notice that each predicate can be added to $\P$ at most once, and this, along with the ordering of predicates, makes the canonical form unique. Further, by viewing the $i$th predicate in $\P$ to be the $\X, \Y$ or $\Z$ pivot for qubit $i$, this procedure is functionally equivalent to row echelonization for matrices and can be computed using $O(n^3)$ operations. %

Note that this normalization process is functionally similar to the reduction of a stabilizer code to a standard form; see, for example, \cite[\S{10.5.7}]{Nielsen2010}. 

\begin{example}[Applying~\Cref{alg:normalize}]
\label{eg:normalize}
Consider the following predicate:
\[ \X \otimes \X  \otimes \I \cap \Z  \otimes \Z  \otimes \I \cap \Z  \otimes \Z \otimes \Z. \]
Conveniently, the first term contains an \X on qubit $1$. However, no subsequent terms have an \X on this qubit, so we move on to qubit $2$.

For the second qubit, no \X's remain in pivot terms, so we take the \Z in the second term, $\Z \otimes \Z \otimes \I$. The third term is now rewritten as:
\begin{align*}
& (\Z \otimes \Z \otimes \Z)(\Z  \otimes \Z \otimes \I)  \\
&\ = \ \Z\Z \otimes \Z\Z \otimes \Z\I \\
&\ = \ \I \otimes \I \otimes \Z.
\end{align*}
For the last qubit, there is only one term with a \X or \Z in the third position, so we are done.

The entire procedure yields the normal form: %
\[\X \otimes \X  \otimes \I \cap \Z  \otimes \Z  \otimes \I \cap \I \otimes \I \otimes \Z. \]
\end{example}

An essential property of the normal form is that it is oblivious to the original ordering of the terms. For instance, in \Cref{eg:normalize}, if we had first swapped the $2^{\text{nd}}$ and $3^{\text{rd}}$ terms then $\Z \otimes \Z \otimes \Z$ would have been the pivot for the second qubit and we would replace the $3^{\text{rd}}$ term with $\I \otimes \I \otimes \Z$. We would then use the third term as our pivot, replacing the second term ($\Z \otimes \Z \otimes \Z$) with $\Z \otimes \Z \otimes \I$.
The entire procedure yields $\X \otimes \X  \otimes \I \cap \Z  \otimes \Z  \otimes \I \cap \I  \otimes \I \otimes \Z$ just as before.

Since all of our normalization operations are justified by the one implication rule, associativity, and commutativity rules, the following simplification rule is admissible:
\begin{equation}\label{rule-normalization}
        \inferrule*[Right=Norm]{\hoare{A}{g}{B}}{\hoare{A}{g}{\normal(B)}}
\end{equation}
where $\textsc{norm}$ is our normalization procedure. This is the rule we will apply in practice before making separability judgments.

We remark that the Pivot Rule can, in fact, be applied to any ordering of the $n$ qubits. For ease of notation, when the  qubit ordering is $i, \ldots, n, 1, \ldots, i-1$, we denote the normalization  as the $\textsc{norm}_i$ rule that satisfies:
\begin{equation}\label{rule-normalization_i}
        \inferrule*[Right=Norm\ensuremath{_i}]{\hoare{A}{g}{B}}{\hoare{A}{g}{\normal_i(B)}}
\end{equation}
Note that if \A fully determines the state, then $\normal_i(\pred{A}) = \normal_j(\pred{A})$ for all $i,j$ and this follows from \cite[Lemma 1 and Theorem 3]{aaronson2004improved}). However, if \A is underdetermined, then $\normal_i(\pred{A})$ will differ based on the pivot $i$.
We will use this variant of the rule with underdetermined states to determine post-measurement predicates, as discussed in \Cref{sec:stab_meas}.

\section{Separability}
\label{sec:separability}

In this section, we present the first application of our Hoare style logic---the ability to make judgments on whether a given sub-system is separable from the remainder of the system. We start with determining whether a single qubit is separable before moving to multi-qubit sub-systems.

\subsection{Single qubit separability}
\label{sec:single-sep}

Following the core semantics that a predicate refers to the $+1$-eigenstate of its semantic operator, we first prove a statement about the separable eigenstates of some operators. For notational simplicity, we state the following proposition with a focus on the first qubit; however, the result holds for any operator of the form $I^{k-1} \otimes U \otimes I^{n-k}$

\begin{proposition}\label{prop:separability}
     For any  $2 \times 2$ unitary, Hermitian matrix $U$, the eigenstates of $U \otimes I^{n-1}$ are all vectors of the form $\ket{u} \otimes \ket{\psi}$ where $\ket{u}$ is an eigenstate of $U$ and $\ket{\psi} \in \complex^{2^{n-1}}$ is an arbitrary state.
\end{proposition}

\begin{proof}
    Let $\ket{\phi}$ be the $\lambda$-eigenstate and $\ket{\phi^{\bot}}$ be the $(-\lambda)$-eigenstate of $U$ where $\lambda \in \{1, -1\}$. Note that $\{\ket{\phi}, \ket{\phi^{\bot}}\}$ forms a single-qubit basis. %

    First, consider states of the form $\ket{\gamma} = \ket{u} \otimes \ket{\psi}$ where $\ket{u} \in \{\ket{\phi}, \ket{\phi^{\bot}}\}$ and $\ket{\psi} \in \mathbb{C}^{2^{n-1}}$. Clearly,
    \[(U \otimes I^{n-1}) \ket{\gamma} = (U \otimes I^{n-1}) (\ket{u} \otimes \ket{\psi}) = (U \ket{u} ) \otimes \ket{\psi} = \lambda_u \ket{u} \otimes \ket{\psi}. \]
    Hence, every state of the form of $\ket{\gamma}$ is an eigenstate of $U \otimes I^{n-1}$. Additionally, note that by similar reasoning, for every separable state $\ket{\gamma} = \ket{v} \otimes \ket{\psi}$, where $\ket{v} \notin \{\ket{\phi}, \ket{\phi^{\bot}}\}$, is not an eigenstate of $U \otimes I^{n-1}$.

    Now we show that any state not in this separable form cannot be an eigenstate of $U \otimes I^{n-1}$. By way of contradiction assume that $\ket{\delta}$ is an eigenstate of $U \otimes I^{n-1}$ with $(U \otimes I^{n-1}) \ket{\delta} = \mu \ket{\delta}$. Expand
    $$\ket{\delta} = \alpha \ket{\phi}\otimes \ket{\psi_1} + \beta\ket{\phi^{\bot}}\otimes \ket{\psi_2}$$
    where $\ket{\psi_1},\ket{\psi_2} \in \mathbb{C}^{2^{n-1}}$. Then we compute
    \begin{eqnarray*}
    (U \otimes I^{n-1}) \ket{\delta} & = & \alpha(U \ket{\phi}) \otimes \ket{\psi_1} +  \beta(U \ket{\phi^{\bot}}) \otimes\ket{\psi_2} \\
    & = & \lambda \alpha \ket{\phi}\otimes \ket{\psi_1} - \lambda \beta \ket{\phi^{\bot}}\otimes\ket{\psi_2}\\
    & = & \mu \alpha \ket{\phi} \otimes \ket{\psi_1} + \mu \beta \ket{\phi^{\bot}} \otimes \ket{\psi_2} %
    \end{eqnarray*}
    where we have used that $\ket{\phi}$ and $\ket{\phi^{\bot}}$ are the $+\lambda$ and $-\lambda$ eigenvalues of $U$ respectively. As the components of the expansion are orthogonal to each other, $\mu$ must satisfy:
    \[
    \mu \alpha = \lambda \alpha \text{ and } \mu \beta = - \lambda \beta.
    \]
    Since $U \otimes I^{n-1}$ is unitary, $\lambda \not= 0$ and we either have (i) $\alpha = 0$, $\mu = - \lambda$, and $\ket{\delta} = \ket{\phi^{\bot}}\otimes \ket{\psi_2}$ or (ii) $\beta = 0$, $\mu = +\lambda$, and $\ket{\delta} = \ket{\phi}\otimes\ket{\psi_1}$. In either case, $\ket{\delta}$ has a separable form as claimed.
\end{proof}

As every Pauli matrix is both Hermitian and unitary,  combining~\Cref{prop:eigenstate,prop:separability}, we immediately obtain the following corollary:
\begin{corollary}
\label{cor:1qubit-sep}
Every term of the form $\I^{i-1}~\otimes~\pred{U}~\otimes~\I^{n-i}$ is separable, for any $U$ $\in$ $\{\pm X, \pm Y, \pm Z\}$. That is, the $i^{th}$ factor satisfies the predicate $\pred{U}$ and is not entangled with the rest of the system.
\end{corollary}

Following Gottesman's notation, let $\pred{U}_k$ be the $n$-qubit predicate where the $k^{th}$ factor satisfies the single-qubit predicate $\pred{U}$ and is separable from the rest of the system. For example, the predicate $\X_1 \equiv \X \otimes \I$ describes the set of two separable qubits where the first qubit is in the \X eigenstate\footnote{For precision, we should say $\X_{1 \in [2]}$ to indicate the size of the system, but this will always be clear from the context.}. The two-qubit product state $\ket{0} \otimes \ket{+}$ satisfies the intersection predicate $\Z_1 \cap \X_2$ to signify that each qubit is separable from the other. \Cref{cor:1qubit-sep} then justifies the following separability-based simplification rules:
\begin{align*}
    \forall \B, \ \I^{k-1} \otimes \B \otimes \I^{n-k} \Leftrightarrow \B_k
\end{align*}

Since $A$ being separable in a larger system $B$ implies that the rest of $B$ is separable from $A$, we can add the following rules for distributing separability judgments across intersections:
\begin{align*}
    \text{If } \T[k] \in \{\B, \I\}, \ \B_k \cap \T \Leftrightarrow \B_k \cap \T_{[n]\setminus \{k\}} &
\end{align*}

Using these rules, we can re-write $\X_1 \cap (\X \otimes \Z \otimes \Z)$ as
$\X_1 \cap (\Z \otimes \Z)_{2,3}$.

\subsection{Multi-qubit separability}
\label{sec:multi-sep}

While~\Cref{cor:1qubit-sep} can be used to identify if a single qubit is separable from the rest of the system, we would also like to make judgments about a multi-qubit subsystem $S \subset \{1, \ldots, n\}$ being separable from $\{1, \ldots, n\} \setminus S$. Generalizing~\Cref{prop:separability} will help us in this regard. However, we only generalize it for the case when the unitaries are Pauli matrices (rather than generic Hermitian matrices). The following fact about Pauli matrices adapted from \textcite[Prop. 10.5]{Nielsen2010} by setting $n \leftarrow k, k \leftarrow 0$, will be useful for the proof.

\begin{fact}
\label{fact:kPauli}
For $k$-qubit Pauli matrices $V \in \{\pm I, \pm X, \pm Y, \pm Z\}^{k}$ such that $V \neq \pm I^k$, the eigenvalue $\lambda \in \{-1, 1\}$ has an eigenspace of dimension $2^{k-1}$. For $k$ independent, pairwise commuting $k$-qubit Pauli matrices $U_{(1)}, \ldots U_{(k)}$, the joint eigenspace for an eigenvalue tuple $(\lambda_1, \ldots, \lambda_k)$ has dimension $1$. %
\end{fact}

This fact can be intuitively argued from the observation that each Pauli matrix divides the total $2^k$-dimensional Hilbert space into two sub-spaces of the same dimension, each corresponding to the $+1$ or $-1$ eigenvalues. The $k$-tuple then identifies a $1$-dimensional subspace at the intersection of the corresponding eigenspaces for $U_{(1)}, \ldots, U_{(k)}$.

\Cref{fact:kPauli} requires the $k$-qubit Pauli matrices to be independent and pairwise commuting. It is straightforward to check independence by ensuring that multiplying any combination of the $k$ matrices together does not yield the $\I^k$ term. Pairwise commutativity can also be directly determined using the following fact, which follows immediately from the fact that distinct Pauli matrices from $\{X,Y,Z\}$ anticommute.

\begin{fact}
\label{fact:commuting}
Given $A = A_1 \otimes \dots \otimes A_k$ and $B = B_1 \otimes \dots \otimes B_k$, where the $A_i$s and  $B_i$s are Pauli matrices, $A$ and $B$ will commute precisely when there is an even number of positions where $A_i$ and $B_i$ are both from $\{X, Y, Z\}$ but $A_i \neq B_i$.
\end{fact}

Given two multi-qubit Pauli matrices $U$ and $V$, we refer to the intersection of their eigenspaces by $U \cap V$ for ease of notation. We can now state the conditions under which a set of Pauli operators could correspond to a separable sub-system.

\begin{proposition}
     \label{prop:ksep}
     For independent, pairwise commutative, non-identity $k$-qubit matrices $U_{(1)}, \ldots, U_{(k)}\allowbreak \in \{\pm I, \pm X, \pm Y, \pm Z\}^{k}$ such that $U_{(i)} \cap U_{(j)} \neq \emptyset$  for all $i \neq j$, the eigenstate of $(I^{n-k} \otimes U_{(1)}) \cap \ldots \cap (I^{n-k} \otimes U_{(k)})$ are all vectors of the form $\ket{u} \otimes \ket{\Psi}$ where $\ket{\Psi}$ is an eigenstate of $U_{(1)}, \ldots, U_{(k)}$.
\end{proposition}

\begin{proof}
    First, it is clear that any state of the form $\ket{u} \otimes \ket{\Psi}$ where $\ket{\Psi}$ is an eigenstate of $U_{(1)}, \dots, U_{(k)}$ is an eigenstate of $I^{n-k} \otimes U_{(1)}, \ldots, I^{n-k} \otimes U_{(k)}$. This implies that it is also an eigenstate of $(I^{n-k} \otimes U_{(1)}) \cap \ldots \cap (I^{n-k} \otimes U_{(k)})$.

    To prove the inverse direction, assume by way of contradiction that there exists an entangled $n$-qubit state $\ket{\delta}$ that is an eigenstate of $(I \otimes U_{(i)}$ with eigenvalue $\lambda_i \in \{-1, 1\}$ for each $i \in \{1, \ldots, k\}$. %
    Let the $n$-qubit state $\ket{\delta}$ be written in terms of its \emph{Schmidt} (singular value) decomposition across the $(n-k, k)$ qubit bipartition as
    \begin{align*}
        \ket{\delta} = \sum_{j=1}^K \alpha_j \ket{\phi_j} \otimes \ket{\gamma_j}
    \end{align*}
    where $\{\ket{\phi_i}\}_i$ and $\{\ket{\gamma_i}\}_i$ are orthonormal vectors in each of their respective subsystems.
    \begin{align}
        \forall i \in \{1, \ldots, k\} \quad (I \otimes U_{(i)}) \ket{\delta} & = \sum_j \alpha_j (I \ket{\phi_j}) \otimes (U_{(i)} \ket{\gamma_j}) \notag \\
        & = \lambda_i \sum_j \alpha_j \ket{\phi_j} \otimes \ket{\gamma_j} \notag \\
        & = \sum_j \alpha_j \ket{\phi_j} \otimes (\lambda_i \ket{\gamma_j}) \notag \\
        &\Rightarrow \forall i, j, \quad U_{(i)} \ket{\gamma_j} = \lambda_i \ket{\gamma_j}. \notag \\
        &\Rightarrow \forall i, j, \quad \lambda_i U_{(i)} \ket{\gamma_j} = \ket{\gamma_j} \quad \text{Since, } \lambda_i \in \{-1, 1\}.
        \label{eq:eigen}
    \end{align}
As $\{\ket{\gamma_i}\}_i$ forms a set of orthonormal vectors, the span of these vectors is contained in the eigenspace for the eigenvalue tuple $(+1, +1, \ldots, +1)$ corresponding to $\lambda_1 U_{(1)}, \ldots, \lambda_k U_{(k)}$ respectively. Additionally, when $U_{(i)}$ is a $k$-qubit Pauli matrix, $\lambda_i U_{(i)}$ is also in $\{\pm I, \pm X, \pm Y,\allowbreak \pm Z\}^{k}$.
Then, from~\Cref{fact:kPauli}, the joint eigenspace for the all-$1$s tuple has dimension $1$. Specifically, there exists only a single $\ket{\gamma}$ that satisfies~\Cref{eq:eigen}. Hence, $K = 1$ contradicting the assumption that $\ket{\delta}$ is entangled across the $(n-k, k)$ qubit bi-partition.
\end{proof}

Extending the $\pred{U}_i$ notation to the multi-qubit setting where  $K \subset \{1, \dots, n\}$ and $0 < |K| < n$, let $(\pred{U})_K$ be the predicate such that qubits in $K$ are separable from the $\{1, \dots, n\} \setminus K$ sub-system.
Formally, we define
 $\pred{U}_K := \left(\cap_{j} \pred{U}_{(j)} \right)_K$ where each $\pred{U}_{(j)}$ is a non-trivial $|K|$-qubit non-identity Pauli string. %

For example, consider a $2$-qubit predicate $(\X \otimes \X \cap \Z \otimes \Z)$ whose joint eigenspace is spanned by the two maximally entangled Bell states $\{\ket{\Phi^+}, \ket{\Psi^-}\}$. In an $n$-qubit state satisfying this predicate on the first and third qubits, these qubits being maximally entangled are disjoint from the rest of the system, and hence, they satisfy the predicate
$$(\X \otimes \X \cap \Z \otimes \Z)_{1,3} = (\X \otimes \I \otimes \X \otimes \I^{n-3}) \cap (\Z \otimes \I \otimes \Z \otimes \I^{n-3}).$$
If the second and fourth qubits are similarly entangled, the system satisfies the predicate $(\X \otimes \X \cap \Z \otimes \Z)_{1,3} \cap (\X \otimes \X \cap \Z \otimes \Z)_{2,4}$. This idea to gather the nontrivial factors within a subsystem is not unique to our work and has been previously employed by \textcite{Honda2015} to determine the entangled components in his static analysis framework. %

Combining this representation with~\Cref{prop:eigenstate,prop:ksep}, we obtain the following corollary:

\begin{corollary}
\label{cor:ksep}
Let $K \subset  \{1, \dots, n\}$ with $|K| = k$ and $\overline{K} := \{1, \dots, n\} \setminus K$. Every intersection predicate that contains the term $\bigcap_{j=1}^k \left(\pred{U}_{(j)} \otimes \I^{n-k} \right)$ where each of the $\pred{U}_{(j)}$s acts on $K$, is pairwise commuting and independent as a subterm is separable across the bipartition $(K, \overline{K})$. That is, the factors in $K$ are separable from the $\overline{K}$ subsystem.
\end{corollary}

Given an  $n$-qubit intersection predicate in normal form (\cref{sec:normal}) with $m$ independent terms $\A_{(1)} \cap \ldots \cap \A_{(m)}$, finding if a subsystem of qubits $K \subset \{1, \dots, n\}$ with $|K| = k < m$, is separable from the remaining system can be determined in a straightforward way.
We first verify that every qubit in $K$ has a pivot; otherwise, some qubit in $K$ has $\I$ in all terms, and we can conclude that $K$ is not separable from the remaining system. If every qubit in $K$ has a pivot, we run the following procedure:
\begin{itemize}
    \item Let $\A_{(j_1)}, \ldots, \A_{(j_k)}$ be the $k$ distinct terms that have the pivots for qubits in $K$.
    \item For each $i = 2 \ldots k$, check that $\A_{(j_i)}$ commutes with $\A_{(j_1)}$ using~\Cref{fact:commuting}.\footnote{This ensures that in all terms where the qubits in $\overline{K}$ are pivots, the terms have an $\I$ for all qubits in $K$.} %
    \item For each $i = 1 \ldots k$, check that the term $\A_{(j_i)}$ has an $\I$ for every qubit $\ell \in \overline{K}$.
\end{itemize}

\Cref{cor:ksep} justifies our multi-qubit separability rules when $S = \{j_1, \ldots, j_k\} \subset [n]$
\begin{align*}
        & \B \cap \pred{T}_{(1)}\cap \ldots \cap \pred{T}_{(k)}  \Leftrightarrow \B_{\overline{S}} \cap \left(\Ct_{(1)} \cap \ldots \cap \Ct_{(k)}\right)_S, \\
        & \text{where } \forall_{j\in[k]} \ \pred{T}_{(j)}[S] = \Ct_{(j)} \text{, } \forall_{j \in [k]} \ \pred{T}_{(j)}[\overline{S}] = \I^{n-k} \text{, and } \B[S] = \I^k.
\end{align*}

Here, for an $n$-qubit term $\A$ and a subset of indices $J \subset [n]$, we use $\A[J]$ to denote the factors of $\A$ in $J$. For instance, when $J = \{1, 3\}$ and $\A = \X \otimes \Y \otimes \Z$, $\A[J] = \X \otimes \Z$.

\begin{example}
Continuing from~\Cref{eg:normalize}, consider the predicate
\[\X \otimes \X  \otimes \I \cap \Z  \otimes \Z  \otimes \I \cap \I \otimes \I \otimes \Z. \]
As $(\X \otimes \X)$ and $(\Z \otimes \Z)$ are two independent and commuting operators, the first two terms with $\I$ on the third qubit ensure that we can apply~\Cref{cor:ksep} to determine that the first two qubits are separable from the third. We can write this as:
\[ (\X \otimes \X \cap \Z  \otimes \Z)_{1, 2} \cap \Z_3.\]
\end{example}

\subsection{Application: GHZ state, Entanglement Creation and Disentanglement}
\label{sec:GHZ}

To demonstrate how we can track the possibly entangling and disentangling properties of the $\CNOT$ gate, we can look at the example of creating the GHZ state $\frac{1}{\sqrt{2}}(\ket{000} + \ket{111})$ starting from $\ket{000}$ and then disentangling it. A similar example was considered by \textcite{Honda2015} to demonstrate how his system can track when $\CNOT$ displays either its entangling or disentangling behavior. One crucial difference is that Honda uses the denotational semantics of density matrices which, in practice, would scale poorly with the size of the program being validated.
Our approach is closer to that of \textcite{Perdrix2007,Perdrix2008} in terms of design and scalability but capable of showing separability where the prior systems could not.

We will consider the following GHZ program acting on the initial state $\Z_1 \cap \Z_2 \cap \Z_3$. We first follow the derivation for $\Z_1$:

\begin{coq}
  Definition GHZ :=
    {Z${}_1$} =>
    {Z ⊗ I ⊗ I}
    H 1;
    {X ⊗ I ⊗ I}
    CNOT 1 2;
    {X ⊗ X ⊗ I}
    CNOT 2 3
    {X ⊗ X ⊗ X}
\end{coq}

Repeating the derivation for $\Z_2$ and $\Z_3$, we obtain: %
\begin{align*}
\hoare{\Z_2}{\mathtt{GHZ}}{\Z ⊗ \Z ⊗ \I} \\
\hoare{\Z_3}{\mathtt{GHZ}}{\I ⊗ \Z ⊗ \Z}
\end{align*}

If we now apply \coqe{CNOT 3 1}, we get the following specifications:
\begin{align*}
&\hoare{\Z_1}{\mathtt{GHZ}}{\X ⊗ \X ⊗ \X}\ \mathtt{CNOT~3~1} \ \{\pred{\I ⊗ \X ⊗ \X\}} \\
&\hoare{\Z_2}{\mathtt{GHZ}}{\Z ⊗ \Z ⊗ \I}\ \mathtt{CNOT~3~1} \ \{\pred{\Z ⊗ \Z ⊗ \Z\}} \\
&\hoare{\Z_3}{\mathtt{GHZ}}{\I ⊗ \Z ⊗ \Z}\ \mathtt{CNOT~3~1} \ \{\pred{\I ⊗ \Z ⊗ \Z\}} \\
\end{align*}

If we want to analyze the output of this program on $\pred{Z_1 \cap Z_2 \cap Z_3}$, we can apply the following normalization steps (the first and second row serving as the first and second pivots):
\begin{align*}
&\{ \I ⊗ \X ⊗ \X ∩
    \Z ⊗ \Z ⊗ \Z ∩
    \I ⊗ \Z ⊗ \Z \} \Rightarrow\\
&\{ \I ⊗ \X ⊗ \X ∩
 \Z ⊗ \Z ⊗ \Z ∩
 \I\Z ⊗ \Z\Z ⊗ \Z\Z \}\
 \leadsto \\
& \{
 \I ⊗ \X ⊗ \X ∩
 \Z ⊗ \Z ⊗ \Z ∩
 \Z ⊗ \I ⊗ \I \}
.\end{align*}
Recognizing that the first qubit can now be separated from the other two, we obtain $\Z_1 \cap (\X \otimes \X \cap \Z \otimes \Z)_{2, 3}$, that is, a \Z qubit and a Bell pair.

Returning to the unnormalized program, if we finally apply \coqe{CNOT 3 2}, we get
\begin{align*}
&\hoare{\Z_1}{\mathtt{GHZ;~CNOT~3~1;~CNOT~3~2}}{\I ⊗ \I ⊗ \X} \\
&\hoare{\Z_2}{\mathtt{GHZ;~CNOT~3~1;~CNOT~3~2}}{\Z ⊗ \Z ⊗ \I}  \\
&\hoare{\Z_3}{\mathtt{GHZ;~CNOT~3~1;~CNOT~3~2}}{\I ⊗ \Z ⊗ \I}
\end{align*}
to which we can apply the intersection rule and single-qubit separability rules to obtain
\[
\hoare{\Z_1 \cap \Z_2 \cap \Z_3}{\mathtt{GHZ;~CNOT~3~1;~CNOT~3~2}}{\Z_1 \cap \Z_2 \cap \X_3}
\]
showing that the whole procedure moves the \X generated by the initial Hadamard gate to the third position.

\subsection{Evolving Separable States}\label{section:evolving_seperability}

What happens when we apply an $\mathtt{H}$ gate to the second qubit of a system satisfying $\Z_1 \cap \Z_2 \cap \X_3$? Since $\mathtt{H~2}$ only affects the second qubit, intuitively we want to obtain $\Z_1 \cap \X_2 \cap \X_3$. We can guarantee this with the following rules, which allow us to apply unitary transformations to separability predicates that contain the set of manipulated qubits (here $\Z_2$) or that are entirely disjoint from it (here $\Z_1$ and $\X_3$):

\begin{equation}\label{rule-embed_and_disjoint}
    \inferrule*[Right=embed]{\hoare{A}{g}{B}}{\hoare{A_K}{g_K}{B_K}} \qquad\qquad\qquad\qquad \inferrule*[Right=disjoint]{\mathit{dom}(\mathtt{g}) \cap K = \emptyset}{\hoare{A_K}{g}{A_K}}
\end{equation}
where $\mathtt{g_K}$ is $\mathtt{g}$ with indices replaces by the corresponding indices of $K$ and the domain, or set of free variables, $\mathit{dom}(\mathtt{g})$ is the set of qubits that $\mathtt{g}$ acts upon.

Hence, we can derive:

\begin{equation*}
    \inferrule*{\hoare{Z}{H~1}{X}}{\hoare{Z_2}{H~2}{X_2}}
\end{equation*}
and 
\begin{equation*}
    \inferrule*{\{2\} \cap \{1,3\} = \emptyset}{\hoare{Z_1 \cap X_3}{H~2}{Z_1 \cap X_3}}
\end{equation*}

By contrast, had we applied a $\CNOT$ from qubit $1$ or $3$ to qubit $2$ (a potentially entangling operation), neither of these rules would apply.

\section{Measurement}
\label{sec:stab_meas}

It is challenging to turn Gottesman's semantics for measurement into an efficient deductive system because it looks at its operation on all the basis states rather than simply the evolution of a single Pauli operator. Namely, it adds significant computational complexity, while our prior deductive rules were linear in the number of qubits. Nonetheless, our normalization in~\Cref{sec:normal} parallels that in the stabilizer formalism, and the action of measurement on stabilizer groups is well-understood~\cite{Gottesman1998}. This produces a method for inferring the postconditions for measurement that is quadratic in the number of qubits in the worst case~\cite{aaronson2004improved}. %

\subsection{Disjunctive predicates}

Before discussing how we check measurement, it helps to consider how we can represent post-measurement states. Unlike unitary gate application, which is deterministic, implying that each input predicate has a specified output predicate, not all measurements have deterministic outcomes. While we do not want to use our logic system to verify the probabilities of measurement outcomes, it would be useful to be able to compute the possible post-measurement states for the system. With this, we could still track how the system evolves with subsequent operations depending on the measurement results.

We use a disjunction connective, $\A \uplus \B$, to denote that the system either satisfies the predicate $\A$ or predicate $\B$. This can be treated as representing an ensemble of states each of which satisfies either $\A$ or $\B$, though we will tend to treat this as a simple disjunction over pure states. %
We show how to use this in the context of measurement with this simple example.

\begin{example}[Measuring $\ket{+}$]
\label{eg:meas_plus}
Consider the single qubit in the $\ket{+}$ state on which a computational basis measurement is performed. 
The outcome has equal probability to be $\ket{0}$ or $\ket{1}$, which correspond to the predicates $\Z$ and $-\Z$ respectively. 
\end{example}

Following~\Cref{eg:meas_plus} and noting that $\ket{+i} \in \pred{Y}$ behaves similar to $\ket{+}$, the inference rules for measuring a single qubit are
\begin{align*}
    \axiom{\hoare{X}{\MEAS}{Z \uplus -Z}},\ 
    \axiom{\hoare{Y}{{\MEAS}}{Z \uplus -Z}},\ 
    \axiom{\hoare{Z}{{\MEAS}}{Z}},\ 
    \axiom{\hoare{I}{{\MEAS}}{Z \uplus -Z}}.
\end{align*}
By the \textsc{scale} rule, we also have \hoare{-Z}{\MEAS}{-Z}, while measuring a negative \X, \Y or \I qubit behaves the same as measuring an \X, \Y, or \I (by double negation). 

Applying a gate to a disjunction distributes across the disjunction, and each component term evolves separately. This gives the following rule for disjunctions:
\begin{equation}\label{rule-union}
    \inferrule*[Right=\ensuremath{\uplus}]{\hoare{A}{g}{A'} \and \hoare{B}{g}{B'}}{\hoare{\A \uplus \B}{g}{\A' \uplus \B'}}
\end{equation} 
Note that when \A and \B are disjoint, \pred{A'} and \pred{B'} will be disjoint as well.

As with intersections, the ordering of the terms does not matter, with commutativity and associativity holding for disjunctions as well, leading to the implication
\begin{equation}\label{rule-union-perm}
    \biguplus_i \A_i \Rightarrow \biguplus_i \A_{P(i)}
\end{equation}  
for any permutation $P$. Furthermore, the idempotent rule also holds for disjunctions with the implication 
\begin{equation}\label{rule-union-drop}
    \A \uplus \A \Rightarrow \A
\end{equation}

\addtocounter{example}{-1}
\begin{example}[Continued]
    After measuring the qubit in the $\ket{+}$ state, let us apply the Hadamard gate. Using the disjunction rule and the action of $H$, the post-measurement state will satisfy $\X \uplus -\X$. Measuring the state again and applying the \emph{\MEAS} rule to each term in the disjunction (via the $\uplus$ distributivity rule above), the post measurement state will satisfy $(\Z \uplus -\Z) \uplus (-\Z \uplus \Z)$. Using the permutation and idempotent rules for $\uplus$, this simplifies the post-measurement predicate to $(\Z \uplus -\Z)$ thereby satisfying the condition that the state must either be $\ket{0}$ or $\ket{1}$.
\end{example}

Extending disjunctive predicates to multiple qubits in the context of measurement is not as straightforward as combining the single-qubit measurement inference with the tensor rule. Measurement on one qubit can have non-trivial effects on all qubits in such a way that we can only make the trivial inference about the post-measurement state on the remaining $n-1$ qubits i.e., they are +1 eigenstates of $I \otimes \ldots \otimes I$. Hence, our inference rule for a multi-qubit predicate $\pred{A}$ where the $k^\text{th}$ qubit is measured becomes
\begin{equation}\label{rule-multiqubit_measurement_weakened}
    \axiom{\hoare{A}{\MEAS \ k}{Z_k \uplus -Z_k}}.
\end{equation}

In the case of intersection predicates, measurement of a single qubit can potentially affect each term in the intersection and describing the post-measurement is a more involved process as discussed in the next section.

\subsection{Conjunctive predicates and measurement}
\label{sec:meas_conj}

For ease of exposition, we will assume that we are performing a z-basis measurement on the $k^\text{th}$ qubit of an $n$ qubit system. In \S{\ref{sec:normal}} we introduced a normalization procedure for intersection predicates. There, we constructed the normal form by examining each qubit in some specified ordering, say $i=1, \dots, n$, and applied the Pivot Rule (see Step 2 in \S{\ref{sec:normal}}). As there, let us write $\pred{A}_{(1)} \cap \cdots \cap \pred{A}_{(m)}$ for the pre-measurement predicate. Furthermore, we  set the qubit ordering as $i = k, \ldots, n, 1, \ldots, k-1$ and apply the $\textsc{norm}_k$ rule (\Cref{rule-normalization_i}) to obtain the normalized predicate $\normal_k \left(\pred{A}_{(1)} \cap \cdots \cap \pred{A}_{(m)}\right)$. For the post-measurement predicate, note that measuring qubit $k$ will force the outcome to be random, as in the case of multi-qubit predicates. Determining exactly how the intersection terms evolve is described below.

\begin{algorithm}
\caption{Determining post-measurement predicates for $n$-qubit intersection predicates on measuring qubit $k$}
\label{alg:meas_k}
\begin{algorithmic}
    \State \textbf{Input:} An $n$-qubit intersection predicate with $m$ terms $\A = \pred{A}_{(1)} \cap \ldots \cap \pred{A}_{(m)}$; the measured qubit index $k \in [1, \ldots, m]$ 
    \State $\Ut \gets \pred{A}_{(1)} \cap \ldots \cap \pred{A'}_{(m)} = \normal_k(\A)$ %
    \For{$j$ in $1..m$}
        \If{$\pred{A'}_{(j)}[k] \in \{\X, \Y\}$} \Comment{if qubit $k$ has an $X$ or $Y$ pivot}
            \State $\Ut' \gets \Ut \setminus \pred{A'}_{(j)}$
            \State \Return $(\Z_k \cap \Ut') \uplus (-\Z_k \cap \Ut')$
        \EndIf
    \EndFor
    \State \Return $(\Z_k \cap \Ut') \uplus (-\Z_k \cap \Ut')$ \Comment{if qubit $k$ has a $Z$ pivot or no pivot at all.}
\end{algorithmic}
\end{algorithm}

Observe that qubit $k$ may not have a pivot only when $m < n$; that is, the predicate is underdetermined. This will commonly be the case when dealing with the physical qubit predicates for stabilizer-based error-correcting codes. On the other hand, if qubit $k$ has a $\Z$ pivot, but is separable from the rest of the system (see \S{\ref{sec:separability}}), then the outcome is deterministic. Furthermore, since separability ensures that either $\Z_k$ or $-\Z_k$ is already contained in the intersection, the post-measurement state satisfies $\Ut$. Finally, observe that, by construction, the measured qubit in each case satisfies the predicate $\Z$ or $-\Z$ and is separable from the rest of the system.

With the procedure described above, we can now state the inference rules for an $n$-qubit multi-qubit predicate $\pred{A}$ where the $k^\text{th}$ qubit is measured 
\begin{align}\label{rule-multiqubit_measurement}
    & \axiom{\hoare{A}{\MEAS\ k}{Z_k \cap \pred{B} \uplus -Z_k \cap \pred{B}}} \text{~~~~where} \\
    \nonumber &\qquad \text{(i) } \B = \normal_k{(\A)} \text{ if $\exists j, \A_{(j)}[k] = \Z$ or $\forall j, \A_{(j)}[k] = \I$ };\\
    \nonumber &\qquad \text{(ii) } \B = \normal_k{(\pred{A})} \setminus \A_{(j)} \text{ if $\exists j, \A_{(j)}[k] \in \{\X, \Y\}$};
\end{align}

\begin{example}
    Continuing our analysis of the GHZ state from \Cref{sec:GHZ}, the circuit \texttt{GHZ} has the postcondition
$$(\X\otimes \X\otimes \X) \cap (\Z \otimes \Z \otimes \I) \cap (\I \otimes \Z \otimes \Z).$$
To compute the output of \emph{\texttt{GHZ; MEAS 1}}, we enact the above program. Fortunately, our intersection already has the requisite form, with the first term being the only one with an $\X$ in the initial position. We remove the term $(\X \otimes \X \otimes \X)$ and replace the intersection with
$$(\Z_1 \cap (\Z \otimes \Z \otimes \I) \cap (\I \otimes \Z \otimes \Z)) \uplus (-\Z_1 \cap (\Z \otimes \Z \otimes \I) \cap (\I \otimes \Z \otimes \Z))$$

We normalize each branch of the disjunction as per~\Cref{sec:normal} to obtain
$$(\Z_1 \cap \Z_2 \cap (\I \otimes \Z \otimes \Z)) \uplus (-\Z_1 \cap -\Z_2 \cap (\I \otimes \Z \otimes \Z)).$$

Finally, the last term can also be simplified to give
$(\Z_1 \cap \Z_2 \cap \Z_3) \uplus (-\Z_1 \cap -\Z_2 \cap -\Z_3).$ 
\end{example}

\subsection{Example: Analyzing teleportation}

In this section, we analyze teleporting a qubit in the $\ket{i}$ state using the Bell state $\Phi^+$. We start with the three qubits represented by the predicate $\Y_1 \cap (\X \otimes \X)_{2, 3} \cap (\Z \otimes \Z)_{2, 3}$.

\begin{figure}[ht]

\begin{center} \begin{tikzpicture}[scale=0.7,y=-1cm]

    \draw (-0.4,1) node[left] {$\ket{q}$} -- (4, 1);
    \draw [double] (4, 1) -- (8, 1); %
    \draw (-0.4,2) -- (4, 2);
    \draw [double] (4, 2) -- (8, 2); %
    \draw (-0.4,3) -- (8, 3) node[right] {$\ket{q}$};
    \draw (-0.4,2.5) node[left] {$\ket{\Phi^+}$};  %

        \cnot{1}{1}{2}
	\unitary{H}{2}{1}

        \meas{4}{1}
        \meas{4}{2}

        \cnot{6}{2}{3}
        \cz{7}{1}{3}

\end{tikzpicture} \end{center}

\caption{Quantum Teleportation}
\label{fig:teleport}
\end{figure}

\begin{align*}
&\qquad \Y_1 \cap (\X \otimes \X)_{2, 3} \cap (\Z \otimes \Z)_{2, 3}\\
\Rightarrow &\qquad (\Y \otimes \I \otimes \I)\cap (\I \otimes \X \otimes \X) \cap (\I \otimes \Z \otimes \Z)\\
\mathtt{CNOT~1~2;} &\qquad (\Y \otimes \X \otimes \I)\cap (\I \otimes \X \otimes \X) \cap (\Z \otimes \Z \otimes \Z)\\
\mathtt{H~1;} &\qquad (-\Y \otimes \X \otimes \I)\cap (\I \otimes \X \otimes \X) \cap (\X \otimes \Z \otimes \Z)\\
\Rightarrow&\qquad (\Z \otimes \Y \otimes \Z) \cap (\I \otimes \X \otimes \X) \cap (\X \otimes \Z \otimes \Z)\\
\mathtt{MEAS~1;} &\qquad (\Z \otimes \Y \otimes \Z) \cap (\I \otimes \X \otimes \X) \cap (\Z_1 \uplus -\Z_1)\\
\Rightarrow &\qquad (\Z_1 \cap (\Y \otimes \Z \cap \X \otimes \X)_{2,3}) \uplus (-\Z_1 \cap (-\Y\otimes \Z \cap \X \otimes \X)_{2,3})\\
\Rightarrow &\qquad (\Z_1 \cap (\Z \otimes \Y \cap \X \otimes \X)_{2,3}) \uplus (-\Z_1 \cap (-\Z \otimes \Y \cap \X \otimes \X)_{2,3})\\
\mathtt{MEAS~2;} &\qquad (\Z_1 \cap (\Z_2 \uplus -\Z_2) \cap (\Z \otimes \Y)_{2,3}) \uplus (-\Z_1 \cap (\Z_2 \uplus -\Z_2) \cap (-\Z \otimes \Y)_{2,3})\\
\Rightarrow &\qquad (\Z_1 \cap \Z_2 \cap \Y_3) \uplus  (-\Z_1 \cap \Z_2 \cap -\Y_3)  \uplus (\Z_1 \cap -\Z_2 \cap -\Y_3) \uplus (-\Z_1 \cap -\Z_2 \cap \Y_3)
\end{align*}

To show that $\CNOT$  and $\CZ$ implement the desired classical Pauli correction, we want to prove:
\begin{equation*}
    \begin{array}{rcl@{\qquad\qquad}rcl}
    \{\pred{Z_1 \cap P_2}\} & \CNOT & \{\pred{Z_1 \cap P_2}\} & \{\pred{Z_1 \cap P_2}\} & \CZ & \{\pred{Z_1 \cap P_2} \} \\
    \{\pred{-Z_1 \cap X_2}\} & \CNOT & \{\pred{-Z_1 \cap X_2}\} & \{\pred{-Z_1 \cap X_2}\} & \CZ & \{\pred{-Z_1 \cap -X_2} \} \\
    \{\pred{-Z_1 \cap Y_2}\} & \CNOT & \{\pred{-Z_1 \cap -Y_2}\} & \{\pred{-Z_1 \cap Y_2}\} & \CZ & \{\pred{-\Z_1 \cap -Y_2} \} \\
    \{\pred{-Z_1 \cap Z_2}\} & \CNOT & \{\pred{-Z_1 \cap -Z_2}\} & \{\pred{-Z_1 \cap Z_2}\} & \CZ & \{\pred{-Z_1 \cap Z_2} \} \\
    \end{array}
\end{equation*}
These are straightforward to show using their corresponding rules; for instance, for $\CNOT$:
\begin{align*}
    &\qquad -\Z_1 \cap \Y_2\\
    \Rightarrow & \qquad(-\Z\otimes \I) \cap (\I \otimes \Y)\\
    \CNOT; & \qquad (-\Z\otimes \I) \cap (\Z \otimes \Y)\\
    \Rightarrow & \qquad (-\Z\otimes \I) \cap (\I \otimes -\Y) & \text{(intersection multiplication rule)} \\
    \Rightarrow & \qquad -\Z_1 \cap -\Y_2
\end{align*}

We can now complete the proof:
\begin{align*}
&\qquad (\Z_1 \cap \Z_2 \cap \Y_3) \uplus  (-\Z_1 \cap \Z_2 \cap -\Y_3)  \uplus (\Z_1 \cap -\Z_2 \cap -\Y_3) \uplus (-\Z_1 \cap -\Z_2 \cap \Y_3)\\
\mathtt{CNOT~2~3;} &\qquad (\Z_1 \cap \Z_2 \cap \Y_3) \uplus  (-\Z_1 \cap \Z_2 \cap -\Y_3)  \uplus (\Z_1 \cap -\Z_2 \cap \Y_3) \uplus (-\Z_1 \cap -\Z_2 \cap -\Y_3)\\
\mathtt{CZ~1~3;} &\qquad (\Z_1 \cap \Z_2 \cap \Y_3) \uplus  (-\Z_1 \cap \Z_2 \cap \Y_3)  \uplus (\Z_1 \cap -\Z_2 \cap \Y_3) \uplus (-\Z_1 \cap -\Z_2 \cap \Y_3)\\
\Rightarrow &\qquad ((\Z_1 \cap \Z_2) \uplus  (-\Z_1 \cap \Z_2)  \uplus (\Z_1 \cap -\Z_2) \uplus (-\Z_1 \cap -\Z_2)) \cap \Y_3
\end{align*}

\section{Soundness and Expressivity}\label{section:soundness}

In this section, we prove the soundness of our logic and discuss what sort of states can be characterized by the class of predicates we have presented.

\subsection{Soundness}

In order to state soundness, we first give a standard denotational semantics to our programs.
We can treat unitary applications as transformations of a vector state, where only measurement is treated non-deterministically.
\begin{align*}
    \denote{\mathtt{U~i}} \ket\psi &= \{ U_i \ket{\psi} \} \\ %
    \denote{\mathtt{CNOT~i~j}} \ket\psi &= \{ \mathit{CNOT}_{ij} \ket{\psi} \} \\
    \denote{\mathtt{MEAS~i}} (\alpha \ket{0}_i \ket{\psi_0} + \beta \ket{1}_i \ket{\psi_1}) &= \begin{cases}
        \{ \alpha\ket{0}_i \ket{\psi_0} \} & \text{ if } \beta = 0 \\
        \{ \beta\ket{1}_i \ket{\psi_1} \} & \text{ if } \alpha = 0 \\
        \{ \ket{0}_i \ket{\psi_0}, \ \ket{1}_i \ket{\psi_1} \} & \text{otherwise}
    \end{cases}  \\
    \denote{\mathtt{p1 ; p2}}\ket\psi &= \bigcup_{\ket\psi' \in \denote{p_1}\ket\psi} \denote{p_2} \ket\psi' 
\end{align*}

Here we use $U_i$ to represent applying the single-qubit unitary $U$ to the $i^\text{th}$ qubit of the system and similarly for $\mathit{CNOT}$. Measurement acts as the identity when the measured qubit is $\ket 0$ or $\ket{1}$. Otherwise, it takes a state to one of two rescaled output states; in our system we are not concerned with the associated probabilities. Note that again $\ket{x}_i\ket{\psi}$ represents a Kronecker product with $\ket{x}_i$ in the  $i^\text{th}$ location, following Ying~\cite{ying2012}. The semantics on vectors lift to sets over vectors in the expected way.

\begin{theorem}[Soundness]\label{thm:soundness}
  The Hoare-like logic given in the previous sections (\Cref{eq:typeH} through \Cref{rule-multiqubit_measurement}) is sound with respect to the $+1$ eigenvector semantics. That is, for any \texttt{p}, $P$, and $Q$, if we can derive \hoare{P}{p}{Q} then $\mathtt{p}$ takes any $+1$ eigenvector of $P$ to a $+1$ eigenvector of $Q$. More formally, if  \hoare{P}{p}{Q} then:
  \[
  P\ket\psi = \ket\psi \implies Q (\denote{p} \ket\psi) = \denote{p} \ket\psi
  \]
\end{theorem}

\begin{proof}

~

\paragraph{Unitary rules} The soundness of our system derives from~\Cref{prop:eigenstate} which states that a unitary $U$ takes a +1-eigenstate of $A$ to a +1-eigenstate of $B$ only if $UAU^{\dagger}=B$. For example, the rules for $H$ and $S$ (\Cref{rule:HS}) are justified by the equalities $HXH^\dag = Z$, $HZH^\dag = X$, $SXS^\dag = Y$, and $SZS^\dag = Z$, which correspond to their actions on the eigenstates $\ket{0}, \ket{1}, \ket{+},$ and $\ket{-}$. The rules for $\mathit{CNOT}$ (\Cref{rule:cnot}) follow similarly by matrix multiplication, and we define other gates in terms of the Clifford gates and derive their logical triples using the rules below.

For the scaling, multiplication and tensor rules, we can assume that $p$ is unitary, since the scalar, matrix, and tensor products are only applicable to Pauli predicates (the operations are not defined on the $\pred{A} \uplus \pred{B}$ induced by non-trivial measurement.) 
For the scaling rule (\cref{eq:rule-scale}), if $p A p^{\dagger} = B$ for unitary $p$, then $p (cA) p^{\dagger} = c pAp^{\dagger} = c B$ for any complex number $c$. 
For multiplication (\cref{eq:rule-mul}), we start with $pAp^{\dagger} = B$ and $pCp^{\dagger} = D$. Then, $p (AC) p^{\dagger} = p A p^{\dagger}p C p^{\dagger}$ for unitary $p$ and the latter simplifies to $B D$ making the rule sound. 

For the tensor rule (\cref{rule-tensor1}), note that applying a single-qubit unitary to specific qubits in an $n$-qubit system is equivalent to applying a larger unitary that is $I^{\otimes (i-1)} \otimes U_i \otimes I^{\otimes (n-i)}$. Hence this leaves all factors in the tensor predicate $\T$ unchanged except the $i^{th}$ one which satisfies $UAU^{\dagger} = B$. The 2-qubit unitary $U_{ij}$ acts as prescribed on qubits $i$ and $j$ and acts as $I$ on the other qubits. Then, focusing only on the terms associated with qubits $i$ and $j$, and using $U (A_i \otimes B_j) U^{\dagger} =(C_i \otimes D_j)$, ~\cref{rule-tensor2} is sound.

The identity rule (\cref{rule-identity}) is trivial since $UIU^{\dagger} = UU^{\dagger} = I$ for any unitary $U$. 

\paragraph{Structural rules} We can now turn to the structural rules, no longer assuming unitarity. For the sequencing rule (\cref{rule-seq}), if a function $p_1$ takes a +1-eigenstate of $A$ to a +1-eigenstate of $B$ and $p_2$ takes any +1-eigenstate of $B$ to a +1-eigenstate of $C$ then it follows immediately that $p_2 \circ p_1$ takes any eigenstate of $A$ to an eigenstate of $C$. This justifies the rule for $\mathtt{p_1 ; p_2}$.

For the intersection rule (\cref{rule-intersection}), we interpret $(\A \cap \Ct)(\ket{\psi})$ to say that $\ket{\psi}$ is +1-eigenstate of both operators $A$ and $C$. If the triples $\hoare{A}{p}{B}$ and $\hoare{C}{p}{D}$ hold then $\ket{\phi} = \denote{p} \ket{\psi}$ is a $+1$-eigenstate of both $B$ and $D$. Hence, $(\B \cap \D) (\ket{\phi})$ must hold. The same reasoning applies to the union rule (\cref{rule-union}), with disjunction replacing conjunction.

For the consequence rule (\cref{rule-consequence}), assume that $\A' \Rightarrow \A$, $\hoare{A}{p}{B}$, and $\B \Rightarrow \B'$. Now suppose that $\ket{\psi}$ is a +1 eigenstate of $A'$. Then $\ket{\psi}$ is a +1 eigenstate of $A$ (by the first implication) and hence $\denote{p} \ket{\psi}$ is an eigenstate of both $B$ (by the Hoare triple) and $B'$ (by the second implication). Hence, the triple $\hoare{A'}{p}{B'}$ holds.

One nontrivial implication rule involving intersections is in~\cref{rule-imp_mul}, which states that $\A \cap \B \Leftrightarrow \A \cap \A\B$. The argument that this rule is sound is slightly subtle: if $(\A \cap \B) (\ket\psi)$ then $\ket\psi$ is a $+1$-eigenstate of both (multi-qubit) Pauli operators $A$ and $B$. This also implies that $\ket\psi$ is a $+1$-eigenstate of the operator $AB$. The converse is also true (as $A^2 = I$), and so semantically, $\pred{A \cap B}$ and $\pred{A \cap AB}$ refer to the same set of states. The other rules in \cref{rule-imp_base} follow directly from the definition of subsets, as do the corresponding union rules (\Cref{rule-union-perm,rule-union-drop}).

The soundness of the normalization rules~(\cref{rule-normalization,rule-normalization_i}) follows from the soundness of the above implication rule. Recall that the procedure in~\cref{sec:normal} repeatedly replaces the term $\A_{(k)}$ with $\A_{(j)}\A_{(k)}$ in the Pivot Rule. Otherwise, the only modifications involve dropping repeated terms and reordering the terms as needed, which are sound due to weakening and the other implication rules.

The validity of our separability judgments follows from the fact that if every +1-eigenstate of an operator is separable, then the operator must satisfy additional structural constraints. These constraints are captured by~\Cref{prop:separability,prop:ksep} in~\Cref{sec:separability}. Typically, a predicate is  normalized before it is determined to be separable. Then, using the soundness of normalization rules from above, the whole procedure is sound.

\paragraph{Measurement} The semantics of the post-measurement evolution of the +1-eigenstates of predicates in the Heisenberg formalism~\cite[Section 7]{Gottesman1998} forms the basis of the soundness for our measurement rules. A similar argument using the terminology of our logic is sketched below. Let $\pred{A}_{(1)} \cap \cdots \cap \pred{A}_{(m)} (\ket{\psi})$. Then, by measuring qubit $i$ in the $Z$ basis, the post-measurement state can be described as $\Pi_Z^\pm \ket{\psi}$ depending on the $\pm 1$ outcome where $\Pi_Z^\pm = \frac{1}{2} (I \pm Z_i)$. When the outcome is +1 (arguments for the -1 outcome are similar), $\Z_i (\Pi_Z^+ \ket{\psi})$ implying that $\Z_i$ should be in the post-measurement predicate. Additionally, $\forall k \in [1, \ldots, m]$ if $A_{(k)}$ commutes with $Z_j$, they share simultaneous +1-eigenstates i.e., $\Z_i \cap \pred{A}_{(k)} (\Pi_Z^+ \ket{\psi})$. Conversely, if there exists $\ell$ such that $A_{(\ell)}$ anti-commutes with $Z_j$, then $\Pi_Z^+ \ket{\psi}$ is not a +1-eigenstate of $A_{(\ell)}$ i.e., $\pred{A}_{(\ell)}$ cannot be present in the post-measurement predicate. Going back to~\Cref{alg:meas_k}, recall that the $\textsc{norm}_i$ rule in Step 1 ensures that there is at most one term in the pre-measurement predicate that anti-commutes with $Z_i$ as it chooses qubit $i$ as a pivot for exactly one term. Step 2 checks if this term anti-commutes with $Z_i$ and removes it from the post-measurement predicate. Step 3 ensures all terms commuting with $Z_j$ are included in the post-measurement predicate. Finally, the disjunction $\pred{Z}_i \cap \B \uplus -\pred{Z}_i \cap \B$ is used to capture the nondeterminism of measurement. Hence, our measurement rules are shown to be sound. 
\end{proof}

\subsection{Expressivity}

We can also show that our predicates are sufficiently expressive to precisely describe any stabilizer state.

\begin{proposition}
\label{prop:prepare-Clifford}
    Let $\ket{\psi}$ be an $n$-qubit state. Then there exist $n$ commuting, independent $n$-qubit Pauli operators {$P_{(1)}, \ldots, P_{(n)}$} such that $\pred{P}_{(1)} \cap \cdots \cap \pred{P}_{(n)}(\ket{\psi})$ if and only if $\ket{\psi}$ can be prepared from $\ket{0 \dots 0}$ with a Clifford circuit. That is, for any set of commuting Pauli operators $P_{(1)}, \dots, P_{(n)}$ there exists a Clifford operator $C: \Z_j \to \pred{P}_{(j)}$ for $j \in \{1,\dots,n\}$.
\end{proposition}

\begin{proof}
    Given $\pred{P}_{(1)} \cap \cdots \cap \pred{P}_{(n)}(\ket{\psi})$, we can view the predicate as an $n$-qubit stabilizer with $n$ generators given by the commuting, independent Pauli operators ${P}_{(1)}, \ldots, {P}_{(n)}$. Then, we can follow the procedure by Cleve and Gottesman \cite[Sections III, IV]{cleve1997efficient}, with $d=n$, to efficiently construct a Clifford circuit from the normal form of the intersection predicate that prepares $\ket{\psi}$ from $\ket{0\ldots0}$. Conversely, recall that any Clifford circuit can be expressed as a sequence of $S$, $\mathit{CNOT}$ and $H$ gates and our language can express how any Pauli operator evolves under these gates (\Cref{rule:HS,rule:cnot}). Then, given a Clifford circuit $C$ to prepare the $n$-qubit state $\ket{\psi}$ from the $\ket{0\ldots0}$ state, we start with the assumption $\Z_1 \cap \ldots \cap \Z_n(\ket{0\ldots0})$, and evolve the predicate according to the triple for each gate in $C$ to obtain the consequent $\pred{P}_{(1)} \cap \cdots \cap \pred{P}_{(n)}(\ket{\psi})$. Since the operators $Z_1, \ldots, Z_n$ are mutually commuting and independent, the unitarity of $C$ ensures that the operators $P_{(1)}, \ldots, P_{(n)}$ satisfy the same properties. 
\end{proof}

Recall that the joint eigenspace of $n$ independent, commuting Pauli operators has dimension $1$ (see \Cref{fact:kPauli}), implying that $\ket{\psi}$ will be uniquely determined by $\pred{P}_{(1)} \cap \cdots \cap \pred{P}_{(n)}$ up to a global phase.

\section{Application: Error-correcting Codes}
\label{sec:qecc}

We can also use our logic system to analyze error-correcting codes. In this section, we consider the $7$-qubit code \textcite{steane96qecc}. Recall that the Steane code encodes a $7$ physical qubits into a single logical qubit and has the ability to detect errors on $2$ qubits and correct all single-qubit errors. The stabilizers and logical operators for the Steane code are generated by:
\[\begin{array}{r@{\:}c@{\:}lr@{\:}c@{\:}l @{\qquad} r@{\:}c@{\:}l}
    g_1 &=& {IIIXXXX} & g_2 &=& {IXXIIXX} & \overline{X} &=& {XXXXXXX} \\
    g_3 &=& {XIXIXIX} & g_4 &=& {IIIZZZZ} & \overline{Z} &=& {ZZZZZZZ} \\
    g_5 &=& {IZZIIZZ} & g_6 &=& {ZIZIZIZ} & \overline{Y} &=& {-YYYYYYY}
\end{array} \]
where concatenation refers to the tensor product.

We use $\pred{I_L}$, to mean any state that lies in the codespace of the Steane code:
$$\pred{I_L} := \pred{g_1 \cap \ldots \cap g_6}.$$ 
This will also be helpful in realizing other logical predicates.

We realize $\ket{0}$ in this setup through the logical state $\ket{0_L}$ defined by projecting the all $0$s state using the stabilizer generators of the code:
\begin{align*}
    \ket{0_L} \propto \frac{1}{2^6} \Pi_{i=1}^6 (I + g_i) \ket{0000000}
\end{align*}
By virtue of being the logical $0$ state, it should also be stabilized by the logical-$\sigma_Z$ equivalent $\overline{Z}$. In other words, $\ket{0_L}$ is uniquely stabilized by $g_1, \ldots, g_6$ and $\overline{Z}$. In our system, this means that $\pred{Z_L}(\ket{0_L})$ where $\pred{Z_L}$ is the $7$-term intersection predicate
\begin{align}
\begin{aligned}
    \pred{Z_L} & := \pred{g_1} \cap \ldots \cap \pred{g_6} \cap \pred{\overline{Z}} \label{eq:log-Z}\\
    & = \I \otimes \I \otimes \I \otimes \X \otimes \X \otimes \X \otimes \X \\
    & \cap \ldots \cap \Z \otimes \I \otimes \Z \otimes \I \otimes \Z \otimes \I \otimes \Z \\
    & \cap \Z \otimes \Z \otimes \Z \otimes \Z \otimes \Z \otimes \Z \otimes \Z
\end{aligned}
\end{align}

By a similar argument, $\pred{X_L}(\ket{+_L})$ where $\pred{X_L} = \pred{g_1} \cap \ldots \cap \pred{g_6} \cap \pred{\overline{X}}$. All states in the Steane code space are stabilized by $g_1, \ldots, g_6$. Then, we can associate the following predicate to the logical Steane code space as
\begin{align*}
    \pred{I_L} = \pred{g_1} \cap \ldots \cap \pred{g_6} \qquad \text { and } \qquad \pred{Z_L} = \pred{I_L} \cap \pred{\overline{Z}}; \quad \pred{X_L} = \pred{I_L} \cap \pred{\overline{X}}.
\end{align*}
Being consistent with the equation above, we can conclude that $\pred{Y_L}(\ket{+i_L})$ where %
\begin{align*}
    \pred{Y_L} := \pred{I_L} \cap \pred{\overline{Y}} \qquad \text{ where } \qquad \overline{Y} = i \overline{X}\,\overline{Z}.
\end{align*}
Let $U_L$ be a unitary that performs a logical operation on the encoded qubit, that is, it preserves the Steane codespace $\hoare{I_L}{U_L}{I_L}$. If $U_L$ satisfies the triples $\hoare{X_L}{U_L}{I_L \cap \overline{P}}$ and $\hoare{Z_L}{U_L}{I_L \cap \overline{Q}}$ where $P,Q \in \{\pm{X},\pm{Y},\pm{Z}\}$ then $U_L$ necessarily takes $\overline{X}$ to $\overline{P}$ and $\overline{Z}$ to $\overline{Q}$. Letting $R = iPQ \in \{\pm{X},\pm{Y},\pm{Z}\}$, using our multiplication rule, we obtain $\hoare{I_L \cap \overline{Y}}{U_L}{I_L \cap \overline{R}}$, meaning that $\hoare{Y_L}{U_L}{R_L}$. Hence the following rule is admissable:
\begin{equation*}
\inferrule*[right=\ensuremath{Y_L}]{\hoare{X_L}{U}{A_L} \\ \hoare{Z_L}{U}{B_L}}{\hoare{Y_L}{U}{(iAB)_L}}
\end{equation*}
Note that the preconditions imply that $U$ is a logical Clifford on the encoded qubit.

\begin{figure}[ht!]
\begin{center} \begin{tikzpicture}[scale=0.5]

    \draw (-0.4,7) node[left] {$y$} -- (10.5,7);
    \draw (-0.4,6) node[left] {$x_1$} -- (10.5,6);
    \draw (-0.4,5) node[left] {$x_2$} -- (10.5,5);
    \draw (-0.4,4) node[left] {$x_3$} -- (10.5,4);
    \draw (-0.4,3) node[left] {$x_4$} -- (10.5,3);
    \draw (-0.4,2) node[left] {$x_5$} -- (10.5,2);
    \draw (-0.4,1) node[left] {$x_6$} -- (10.5,1);
    \draw [decorate,decoration={brace,amplitude=10pt}, xshift=-4pt,yshift=0pt]
(11,7.5) -- (11,0.5) node [black,midway,xshift=15pt] {$a$};

     \draw[densely dotted] (1,8) -- (1,0.5);
     \draw[densely dotted] (3,8) -- (3,0.5);
     \draw[densely dotted] (5,8) -- (5,0.5);
     \draw[densely dotted] (7,8) -- (7,0.5);
     \draw[densely dotted] (9,8) -- (9,0.5);
     \draw (0.5,8) node {1};
     \draw (2,8) node {2};
     \draw (4,8) node {3};
     \draw (6,8) node {4};
     \draw (8,8) node {5};

    \draw (2,7) -- (2,5);
 	\ctrl{2}{7}
 	\qnot{2}{6}
 	\qnot{2}{5}
 	\unitary{H}{2}{3}
 	\unitary{H}{2}{2}
 	\unitary{H}{2}{1}

 	\draw (4,7) -- (4, 1);
 	\ctrl{4}{1}
 	\qnot{4}{4}
 	\qnot{4}{6}
 	\qnot{4}{7}

    \draw (6,7) -- (6, 2);
 	\ctrl{6}{2}
 	\qnot{6}{4}
 	\qnot{6}{5}
 	\qnot{6}{7}

    \draw (8,6) -- (8, 3);
 	\ctrl{8}{3}
 	\qnot{8}{4}
 	\qnot{8}{5}
 	\qnot{8}{6}

\end{tikzpicture} \end{center}
\caption{Encoding circuit for the Steane $[[7,1,3]]$ code}
\label{fig:encode}
\end{figure}

Consider the Steane code unitary encoding circuit $\mathtt{Enc\!-\!St}$ given in~\Cref{fig:encode} where a data qubit $y$ is converted into a logical qubit $a$. By construction, it takes $\ket{a} \otimes \ket{000000} \to \ket{a_L}$ for $a \in \{0, 1, +, -\}$. Consider $a = 0$ for instance. Then we can describe the action of $\mathtt{Enc\!-\!St}$ in our system as follows:
\begin{enumerate}
    \item start with the precondition $\Z_{y} \cap \Z_{x_1} \cap \ldots \cap \Z_{x_6}$;
    \item apply each gate from~\Cref{fig:encode} using the axioms for $H$ and $\mathit{CNOT}$;
    \item normalize the output.
\end{enumerate}
A straightforward computation (an exercise left to the reader) will show that we indeed obtain $\normal(\pred{Z_L})$ as the output. Extending this argument, we characterize $\mathtt{Enc-St}$ as:
\begin{align*}
    & \hoare{Z_y \cap Z_{x_1} \cap \ldots \cap Z_{x_6}}{\mathtt{Enc\!-\!St}}{\normal(Z_L)}\\
    & \hoare{X_y \cap Z_{x_1} \cap \ldots \cap Z_{x_6}}{\mathtt{Enc\!-\!St}}{\normal(X_L)}
\end{align*}

Another application of our system is verifying that a specified circuit on the encoding qubits applies the claimed operation on the logical qubits. We illustrate this by verifying the \emph{transversality} of a gate with respect to a code. A logical operation $U_L$ is transversal with respect to a code if it can be implemented as $U$ applied in parallel on each physical qubit of the code block, making transversal gates the easiest to implement in the context of error-correction. For instance, it is straightforward to verify that $\mathtt{H_L} := \mathtt{H \ y; \ H \ {x_1}; \ H \ {x_2}; \ H \ {x_3}; \ H \ {x_4};}\allowbreak \mathtt{H \ {x_5}; \ H \ {x_6}}$ is transversal for the Steane code i.e.,
\begin{align}
    \hoare{\X_L}{H_L}{\Z_L} \qquad \text{and } \qquad \hoare{\Z_L}{H_L}{\X_L}
\end{align}
Clearly, $\hoare{\pred{g_1} \cap \pred{g_2} \cap \pred{g_3}}{H_L}{\pred{g_4} \cap \pred{g_5} \cap \pred{g_6}}$ and vice-versa. Hence, $\hoare{I_L}{H_L}{I_L}$. Further, $\hoare{\overline{X}}{H_L}{\overline{Z}}$ and vice-versa. Therefore, ${H_L}$ takes $\pred{Z_L} = \pred{I_L} \cap \pred{\overline{Z}}$ to $\pred{I_L} \cap \pred{\overline{X}} = \X_L$ and vice-versa. %

In a similar vein, we can prove that the operation $U =  \mathtt{S \ y; \ S \ {x_1}; \ S \ {x_2}; \ S \ {x_3}; \ S \ {x_4};}\allowbreak \mathtt{S \ {x_5}; \ S \ {x_6}}$ is not the logical-$S$ gate ${S_L}$. Firstly, the triple for the logical-$S$ should satisfy
\begin{align*}
    \hoare{\Z_L}{S_L}{\Z_L} \qquad \text{and } \qquad \hoare{\X_L}{S_L}{\Y_L}
\end{align*}
Now, $\hoare{\pred{g_4} \cap \pred{g_5} \cap \pred{g_6} \cap \pred{\overline{Z}}}{U}{\pred{g_4} \cap \pred{g_5} \cap \pred{g_6} \cap \pred{\overline{Z}}}$ as the $S$ acts only on $\Z$ or $\I$. In the case of $\{\pred{g_1},  \pred{g_2}, \pred{g_3}, \pred{\overline{X}}\}$, the $\X$s are converted to $\Y$s such that the predicates are changed on output. Clearly, $\hoare{\overline{X}}{U}{Y \otimes Y \otimes Y \otimes Y \otimes Y \otimes Y \otimes Y}$; note that this postcondition is equal to $-\pred{\overline{Y}}$.
Let us take $\pred{g_1} \cap \pred{g_4}$ to see how the remaining stabilizers would evolve:
\begin{equation*}
    \inferrule*[Right=cons]{\hoare{g_1 \cap g_4}{U}{(\I \otimes \I \otimes \I \otimes \Y \otimes \Y \otimes \Y \otimes \Y) \cap g_4}}
    {\inferrule*{\hoare{g_1 \cap g_4}{U}{(\I \otimes \I \otimes \I  \otimes \Y \otimes \Y \otimes \Y \otimes \Y)g_4 \cap g_4}}{\hoare{g_1 \cap g_4}{U}{g_1 \cap g_4}}}
\end{equation*}
\begin{align*}
    \{\pred{g_1} \cap \pred{g_4}\}\ \mathtt{U}\ & \{ \I \otimes \I \otimes \I \otimes \Y \otimes \Y \otimes \Y \otimes \Y) \cap \pred{g_4} \} \Rightarrow \\
    & \{ \I \otimes \I \otimes \I  \otimes \Y \otimes \Y \otimes \Y \otimes \Y) \pred{g_4} \cap \pred{g_4} \} \leadsto \\
    & \{ \pred{g_1} \cap \pred{g_4} \}
\end{align*}

Extending this reasoning to $(\pred{g_2 \cap g_5})$ and $(\pred{g_3 \cap g_6})$, $\hoare{I_L}{U}{I_L}$. Putting the pieces together, $\hoare{Z_L}{U}{Z_L}$ but $U$ takes $\pred{X_L}$ to $\pred{I_L} \cap - \pred{\overline{Y}} = -\Y_L$. By contrast, defining
\begin{align*}
    \mathtt{S_L} := \mathtt{Z \ y; \ S \ y ; \ Z \ x_1; S \ {x_1} \ Z \ {x_2}; \ S \ {x_2}; \ Z \ x_3; \ S \ {x_3}; \ Z \ x_4; \ S \ {x_4}; \ Z \ x_5; \ S \ {x_5}; \ Z \ x_6; \ S \ {x_6}}
\end{align*}
gives us the desired behavior.

We would also like to show that the $T$-gate is not transversal for the Steane code. However, with $T$ not being a Clifford gate, we find that Pauli predicates are insufficient to describe it fully. For this, we consider the additive extension to our logic system in subsequent sections and demonstrate this in~\Cref{eg:t-transversal}.

\subsection{Logical Multi-qubit predicates}

Extending the discussion on logical qubits and quantum error correcting codes to multi-qubit logical states requires us to add some additional rules to our system.

\paragraph{Separable states}
Describing states where each qubit is separable will be the most straightforward of these. A simple example is with the state $\ket{01}$, which satisfies $\X_1 \cap \Z_2$. Correspondingly, the state $\ket{0_L 1_L}$ satisfies $(\pred{X_L})_{\underline{1}}\cap (\pred{Z_L})_{\underline{2}}$ where $\underline{1}, \underline{2}$ represent the logical qubits. From the point of the physical qubits $\underline{1}, \underline{2}$ denote the sets of physical qubits that encode each logical qubit. For instance, for the $7$-qubit Steane code, $\underline{1} := (y, x_1, \ldots, x_6)$ and  $\underline{2} := (y', x'_1, \ldots, x'_6)$. Formally, we get,
\begin{align*}
    (\pred{X_L})_{\underline{1}} \cap (\pred{Z_L})_{\underline{2}} = (\pred{g_1} \cap \ldots \pred{g_6} \cap \pred{\overline{X}})_{\underline{1}} \cap (\pred{g_1} \cap \ldots \pred{g_6} \cap \pred{\overline{Z}})_{\underline{2}}
\end{align*}

\paragraph{Entangled multi-qubit states}
To express the predicate for two logical qubits as, say, $\pred{X_L} \otimes_L \pred{Z_L}$, and effectively tracking their evolution requires more advanced notions such as a logical tensor product $\otimes_L$ between logical predicates and further rules on how tensor products behave with intersections. For the sake of this example, we consider a logical tensor operation $\otimes_L$ that acts as follows:
\begin{itemize}
    \item Consider two basic logical predicates $\pred{A_L}, \pred{B_L}$ for $A, B \in \{I, X, Y, Z\}$
    \item We have $\pred{A_L} = \pred{I_L} \cap \pred{\overline{A}}$ and $\pred{B_L} = \pred{I_L} \cap \pred{\overline{B}}$ where $\pred{\overline{I}} := \I^7$ is an identity.
    \item $(\pred{A_L}) \otimes_L (\pred{B_L}) \ {:=} \ (\pred{I_L})_{\underline{1}} \cap (\pred{I_L})_{\underline{2}} \cap (\pred{\overline{A}} \otimes \pred{\overline{B}})$
\end{itemize}

Now, the question becomes: can we show the transversality of $\CNOT$ with respect to the Steane code? We can begin by defining:
\begin{align*}
    \mathtt{CNOT_L \ \underline{1} \ \underline{2}} & := \mathtt{CNOT \ y \ y'; \ CNOT \ x_1 \ x'_1; \ CNOT x_2 \ x'_2; \ CNOT \ x_3 \ x'_3;} \\
    & \mathtt{CNOT \ x_4 \ x'_4; \ CNOT \ x_5 \ x'_5; \ CNOT \ x_6 \ x'_6; }.
\end{align*}

Using the behavior of $\CNOT$ from~\Cref{table:2-qubit-types}, we need to show that
\begin{gather}
\label{eq:cnotl}
    \begin{aligned}
    & \hoare{\X_L \otimes_L \I_L}{CNOT_L}{\X_L \otimes \X_L}
    & \hoare{\I_L \otimes_L \X_L}{CNOT_L}{\I_L \otimes \X_L} \\
    & \hoare{\I_L \otimes_L \Z_L}{CNOT_L}{\Z_L \otimes \Z_L}
    & \hoare{\Z_L \otimes_L \I_L}{CNOT_L}{\Z_L \otimes \I_L}
    \end{aligned}
\end{gather}

It will be easier to derive the action of $\CNOT_L$ by understanding its actions on each of the stabilizers and logical operators for the Steane code as all logical predicates use these as the building blocks. Applying $\CNOT_L$ to each of the operators gate-wise, we get:
\begin{itemize}
    \item For the $X$-terms, i.e., for $\A \in \{\pred{g_1}, \pred{g_2}, \pred{g_3}, \pred{\overline{X}}\}$
    \begin{align}
    \label{eq:Xtype}
    \hoare{\A \otimes \pred{\overline{I}}}{CNOT_L}{\A \otimes \A} \quad \text{and } \quad \hoare{\pred{\overline{I}} \otimes \A}{CNOT_L}{\pred{\overline{I}} \otimes \A}
    \end{align}
    \item For the $Z$-term, i.e., for $\A \in \{\pred{g_4}, \pred{g_5}, \pred{g_6}, \pred{\overline{Z}}\}$
    \begin{align}
    \label{eq:Ztype}
    \hoare{\A \otimes \pred{\overline{I}}}{CNOT_L}{\A \otimes \pred{\overline{I}}} \quad \text{and } \quad \hoare{\pred{\overline{I}} \otimes \A}{CNOT_L}{\A \otimes \A}
    \end{align}
\end{itemize}

As the term $(\pred{I_L})_{\underline{1}} \cap (\pred{I_L})_{\underline{2}}$ appears in every predicate for entangled states, we first derive the action of $\CNOT_L$ on it.
\begin{align}
\label{eq:St7type}
    \inferrule*[Right=cons]{\hoare{(\pred{I_L})_{\underline{1}} \cap (\pred{I_L})_{\underline{2}}}{CNOT_L}{\bigcap_{i=1}^6 (\pred{g_i} \otimes \pred{g_i}) \bigcap_{i=4}^6 (\pred{g_i} \otimes \pred{\overline{I}}) \bigcap_{i=1}^3 (\pred{\overline{I}} \otimes \pred{g_i})}}
    {\inferrule*{\hoare{(\pred{I_L})_{\underline{1}} \cap (\pred{I_L})_{\underline{2}}}{CNOT_L}{\bigcap_{i=1}^6 (\pred{g_i} \otimes \pred{\overline{I}}) \bigcap_{i=1}^6 (\pred{\overline{I}} \otimes \pred{g_i})}}
    {\hoare{(\pred{I_L})_{\underline{1}} \cap (\pred{I_L})_{\underline{2}}}{CNOT_L}{(\pred{I_L} \otimes \pred{\overline{I}}) \cap (\pred{\overline{I}} \otimes \pred{I_L})}}
    }
\end{align}

Now, using~\cref{eq:Xtype,eq:Ztype,eq:St7type}, we can derive the action of $\CNOT_L$ on the remaining logical predicates. Taking $\pred{X_L} \otimes \pred{I_L} = (\pred{I_L})_{\underline{1}} \cap (\pred{I_L})_{\underline{2}} \cap (\pred{\overline{X}} \otimes \pred{\overline{I}}) $ as an example,
\begin{align*}
    \inferrule*[Right=cons]{\hoare{\X_L \otimes \I_L}{CNOT_L}{(\pred{I_L})_{\underline{1}} \cap (\pred{I_L})_{\underline{2}} \cap (\pred{\overline{X}} \otimes \pred{\overline{X}})}}
    {\hoare{\X_L \otimes \I_L}{CNOT_L}{\X_L \otimes \X_L}
    }
\end{align*}

Similarly, for the other three cases, we can directly get
\begin{align*}
    & \hoare{\Z_L \otimes \I_L}{CNOT_L}{\Z_L \otimes \I_L}\\
    & \hoare{\I_L \otimes \X_L}{CNOT_L}{\I_L \otimes \X_L}\\
    & \hoare{\I_L \otimes \Z_L}{CNOT_L}{\Z_L \otimes \Z_L}
\end{align*}
thereby satisfying~\Cref{eq:cnotl} and confirming the transversality of $\CNOT$ for the Steane code.

\section{Additive Predicates}\label{sec:additive-types}

\subsection{The Clifford + T set}
\label{adding-T}

Up to this point, we have focused on Hoare triples of Clifford operations, which are not universal for quantum computation. The easiest path from the Clifford set to a universal set is adding the $T$ operator to our language. Appealing to Gottesman's original Heisenberg formalism $T\sigma_z T^\dag = \sigma_z$, and so we can use the triple $\hoare{\Z}{T}{\Z}$. Unfortunately,
$$T\sigma_x T^\dag = \sqt\left(\begin{smallmatrix}0 & 1-i \\ 1+i & 0\end{smallmatrix}\right)$$
is not in the Pauli group and hence not expressible in our logic using Pauli predicates. Consequently to treat $T$ in our framework requires addition of the triple
\begin{equation}\label{eq:T-triple}
    \{\pred{X}\}\ \mathtt{T}\ \{ \tfrac{1}{\sqrt{2}}(\pred{X} + \pred{Y})\}.
\end{equation}
to our system as soundness requires our post-condition be derived from $\sqt\left(\begin{smallmatrix}0 & 1-i \\ 1+i & 0\end{smallmatrix}\right) = \sqt(\sigma_x + \sigma_y)$. So, we extend our judgments to distribute over addition and hence expand them to deal with \emph{additive} predicates $\A + \B$. To incorporate terms involving added predicates, we extend our grammar to include words of the form $\Gt + \Gt$, where $\Gt$ is the language of Pauli predicates. Throughout, we will use the shorthand $\pred{A} - \pred{B} = \pred{A} + (-\pred{B})$. The introduction rule for addition is:
\begin{equation}\label{rule-addition}
\inferrule*[Right=add]{\hoarepre{U}{A}{C} \and \hoarepre{U}{B}{D}}{\hoarepre{U}{A + B}{C + D}}
\end{equation}
The soundness of this rule is immediate: $U(A + B)U^\dag = UAU^\dag + UBU^\dag$.

Note that this will so frequently be combined with our \textsc{scale} rule (which can now see a broader range of coefficients $c$) to deal with additive predicates that we will tend to apply them together.

\begin{example}[$T$ on precondition $\Y$]
We prove the behavior of $T$ on precondition $\Y$ explicitly:
\begin{equation*}
\inferrule*[right=mul+scale]{\axiom{\hoare{X}{T}{\sqt(X + Y)}} \and
    \axiom{\hoare{Z}{T}{Z}}}
    {\hoare{Y}{T}{\sqt(Y - X)}}
\end{equation*}
\end{example}

\begin{example}[Composing $T$]
We prove that $T;T$ yields the same triples as $S$. This is trivially true on $\Z$ since both
$S$ and $T$ take precondition $\Z$ to postcondition $\Z$. We prove the $T;T$ reproduces the correct postcondition on $\X$ explicitly:
\begin{equation*}
\inferrule*[right=seq]{%
    \axiom{\hoare{X}{T}{\sqt(X + Y)}} \and
    \inferrule*[right=add+scale]{%
        \axiom{\hoare{X}{T}{\sqt(X + Y)}} \and
        \axiom{\hoare{Y}{T}{\sqt(Y-X)}}
}{\hoare{\sqt(X + Y)}{T}{\sqt(\sqt(X + Y + Y - X))}}
}{\hoare{X}{T;T}{Y}}
\end{equation*}
\end{example}

\begin{example}[$T^\dag$ Triples]
\label{eg:tdag}
Finally, it is useful to know the triples of $T^\dag$, which we'll define as $Z;S;T$. Again, one trivially has $\hoare{Z}{T^\dag}{Z}$, so we simply prove its behavior on $\X$ as follows:
\begin{equation*}
\inferrule*[right=seq]{
    \inferrule*[right=seq]{\axiom{\hoare{X}{Z}{-X}} \and \axiom{\hoare{-X}{S}{-Y}}}
        {\hoare{X}{Z;S}{-Y}} \and
    \axiom{\hoare{-Y}{T}{\sqt(X - Y)}}
}{\hoare{X}{Z;S;T}{\sqt(X - Y)}}
\end{equation*}
\end{example}

\begin{remark}
    We could alternatively include a general rule for $U^\dag$ for any unitary gate (or, indeed, unitary circuit) $U$:
\begin{equation*}
\inferrule*[right=dag]{\hoare{P}{U}{Q}}{\hoare{Q}{U^\dag}{P}}
\end{equation*}
This follows from the simple observation that $UPU^\dag = Q$ implies 
  \[U^\dag Q (U^\dag)^\dag = U^\dag UPU^\dag U = P\] 
We could then derive the behavior of $T^\dag$ from the triples $\hoare{Z}{T}{Z}$ and $\hoare{\frac{1}{\sqrt{2}}(X - Y}{T}{X}$.

This relates closely to the reversibility of the logic in general, where we could easily give $T$ the rule $\hoare{\frac{1}{\sqrt{2}}(X - Y)}{T}{X}$ from the equality $T^\dag X T = \frac{1}{\sqrt{2}}(X - Y)$ if we wanted to do backward reasoning. (Note that our rules for $H$ would remain the same since $H^\dag = H$, while we would want new backwards rules for $S$ and $\mathit{CNOT}$.) However, this would require new rules for measurement.
\label{rule:dag}
\end{remark}

\begin{example}[Steane code non-transversality of $T$]
\label{eg:t-transversal}
Now that we have the triple for $T$, the logical-$T$ gate for the Steane code $T_L$ should satisfy
\begin{align*}
    \hoare{Z_L}{T_L}{Z_L} \text{ and } \hoare{X_L}{T_L}{\tfrac{1}{\sqrt{2}} (\X_L + \Y_L)}
\end{align*}
where we use the descriptions from~\Cref{sec:qecc} for \pred{Z_L}, \pred{X_L} and \pred{Y_L}. However, we can easily show that the operation $U := \mathtt{T \ y; \ T \ {x_1}; \ T \ {x_2}; \ T \ {x_3}; \ T \ {x_4}; \ T \ {x_5}; \ T \ {x_6}}$ does not satisfy this behavior. In fact, $U$ acting on $\pred{I_L}$ changes the output as any stabilizer containing an $\X$ is converted into a non-trivial additive predicate. Then, $T$ applied to $\pred{I_L}$ becomes a predicate containing states outside both the Steane code space as well as the larger stabilizer state space. For instance
\begin{align*}
    \{\pred{g_1}\}\ U\ \{ \I \otimes \I \otimes \I \otimes \tfrac{1}{\sqrt{2}}(\X + \Y) \otimes \tfrac{1}{\sqrt{2}}(\X + \Y) \otimes \tfrac{1}{\sqrt{2}}(\X + \Y) \otimes \tfrac{1}{\sqrt{2}}(\X + \Y) \},
\end{align*}
which is clearly not a simple tensor product of Paulis as a stabilizer is expected to be.
\end{example}

\begin{proposition}\label{prop:prepare-Clifford+oneT}
    Let $\ket{\psi}$ be an $n$-qubit state satisfying $\left(\tfrac{1}{\sqrt{2}}(\pred{P}_0 + \pred{P}_1) \cap \pred{P}_2 \cdots \cap \pred{P}_n\right)$ with $P_0,P_1$ anticommuting. Then $\ket{\psi}$ can be prepared from $\ket{0 \dots 0}$ with a Clifford plus one $T$-gate circuit.
\end{proposition}
\begin{proof}
    As $P_0,P_1$ are anticommuting, by \Cref{theorem:2-transitive} there exists a Clifford circuit $C$ such that $\{\pred{P}_0\}\ C\ \{-\X\otimes \I \otimes \cdots \I\}$ and $\{\pred{P}_1\}\ C\ \{\Y\otimes \I \otimes \cdots \I\}$. Hence
    $$\left[\left(\tfrac{1}{\sqrt{2}}(\X+\Y)\otimes \I^{\otimes (n-1)}\right) \cap \pred{P}'_2 \cdots \cap \pred{P}'_n\right] (C\ket\psi).$$
    Note that each $P'_j$ ($j=2, \dots, n$) must commute with $\tfrac{1}{\sqrt{2}}(X+Y)\otimes I^{\otimes (n-1)}$ and hence we may write $P_j = I\otimes Q_j$ or $P_j = \sigma_z \otimes Q_j$ for some $(n-1)$-qubit Pauli operator $Q_j$. In either case, apply a $T^\dagger$-gate to the first qubit gives $T^\dagger_1:\pred{P}'_j \to \pred{P}'_j$. Consequently,
    $$\left[\left(\X \otimes \I^{\otimes (n-1)}\right) \cap \pred{P}'_2 \cdots \cap \pred{P}'_n\right](T^\dagger_1 C\ket\psi).$$
    Now, we can apply \Cref{prop:prepare-Clifford} and obtain a Clifford $C'$ such that $C' T^\dagger_1 C\ket\psi$ satisfies $\Z_1 \cap \cdots \cap \Z_n$. Therefore $(C' T^\dagger_1 C)^\dagger$ is our desired circuit.
\end{proof}

Note that there is a converse of this result. One can always squander the single $T$-gate by applying it directly to $\ket{0 \dots 0}$. But presuming this is not the case, and we prepare a state on which $T$ acts nontrivially, then additional Clifford gates can only produce predicates of the form $\frac{1}{\sqrt{2}}(\pred{P}_0 + \pred{P}_1) \cap \pred{P}_2 \cdots \cap \pred{P}_n$ with $P_0,P_1$ anticommuting.

\subsection{Example: Hoare triple for Toffoli}
\label{sec:toffoli}

Now that we have a judgment for $T$, we can use it to derive a valid Hoare triple for Toffoli via the latter's standard decomposition into $T$, $H$ and $\mathit{CNOT}$ gates:
\begin{coq}
TOFFOLI a b c :=
  H c; CNOT b c; T† c; CNOT a c; T c; CNOT b c; T† c;
  CNOT a c; T b; T c; H c; CNOT a b; T a; T† b; CNOT a b.
\end{coq}

Showing that \hoare{\Z_1 \cap \Z_2}{\mathtt{TOFFOLI}}{\Z_1 \cap \Z_2} proves remarkably straightforward:

{\small
\begin{align*}
  & \qquad \{ Z ⊗ I ⊗ I ∩ I ⊗ Z ⊗ I \} \\
  \mathtt{H~c;} & \qquad \{ Z ⊗ I ⊗ I ∩ I ⊗ Z ⊗ I \} \\
  \mathtt{CNOT~b~c;~T^\dagger~c;} & \qquad     \{  Z ⊗ I ⊗ I ∩ I ⊗ Z ⊗ I\} \\
  \mathtt{CNOT~a~c;~T~c;}         & \qquad     \{  Z ⊗ I ⊗ I ∩ I ⊗ Z ⊗ I\} \\
  \mathtt{CNOT~b~c;~T^\dagger c;} & \qquad     \{  Z ⊗ I ⊗ I ∩ I ⊗ Z ⊗ I\} \\
  \mathtt{CNOT~a~c;~T~b;~T~c;}    & \qquad \{  Z ⊗ I ⊗ I ∩ I ⊗ Z ⊗ I\} \\
  \mathtt{H~c;}                   & \qquad \{ Z ⊗ I ⊗ I ∩ I ⊗ Z ⊗ I\} \\
  \mathtt{CNOT~a~b;~T~a;~T^\dag b;}   & \qquad \{ Z ⊗ I ⊗ I ∩ Z ⊗ Z ⊗ I\} \\
  \mathtt{CNOT~a~b.}              & \qquad \{ Z ⊗ I ⊗ I ∩ I ⊗ Z ⊗ I\}
\end{align*}
}

Note that the trailing $T$ and $T^\dag$s are all applied to $\I$ or $\Z$ and therefore have no effect on the predicates. (We combine them with the previous commands for conciseness.) Noticeably, the derivation that \hoare{\X_3}{\mathtt{TOFFOLI}}{\X_3} also proves trivial (since $\mathtt{H~c}$ immediately converts the $\X$ to a $\Z$), showing that a $\ket +$ in the third position is not entangled by a Toffoli gate. By contrast, Toffoli's action on $\Z_3$ does get a bit messy\footnote{We leave off the coefficients for readability's sake, but this derivation, while wholly mechanical, can be difficult to follow to the end. The reader may wish to skip to the final steps of the deduction.}:

\allowdisplaybreaks
{\small
\begin{align*}
              & \qquad \{I ⊗ I⊗ Z \} \\
  \mathtt{H~c;}            & \qquad \{I ⊗ I ⊗ X \} \\
  \mathtt{CNOT~b~c;}       & \qquad \{I ⊗ I ⊗ X \} \\
  \mathtt{T^\dag~c;}           & \qquad \{I ⊗ I ⊗ X - I ⊗ I ⊗ Y \} \\
  \mathtt{CNOT~a~c;}       & \qquad \{I ⊗ I ⊗ X - Z ⊗ I ⊗ Y\} \\
  \mathtt{T~c;}            & \qquad \{I ⊗ I ⊗ X + I ⊗ I ⊗ Y - Z ⊗ I ⊗ Y + Z ⊗ I ⊗ X \} \\
  \mathtt{CNOT~b~c;}       & \qquad \{I ⊗ I ⊗ X + I ⊗ Z ⊗ Y - Z ⊗ Z ⊗ Y + Z ⊗ I ⊗ X \} \\
  \mathtt{T^\dag~c;}           & \qquad \{I ⊗ I ⊗ X - I ⊗ I ⊗ Y + I ⊗ Z ⊗ X + I ⊗ Z ⊗ Y ~ - \\
                          & \qquad ~Z ⊗ Z ⊗ X - Z ⊗ Z ⊗ Y + Z ⊗ I ⊗ X - Z ⊗ I ⊗ Y \} \\
  \mathtt{CNOT~a~c;~T~b;} & \qquad \{I ⊗ I ⊗ X - Z ⊗ I ⊗ Y + I ⊗ Z ⊗ X + Z ⊗ Z ⊗ Y ~ -  \\
                          & \qquad ~Z ⊗ Z ⊗ X - I ⊗ Z ⊗ Y + Z ⊗ I ⊗ X - I ⊗ I ⊗ Y \} \\
  \mathtt{T~c;}            & \qquad \{I ⊗ I ⊗ X + I ⊗ I ⊗ Y - Z ⊗ I ⊗ Y + Z ⊗ I ⊗ X ~ + \\
                       & \qquad ~I ⊗ Z ⊗ X + I ⊗ Z ⊗ Y + Z ⊗ Z ⊗ Y - Z ⊗ Z ⊗ X ~ -\\
			           & \qquad ~Z ⊗ Z ⊗ X - Z ⊗ Z ⊗ Y - I ⊗ Z ⊗ Y + I ⊗ Z ⊗ X ~ +\\
			           & \qquad ~Z ⊗ I ⊗ X + Z ⊗ I ⊗ Y - I ⊗ I ⊗ Y + I ⊗ I ⊗ X\} \\
  \mathtt{H~c;}        & \qquad \{ I ⊗ I ⊗ Z - I ⊗ I ⊗ Y + Z ⊗ I ⊗ Y + Z ⊗ I ⊗ Z ~ + \\
                       & \qquad ~I ⊗ Z ⊗ Z - I ⊗ Z ⊗ Y - Z ⊗ Z ⊗ Y - Z ⊗ Z ⊗ Z ~ - \\
		               & \qquad ~Z ⊗ Z ⊗ Z + Z ⊗ Z ⊗ Y + I ⊗ Z ⊗ Y + I ⊗ Z ⊗ Z + \\
			           & \qquad ~Z ⊗ I ⊗ Z - Z ⊗ I ⊗ Y + I ⊗ I ⊗ Y + I ⊗ I ⊗ Z \}\\
  \mathtt{CNOT~a~b;}   & \qquad \{ I ⊗ I ⊗ Z - I ⊗ I ⊗ Y + Z ⊗ I ⊗ Y + Z ⊗ I ⊗ Z ~ + \\
                       & \qquad ~Z ⊗ Z ⊗ Z - Z ⊗ Z ⊗ Y - I ⊗ Z ⊗ Y - I ⊗ Z ⊗ Z ~ - \\
		   	           & \qquad ~I ⊗ Z ⊗ Z + I ⊗ Z ⊗ Y + Z ⊗ Z ⊗ Y + Z ⊗ Z ⊗ Z + \\
			           & \qquad ~Z ⊗ I ⊗ Z - Z ⊗ I ⊗ Y + I ⊗ I ⊗ Y + I ⊗ I ⊗ Z \}\\
  \mathtt{T~a;~T^\dag~b;}  & \qquad \{I ⊗ I ⊗ Z - I ⊗ I ⊗ Y + Z ⊗ I ⊗ Y + Z ⊗ I ⊗ Z ~ + \\
                           & \qquad ~Z ⊗ Z ⊗ Z - Z ⊗ Z ⊗ Y - I ⊗ Z ⊗ Y - I ⊗ Z ⊗ Z ~ - \\
	                       & \qquad ~I ⊗ Z ⊗ Z + I ⊗ Z ⊗ Y + Z ⊗ Z ⊗ Y + Z ⊗ Z ⊗ Z ~ + \\
	                       & \qquad ~Z ⊗ I ⊗ Z - Z ⊗ I ⊗ Y + I ⊗ I ⊗ Y + I ⊗ I ⊗ Z \}\\
  \mathtt{CNOT~a~b;}          & \qquad \{I ⊗ I ⊗ Z - I ⊗ I ⊗ Y + Z ⊗ I ⊗ Y + Z ⊗ I ⊗ Z + \\
                           & \qquad ~I ⊗ Z ⊗ Z - I ⊗ Z ⊗ Y - Z ⊗ Z ⊗ Y - Z ⊗ Z ⊗ Z ~ -  \\
			               & \qquad ~Z ⊗ Z ⊗ Z + Z ⊗ Z ⊗ Y + I ⊗ Z ⊗ Y + I ⊗ Z ⊗ Z ~ + \\
			               & \qquad ~Z ⊗ I ⊗ Z - Z ⊗ I ⊗ Y + I ⊗ I ⊗ Y + I ⊗ I ⊗ Z\} \\
        \Rightarrow    & \qquad \{ I ⊗ I ⊗ Z + Z ⊗ I ⊗ Z + I ⊗ Z ⊗ Z - Z ⊗ Z ⊗ Z ~ -\\
                           & \qquad ~ Z ⊗ Z ⊗ Z + I ⊗ Z ⊗ Z + Z ⊗ I ⊗ Z + I ⊗ I ⊗ Z\} \\
        \Rightarrow    & \qquad \{I ⊗ I ⊗ Z + Z ⊗ I ⊗ Z + I ⊗ Z ⊗ Z - Z ⊗ Z ⊗ Z\}
  \end{align*}
}

Note that, despite the presence of seven $T$ or $T^\dag$-gates (and hence a potential of 128 summands appearing in the additive predicate), only four of them enlarged the term -- all $T$ or $T^\dag$ gates applied to either of the first two qubits failed to change the predicates. The $\X$s all left the picture once we applied the second Hadamard, and the $\Y$s all canceled out, leaving only $\Z$s and $\I$s. In fact, every summand has a $\Z$ in the last position, and the first predicate leaves us no way to introduce a negative on the $\Z$.

Hence, we can conclude that Toffoli satisfies the triple 
\[ \hoare{\Z_1 \cap \Z_2 \cap \Z_3}{\mathtt{TOFFOLI}}{\Z_1 \cap \Z_2 \cap \Z_3} \] %

\subsection{General additive predicates}

Clifford+$T$, as studied in the previous section, is universal in that all unitaries can be approximated as a composition of such gates. However, one often wishes to do an exact analysis %
or simply use a different universal gate set. Recall that we say $\ket\psi$ satisfies a predicate $\pred{P}$ if $\ket\psi$ is a $+1$-eigenstate of the (multi-qubit) Pauli operator $P$.
The critical feature of Pauli operators is they are unitary, Hermitian, and (except in the case of the identity) trace zero. Thus, unitary, Hermitian, and trace zero operators form a natural basis to extend Pauli predicates. Note that the Pauli operators form a linear basis of the vector space of all linear operators, and so for any unitary and Hermitian operator $M$, we can write $M = \sum_{j} c_j P_j$ for some (real) coefficients $c_j$ and Pauli operators $P_j$.

\begin{definition}
An \emph{additive predicate} is an expression of the form $\pred{M} = \sum_{j} c_j \pred{P}_j$ where $c_j \in \mathbb{R}$ and $\pred{P}_j$ are Pauli predicates, such that the operator $M = \sum_{j} c_j P_j$ is unitary, Hermitian, and trace zero. We say a state $\ket\psi$ satisfies $\pred{M}$, written $\pred{M}(\ket\psi)$, if $M\ket\psi = \ket\psi$.
\end{definition}

\begin{lemma}
    A one-qubit additive predicate has the form $\pred{M} = a\X + b\Y + c\Z$ with $a^2 + b^2 + c^2 = 1$.
\end{lemma}
\begin{proof}
Any one-qubit operator may be written as $M = tI + a\sigma_x + b\sigma_y + c\sigma_z$. As $M$ is Hermitian $t,a,b,c\in\mathbb{R}$. But $M$ is also unitary so
$$I = M^2 = (t^2 + a^2 + b^2 + c^2)I + 2ta\sigma_x + 2tb\sigma_y + 2tc\sigma_z.$$
Hence $t=0$, and therefore $M$ has trace zero and $a^2 + b^2 + c^2 = 1$.
\end{proof} 

This lemma shows that $1$-qubit additive predicates are particularly simple in that they form a representation of the familiar Bloch sphere.

\begin{example}\label{eg:Ry}
    Suppose we want to add the unitary $R_y(\frac{\pi}{3})$ to our language. As $R_y(\frac{\pi}{3}) \sigma_z R_y(\frac{\pi}{3})^\dag = \frac{\sqrt{3}}{2} \sigma_x + \frac{1}{2} \sigma_z$ and $R_y(\frac{\pi}{3}) \sigma_x R_y(\frac{\pi}{3})^\dag = \frac{1}{{2}} \sigma_x - \frac{\sqrt{3}}{2} \sigma_z$, we would add the triples $$\hoare{Z}{R_y(\frac{\pi}{3})}{\frac{\sqrt{3}}{2} X + \frac{1}{2}Z} \text{ and } \hoare{X}{R_y(\frac{\pi}{3})}{\frac{1}{{2}} X - \frac{\sqrt{3}}{2} Z}$$ as axioms in our logic. 
\end{example}

\Cref{prop:separability} on separability was stated at a level of generality that supports additive predicates as follows.
\begin{corollary}
    Let $\I^{(k-1)}\otimes \pred{M} \otimes \I^{(n-k)}$ be an additive predicate, and suppose $\ket\psi$ satisfies $\I^{(k-1)} \otimes \pred{M} \otimes \I^{(n-k)}$. Then the $k^{\text{th}}$ qubit of $\ket{\psi}$ is unentangled from the rest of the system.
\end{corollary}

\subsection{Logic of general unitary maps}

The logic of Clifford operators carries over to general unitaries. With Cliffords, we claimed that a complete description of an $n$-qubit operator could be deduced based on examining the preconditions $\X_j$ and $\Z_j$ as $j=1, \dots, n$. For example, for a general one-qubit unitary $U$ we would add triples $\{\pred{X}\}\:\mathtt{U}\:\{U\pred{X}U^\dagger\}$ and $\{\pred{Z}\}\:\mathtt{U}\:\{U\pred{Z}U^\dagger\}$ where we have abused notation here and written $U\pred{X}U^\dagger$ for the predicate associated to the Hermitian operator $U\sigma_x U^\dagger$, and similarly for $U\pred{Z}U^\dagger$. But as $U\sigma_y U^\dagger = i U\sigma_x U^\dagger U\sigma_z U^\dagger$ we would be able to derive $\{\pred{Y}\}\:\mathtt{U}\:\{U\pred{Y}U^\dagger\}$ within the language by applying our multiplication rules (\Cref{eq:rule-scale,eq:rule-mul}). For instance, in Example~\ref{eg:Ry} above, $$i(\tfrac{1}{{2}} \sigma_x - \tfrac{\sqrt{3}}{2} \sigma_z)(\tfrac{\sqrt{3}}{2} \sigma_x + \tfrac{1}{2}\sigma_z) = i(\tfrac{\sqrt{3}}{4}I + \tfrac{1}{4}\sigma_x\sigma_z - \tfrac{3}{4}\sigma_z\sigma_x -\tfrac{\sqrt{3}}{4}I) = i\sigma_x\sigma_z = \sigma_y,$$ and so one derives \hoare{Y}{R_y(\frac{\pi}{3})}{Y}.

This extends to multi-qubit unitaries as expected. If we know how an $n$-qubit unitary $U$ behaves on preconditions $\X_j$ and $\Z_j$ for any $j=1, \dots, n$, then we can compute $UPU^\dagger$ for any $n$-qubit Pauli operator. From this we can then compute the postcondition of $U$ with any additive predicate $\pred{M}$ as precondition from $UMU^\dagger = \sum_j c_j UP_jU^\dagger$. That is, each unitary can be viewed as a matrix acting on Pauli operators, which Gosset et al. \cite{Gosset2014} refer to the ``channel'' representation of $U$. In particular, we see that as a Clifford operators only permutes Pauli operators (possibly with a sign change) its matrix only contains values $-1$, $0$, or $1$. From above, the channel representation of a $T$-gate has coefficients of the form $\frac{c}{2^{s/2}}$ for $c\in\{-1,0,1\}$ and $s\in\{0,1\}$ (see also \cite[Equation (6.2)]{Gosset2014}). With a straightforward induction argument we prove the following result.

\begin{theorem}\label{thm:T-gate-bound}
    Let $U$ be a unitary circuit on $n$ qubits composed of $t$ number of $T$-gates and an arbitrary number of Clifford gates, and suppose $\hoare{\X_j}{U}{\pred{M}_j}$ and $\hoare{\Z_j}{U}{\pred{N}_j}$ for additive predicates $\pred{M}_j$ and $\pred{N}_j$, for each $j=1,\dots, n$. Then every coefficient of $\pred{M}_j$ and $\pred{N}_j$ is of the form $\frac{c}{2^{s/2}}$ where $c,s \in \mathbb{Z}$ and $s \leq t$.
\end{theorem}
\begin{proof}
    Inductively, if $t = 0$, then $U$ is a Clifford operator, and so each $\pred{M}_j$ and $\pred{N}_j$ is a Pauli predicate, and hence as additive predicates, all their coefficients are in $\{-1,0,1\}$ as desired.

    Suppose the statement is true for all unitary circuits containing at most $t-1$ number of $T$-gates, and suppose $U$ is a unitary circuit with $t$ number of $T$-gates. Suppose $U = C\circ U'$ with $C$ a Clifford operator. We claim  $U'$ has the same assumptions and requirements as $U$: clearly $U'$ also contains $t$ number of $T$-gates, and if $\hoare{\X_j}{U'}{\pred{M}'_j}$ and $\hoare{\Z_j}{U'}{\pred{N}'_j}$ then we must have $\hoare{\pred{M}'_j}{C}{\pred{M}_j}$ and $\hoare{\pred{N}'_j}{C}{\pred{N}_j}$; since $C$ is a Clifford operator the coefficients of $\pred{M}_j$ (respectively $\pred{N}_j$) are the same as those of $\pred{M}'_j$ (respectively $\pred{N}'_j$) up to sign changes and reordering.

    Therefore we may assume $U = T_k\circ U'$, where $T_k$ represents a $T$-gate operating on the $k$-th qubit. For notational convenience, let us assume $k=1$ as the general case will follow identically. As above assume $\hoare{\X_j}{U'}{\pred{M}'_j}$ and $\hoare{\Z_j}{U'}{\pred{N}'_j}$, and let us write
    $$M'_j = I \otimes \left(\sum_J c_{J,0} P_{J,0}\right) + \sigma_x \otimes \left(\sum_J c_{J,1} P_{J,1}\right) + \sigma_y \otimes \left(\sum_J c_{J,2} P_{J,2}\right) + \sigma_z \otimes \left(\sum_J c_{J,0} P_{J,0}\right)$$
    where each $P_{J,l}$ is a $(n-1)$-qubit Pauli operator. Inductively each coefficient satisfies $2^{(t-1)/2}c_{J,l} \in \mathbb{Z}$. Then $T_1:\pred{M}'_j \to \pred{M}_j$ and so
    \begin{align*}
        M_j &= I \otimes \left(\sum_J c_{J,0} P_{J,0}\right) + \tfrac{1}{\sqrt{2}}(\sigma_x + \sigma_y) \otimes \left(\sum_J c_{J,1} P_{J,1}\right)\\
        &\qquad +\ \tfrac{1}{\sqrt{2}}(-\sigma_x + \sigma_y) \otimes \left(\sum_J c_{J,2} P_{J,2}\right) + \sigma_z \otimes \left(\sum_J c_{J,0} P_{J,0}\right)\\
        &= I \otimes \left(\sum_J c_{J,0} P_{J,0}\right) + \sigma_x \otimes \left(\sum_J \frac{c_{J,1} - c_{J,2}}{\sqrt{2}} P_{J,1}\right)\\
        &\qquad +\ \sigma_y \otimes \left(\sum_J \frac{c_{J,1} + c_{J,2}}{\sqrt{2}} P_{J,2}\right) + \sigma_z \otimes \left(\sum_J c_{J,0} P_{J,0}\right).
    \end{align*}
    Finally, $2^{t/2}\cdot\frac{c_{J,1} \pm c_{J,2}}{\sqrt{2}} = 2^{(t-1)/2}(c_{J,1} \pm c_{J,2}) \in \mathbb{Z}$. The same argument works for the $\pred{N}_j$.
\end{proof}

Note that certain unitaries, those that are Hermitian and have trace zero, also define an additive predicate. This was clear for Pauli operators in the context of Pauli predicates: $\X$ is a predicate as well as satisfying triples $\hoare{\X}{\sigma_x}{\X}$ and $\hoare{\Z}{\sigma_x}{-\Z}$. It is straightforward to check when a unitary is also Hermitian (up to a global phase) using Hoare-style triples. A Hermitian unitary has $U = U^\dagger = U^{-1}$ and so $U^2 = I$. Thus $U$ is Hermitian if one can show $\{\X_j\}\:{U}\:\{\X_j\}$ and $\{\Z_j\}\:{U}\:\{\Z_j\}$ for all $j=1,\dots, n$.

\begin{example}
    The Hadamard gate satisfies $\hoare{\X}{H}{\Z}$ and $\hoare{\Z}{H}{\X}$, and is also Hermitian, and so defines an additive predicate $\pred{H}$. It is straightforward to verify $\pred{H} = \sqt(\pred{X} + \pred{Z})$ by writing out $H$ and $\sqt(\sigma_x + \sigma_z)$ in the computational basis and comparing the resulting matrices. However, we can deduce this expression (again up to a global sign change) from the triples above. From the lemma above we know $H = a\sigma_x + b\sigma_y + c\sigma_z$; just using the Pauli relations
    \begin{align*}
        H\sigma_x H &= (a^2 - b^2 - c^2)\sigma_x + 2ab\sigma_y + 2ac\sigma_z\ =\ \sigma_z\\
        H\sigma_z H &= 2ac\sigma_x + 2bc\sigma_y + (c^2 - a^2 - b^2)\sigma_z\ =\ \sigma_x.
    \end{align*}
    And so we obtain the quadratic system
    \begin{align*}
        0 &= a^2 - b^2 - c^2 = ab = bc\\
        1 &= 2ac = a^2 + b^2 + c^2.
    \end{align*}
    This is easy to solve by noting $1 = (a^2 + b^2 + c^2) + (a^2 - b^2 - c^2) = 2a^2$ and so $a = c = \pm\sqt$ and $b = 0$.
\end{example}

While a general unitary $U$ is not Hermitian, we can construct additive predicates associated with $U$ by adding an ancillary qubit. This is based on the real and imaginary parts of $U$ as defined as follows.

\begin{definition}
    Given any operator $U$ define its \emph{real part} as $\mathrm{Re}(U) = \frac{1}{2}(U + U^\dagger)$ and \emph{imaginary part} as $\mathrm{Im}(U) = \frac{1}{2i}(U - U^\dagger)$.
\end{definition}

Clearly, both $\mathrm{Re}(U)$ and $\mathrm{Im}(U)$ are Hermitian; however, neither is generally unitary. Nonetheless, we claim they do satisfy $\mathrm{Re}(U)^2+\mathrm{Im}(U)^2 = I$ and $\mathrm{Re}(U)\cdot\mathrm{Im}(U) = \mathrm{Im}(U)\cdot\mathrm{Re}(U)$, and so look like the blocks in a $2\times 2$ block unitary. Hence we could extend them to a unitary with an additional qubit.

\begin{lemma}
    Let $U$ be unitary and $\mathrm{Re}(U), \mathrm{Im}(U)$ be as above. Let $P$ and $Q$ be any anticommuting Pauli operators (on any number of qubits). Then $P \otimes \mathrm{Re}(U) + Q \otimes \mathrm{Im}(U)$ is unitary, Hermitian, and trace zero.
\end{lemma}
\begin{proof}
    As $P,Q,\mathrm{Re}(U), \mathrm{Im}(U)$ are all Hermitian so is $P \otimes \mathrm{Re}(U) + Q \otimes \mathrm{Im}(U)$. Similarly, since $P,Q$ anticommute, neither is the identity, and hence
    $$\tr(P \otimes \mathrm{Re}(U) + Q \otimes \mathrm{Im}(U)) = \tr(P)\tr(\mathrm{Re}(U)) + \tr(Q)\tr(\mathrm{Im}(U)) = 0.$$
    Finally, compute
    \begin{align*}
        &(P\otimes\mathrm{Re}(U) + Q\otimes\mathrm{Im}(U))^2\\
        &\quad = I\otimes(\mathrm{Re}(U)^2 +\mathrm{Im}(U)^2) + PQ\otimes\mathrm{Re}(U) \mathrm{Im}(U) + QP\otimes \mathrm{Im}(U) \mathrm{Re}(U).
    \end{align*}
    So, to finish, we merely complete our claims from above:
    $$\mathrm{Re}(U)^2+\mathrm{Im}(U)^2 = \frac{1}{4}(U^2 + 2I + (U^\dagger)^2) - \frac{1}{4}(U^2 - 2I + (U^\dagger)^2) = I$$
    and
    $$\mathrm{Re}(U)\mathrm{Im}(U) = \frac{1}{4i}(U^2 - (U^\dagger)^2) = \mathrm{Im}(U)\mathrm{Re}(U).$$
\end{proof}

For example if $P = \sigma_x$ and $Q = \sigma_z$ then
$$P \otimes \mathrm{Re}(U) + Q \otimes \mathrm{Im}(U) = \left(\begin{array}{cc} \mathrm{Im}(U) & \mathrm{Re}(U)\\ \mathrm{Re}(U) & -\mathrm{Im}(U)\end{array}\right).$$

\begin{definition}
    Let $U$ be a $n$-qubit unitary, and $P,Q \in \{\sigma_x, \sigma_y, \sigma_z\}$ be distinct. Then the additive predicate of $U$ relative to $P,Q$ is the $(n+1)$-qubit additive predicate corresponding to $P \otimes \mathrm{Re}(U) + Q \otimes \mathrm{Im}(U)$. We denote this as $\pred{P} \otimes \pred{Re}(U) + \pred{Q} \otimes \pred{Im}(U)$ (despite that neither $\pred{Re}(U)$ and $\pred{Im}(U)$ are predicates themselves).
\end{definition}

\subsection{Example: Triples satisfied by controlled unitaries}

Consider the triples for the controlled-phase gate:
$$\begin{array}{ll}
\{\I\otimes\X\}\: \texttt{control-}\sigma_z\: \{\Z\otimes \X\}, & \{\X\otimes\I\}\: \texttt{control-}\sigma_z\: \{\X\otimes \Z\}, \\
\{\I\otimes\Z\}\: \texttt{control-}\sigma_z\: \{\I\otimes \Z\}, & \{\Z\otimes\I\}\: \texttt{control-}\sigma_z\: \{\Z\otimes \I\}.
\end{array}$$
Note $\{\X\}\: \sigma_z\: \{-\X\}$ and $\{\Z\}\: \sigma_z\: \{\Z\}$. One is naturally led to the question: could we have deduced the appropriate postconditions of $\texttt{control-}\sigma_z$ from those of $\sigma_z$? More generally, can we deduce the triples of \texttt{control-}$U$ from the triples of $U$? Unfortunately, the answer must be no for a general unitary as this would imply ``control-'' is a functor of some sort, which is not the case. Nonetheless, we can deduce the triples of control-$U$ from those of $U$ \emph{and} its additive predicate $\X\otimes\pred{Re}(U) + \Y\otimes\pred{Im}(U)$.

Consider \texttt{control-}$U$, and decompose our Hilbert space along the control bit $\mathfrak{H} = \mathfrak{H}_0 \oplus \mathfrak{H}_1$. Namely, with respect to this decomposition \texttt{control-}$U$ is the matrix operator
$$\left(\begin{array}{cc} I & 0 \\ 0 & U\end{array}\right).$$
The component of our state where the control bit is $\ket{0}$ lives in $\mathfrak{H}_0$ where \texttt{control-}$U$ is trivial, but the component of the state where the control bit is $\ket{1}$ lives in $\mathfrak{H}_1$ where \texttt{control-}$U$ act as $U$. We use this representation to assert the judgments involving controlled operations in the following lemma.

\begin{lemma}\label{lemma:control-U-type}
Let $U$ be any $n$-qubit unitary. Then the following triples are valid for the controlled $U$:
\begin{enumerate}
    \item $\{\Z\otimes \I\}\: \texttt{control-}U\: \{\Z\otimes \I\}$, and
    \item $\{\X\otimes \I\}\: \texttt{control-}U\: \{\X\otimes \pred{Re}(U) + \Y\otimes \pred{Im}(U)\}$.
\end{enumerate}
If $P$ is any $n$-qubit Pauli, and $\{\pred{P}\}\: U\: \{\pred{V}\}$ then
\begin{enumerate}
    \item[(3)] $\{ \I\otimes\pred{P}\}\: \texttt{control-}U\: \{\I\otimes \tfrac{1}{2}(\pred{P} + \pred{V}) + \Z\otimes \tfrac{1}{2}(\pred{P} - \pred{V})\}$.
\end{enumerate}
\end{lemma}
\begin{proof} For (1), we simply note:
$$\left(\begin{array}{cc} I & 0 \\ 0 & U \end{array}\right) \left(\begin{array}{cc} I & 0 \\ 0 & -I \end{array}\right) \left(\begin{array}{cc} I & 0 \\ 0 & U^\dagger \end{array}\right) = \left(\begin{array}{cc} I & 0 \\ 0 & -I \end{array}\right).$$

For (2), again, we compute
$$\left(\begin{array}{cc} I & 0 \\ 0 & U \end{array}\right) \left(\begin{array}{cc} 0 & I \\ I & 0 \end{array}\right) \left(\begin{array}{cc} I & 0 \\ 0 & U^\dagger \end{array}\right) = \left(\begin{array}{cc} 0 & U^\dagger \\ U & 0 \end{array}\right).$$
But now, similar to above, we compute
\begin{align*}
    \left(\begin{array}{cc} 0 & I \\ I & 0 \end{array}\right) \left(\begin{array}{cc} 0 & U^\dagger \\ U & 0 \end{array}\right) &= \left(\begin{array}{cc} U & 0 \\ 0 & U^\dagger \end{array}\right)\\
    &= \left(\tfrac{1}{2}(I+\sigma_z)\otimes U +  \tfrac{1}{2}(I-\sigma_z)\otimes U^\dagger\right).
\end{align*}
Therefore
\begin{align*}
\left(\begin{array}{cc} 0 & U^\dagger \\ U & 0 \end{array}\right) &= (\sigma_x\otimes I)\left(\tfrac{1}{2}(I+\sigma_z)\otimes U +  \tfrac{1}{2}(I-\sigma_z)\otimes U^\dagger\right)\\
    &= (\sigma_x\otimes I)\left(I \otimes \tfrac{1}{2}(U+U^\dagger) + \sigma_z \otimes \tfrac{1}{2}(U - U^\dagger)\right)\\
    &= \sigma_x\otimes \mathrm{Re}(U) + \sigma_y\otimes \mathrm{Im}(U).
\end{align*}

Finally, for (3), we compute
$$\left(\begin{array}{cc} I & 0 \\ 0 & U \end{array}\right) \left(\begin{array}{cc} P & 0 \\ 0 & P \end{array}\right) \left(\begin{array}{cc} I & 0 \\ 0 & U^\dagger \end{array}\right) = \left(\begin{array}{cc} P & 0 \\ 0 & U P U^\dagger \end{array}\right) =  \left(\begin{array}{cc} P & 0 \\ 0 & V \end{array}\right).$$
In the computational basis $\frac{1}{2}(I+\sigma_z) =  \left(\begin{array}{cc} 1 & 0 \\ 0 & 0 \end{array}\right)$ and $\frac{1}{2}(I-\sigma_z) =  \left(\begin{array}{cc} 0 & 0 \\ 0 & 1 \end{array}\right)$. Therefore
\begin{align*}
    \left(\begin{array}{cc} P & 0 \\ 0 & V \end{array}\right) &= \left(\begin{array}{cc} P & 0 \\ 0 & 0 \end{array}\right) + \left(\begin{array}{cc} 0 & 0 \\ 0 & V \end{array}\right)\\
    &= \tfrac{1}{2}(I+\sigma_z)\otimes P +  \tfrac{1}{2}(I-\sigma_z)\otimes V\\
    &= I\otimes \tfrac{1}{2}(P + V) + \sigma_z\otimes \tfrac{1}{2}(P - V).
\end{align*}
\end{proof}

\begin{example}
Recall $\{\Z\}\: S\: \{\Z\}$ and $\{\X\}\: S\: \{\Y\}$. It is straightforward to verify that $\mathrm{Re}(S) = \tfrac{1}{2}(I + \sigma_z)$ and $\mathrm{Im}(S) = \tfrac{1}{2}(I - \sigma_z)$. Thus, we have
\begin{align*}
    \{\Z\otimes\I\}\: \text{\texttt{control-}$S$}\:& \{\Z \otimes \I\}\\
    \{\X\otimes\I\}\: \text{\texttt{control-}$S$}\:& \left\{\left(\X \otimes \tfrac{1}{2}(\I + \Z)\right) + \left(\Y \otimes \tfrac{1}{2}(\I - \Z)\right)\right\}\\
    & \Rightarrow\ \left\{\left(\tfrac{1}{2}(\X + \Y) \otimes \I\right) + \left(\tfrac{1}{2}(\X - \Y) \otimes \Z\right)\right\}\\
    \{\I\otimes \Z\}\: \text{\texttt{control-}$S$}\:& \{\I \otimes \Z\}\\
    \{\I\otimes \X\}\: \text{\texttt{control-}$S$}\:& \left\{\left(\I \otimes \tfrac{1}{2}(\X + \Y)\right) + \left(\Z \otimes \tfrac{1}{2}(\X - \Y)\right)\right\}.
\end{align*}
Note that by \Cref{thm:T-gate-bound}, any unitary Clifford+$T$ circuit that synthesizes \texttt{control-}$S$ requires at least 2 $T$-gates.
\end{example}

\begin{theorem}\label{thm:control-additive-type}
Let $U$ be a $n$-qubit Hermitian unitary with trace zero, and let $\pred{U}$ be its associated additive predicate. Then for each $k\geq 0$, we have \texttt{control$^k$-}$U$ is also a Hermitian unitary of trace zero, and its associated additive predicate is given by
$$\mathbf{C}^k\mathbf{U} = \I^{(k+n)} - \tfrac{1}{2^k}(\I-\Z)^{k} \otimes (\I^{n} - \mathbf{U}).$$
\end{theorem}
\begin{proof}
As above, we write the operator relation
\begin{align*}
    \left(\begin{array}{cc} I & 0 \\ 0 & U \end{array}\right) &= \tfrac{1}{2}(I + \sigma_z)\otimes I + \tfrac{1}{2}(I - \sigma_z)\otimes U\\
    &= I\otimes \tfrac{1}{2}(I + U) + \sigma_z\otimes \tfrac{1}{2}(I - U).
\end{align*}
Now we can prove the theorem by induction. Clearly, the $k=0$ case holds:
$$\mathbf{U} = \mathbf{C}^0\mathbf{U} = \I^{n} - (\I^{n} - \mathbf{U}).$$
Inductively suppose $\mathbf{C}^k\mathbf{U} = \I^{(k+n)} - \tfrac{1}{2^k}(\I-\Z)^{k} \otimes (\I^{n} - \mathbf{U})$ then using the relation above
\begin{align*}
    \mathbf{C}^{k+1}\mathbf{U} &= \I \otimes \tfrac{1}{2}(\I^{ (k+n)} + \mathbf{C}^k\mathbf{U}) + \Z \otimes \tfrac{1}{2}(\I^{ (k+n)} - \mathbf{C}^k\mathbf{U})\\
    &=  \I \otimes \tfrac{1}{2}(\I^{ (k+n)} + \I^{(k+n)} - \tfrac{1}{2^k}(\I-\Z)^{ k} \otimes (\I^{ n} - \mathbf{U}))\\
    &\qquad + \Z \otimes \tfrac{1}{2}(\I^{ (k+n)} - \I^{(k+n)} + \tfrac{1}{2^k}(\I-\Z)^{ k} \otimes (\I^{ n} - \mathbf{U}))\\
    &= \I^{ (k+1+n)} - \tfrac{1}{2^{k+1}}\I \otimes (\I-\Z)^{ k} \otimes (\I^{ n} - \mathbf{U})\\
    &\qquad + \tfrac{1}{2^{k+1}}\Z \otimes (\I-\Z)^{ k} \otimes (\I^{ n} - \mathbf{U})\\
    &= \I^{ (k+1+n)} - \tfrac{1}{2^{k+1}}(\I-\Z)^{ (k+1)} \otimes (\I^{ n} - \mathbf{U})).
\end{align*}
\end{proof}

\begin{corollary}
$\mathbf{C}^{k-1}\mathbf{Z} = \I^{k} - \tfrac{1}{2^{k-1}}(\I-\Z)^{k}.$
\end{corollary}

As a simple example of the utility of the above formulation, we can easily derive a complete set of triples for an arbitrarily multiply controlled $Z$ operator as follows.

\begin{theorem}
We have
\begin{align*}
    \{\Z_j\}\: \text{\texttt{control$^k$-}$\sigma_z$}\:& \{\Z_j\}\\
    \{\X_j\}\: \text{\texttt{control$^k$-}$\sigma_z$}\:& \left\{\X_j - \frac{1}{2^{k-1}}(\I-\Z)^{j-1}\otimes \X \otimes (\I-\Z)^{(k+1-j)}\right\}.
\end{align*}
\end{theorem}
\begin{proof}
As \texttt{control$^k$-}$\sigma_z$ is symmetric, it suffices to prove these statements for $j=1$. For the first, we already have
$$\{\Z\otimes\I^{k}\}\: \mathtt{control-}(\text{\texttt{control$^{k-1}$-}$\sigma_z$})\: \{\Z\otimes\I^{k}\}$$
from (1) of \Cref{lemma:control-U-type}. Now from \Cref{lemma:control-U-type} part (2), and that \texttt{control$^{k-1}$-}$\sigma_z$ is Hermitian, we have
$$\{\X\otimes\I^{k}\}\: \mathtt{control-}(\text{\texttt{control$^{k-1}$-}$\sigma_z$})\: \{\X\otimes \mathbf{C}^{k-1}\Z\}.$$
Then the result follows from the previous corollary.
\end{proof}

\begin{corollary}
    Any unitary Clifford+$T$ circuit that synthesizes \texttt{control$^k$-}$\sigma_z$ contains at least $(2k-2)$ $T$-gates.
\end{corollary}

Note that this corollary does not provide a sharp bound. For example, Gosset et al.~\cite{Gosset2014} provide an algorithm  that explicitly computes the optimal $T$-count of the Toffoli gate to be seven. However, our bound only provides the lower bound of two.

\begin{lemma}
    For any unitary $U$ we have
    \begin{enumerate}
        \item $\mathrm{Re}(\text{\texttt{control$^k$-}$U$}) = \text{\texttt{control$^k$-}$(\mathrm{Re}(U))$}$, and
        \item $\mathrm{Im}(\text{\texttt{control$^k$-}$U$}) = \frac{1}{2^k}(I - \sigma_z)^{k} \otimes \mathrm{Im}(U)$.
    \end{enumerate}
\end{lemma}
\begin{proof}
Note that
$$\frac{1}{2}\left[\left(\begin{array}{cc} I & 0 \\ 0 & U\end{array}\right) + \left(\begin{array}{cc} I & 0 \\ 0 & U^\dagger\end{array}\right)\right] = \left(\begin{array}{cc} I & 0 \\ 0 & \tfrac{1}{2}(U + U^\dagger)\end{array}\right)$$
and so $\mathrm{Re}(\text{\texttt{control-}$U$}) = \text{\texttt{control-}$(\mathrm{Re}(U))$}$. Then (1) follows from straightforward recursion.

Similarly,
$$\frac{1}{2i}\left[\left(\begin{array}{cc} I & 0 \\ 0 & U\end{array}\right) - \left(\begin{array}{cc} I & 0 \\ 0 & U^\dagger\end{array}\right)\right] = \left(\begin{array}{cc} 0 & 0 \\ 0 & \tfrac{1}{2i}(U - U^\dagger)\end{array}\right) = \tfrac{1}{2}(I - \sigma_z)\otimes \mathrm{Im}(U).$$
Therefore (2) also follows from recursion.
\end{proof}

\begin{theorem}
    Let $U$ be any $n$-qubit unitary, and $k>0$. Then for $j = 1, \dots, k$, we have
    \begin{align*}
        \{\Z_j\}\: \text{\texttt{control$^k$-}$U$}\: &\{\Z_j\}, \text{ and}\\
        \{\X_j\}\: \text{\texttt{control$^k$-}$U$}\: &\left\{\X_j - \frac{1}{2^{k-1}}(\I-\Z)^{j-1}\otimes \X \otimes (\I-\Z)^{(k-j)}\otimes \I^{n}\right.\\
        &\qquad +\ \frac{1}{2^{k-1}} (\I-\Z)^{j-1}\otimes \X \otimes (\I-\Z)^{(k-j)}\otimes \pred{Re}(U)\\
        &\left. \qquad +\ \frac{1}{2^{k-1}} (\I-\Z)^{j-1}\otimes \Y \otimes (\I-\Z)^{(k-j)}\otimes \pred{Im}(U)\right\}.
    \end{align*}
    Moreover, for any $n$-qubit Pauli $P$ with $\{\pred{P}\}\: U\: \{\pred{V}\}$ then
    $$\{\I^k\otimes \pred{P}\}\: \text{\texttt{control$^k$-}$U$}\: \left\{\I^k \otimes \pred{P} - \tfrac{1}{2^k}(\I - \Z)^k \otimes (\pred{P} - \pred{V})\right\}.$$
\end{theorem}
\begin{proof}
    Clearly (1) follows immediately from \Cref{lemma:control-U-type} part (1).

    For (2), we will assume $j=1$ for clarity as the general case follows identically. From \Cref{lemma:control-U-type} part (2) we have
    \begin{align*}
        &\{\X \otimes \I^{(k+n-1)}\}\\
        &\text{\texttt{control-(control$^{k-1}$-}$U$)}\\
        &\{\X \otimes \pred{Re}(\text{\texttt{control$^{k-1}$-}$U$}) + \Y \otimes \pred{Im}(\text{\texttt{control$^{k-1}$-}$U$})\}.
    \end{align*}
    From part (1) of the previous lemma we have $\mathrm{Re}(\text{\texttt{control$^{k-1}$-}$U$}) = \text{\texttt{control$^{k-1}$-}$(\mathrm{Re}(U))$}$, and so from \Cref{thm:control-additive-type}
    $$\mathrm{Re}(\text{\texttt{control$^{k-1}$-}$U$}) = I^{k+n-1} - \frac{1}{2^{k-1}}(I - \sigma_z)^{(k-1)}(I^{n} - \mathrm{Re}(U)).$$
    Part (2) of the that lemma simply gives $\mathrm{Im}(\text{\texttt{control$^{k-1}$-}$U$}) = \frac{1}{2^{k-1}}(I - \sigma_z)^{(k-1)} \otimes \mathrm{Im}(U)$. Substituting these into the formula above gives the desired results.

    We prove (3) inductively. For $k=1$, \Cref{lemma:control-U-type} part (3) gives
    \begin{align*}
        \{\I \otimes \pred{P}\}\: \text{\texttt{control-}$U$}\:& \{\I\otimes \tfrac{1}{2}(\pred{P} + \pred{V}) + \Z\otimes \tfrac{1}{2}(\pred{P} - \pred{V})\}\\
        &\quad \Rightarrow\ \{\tfrac{1}{2}\I \otimes \pred{P} + \tfrac{1}{2}\I \otimes \pred{V} + \tfrac{1}{2}\Z \otimes \pred{P} - \tfrac{1}{2}\Z \otimes \pred{V}\}\\
        &\quad \Rightarrow\ \{\I \otimes \pred{P} - \tfrac{1}{2}(\I - \Z)\otimes (\pred{P} - \pred{V})\}.
    \end{align*}
    Now suppose the formula above holds for $k-1$. Then from the typing statement we just derived,
    \begin{align*}
        &\{\I^{(k-1)}\otimes (\I \otimes\pred{P})\}\\
        &\text{\texttt{control$^{k-1}$-}(\texttt{control-}$U$)}\\
        &\left\{ \I^{(k-1)}\otimes (\I \otimes \pred{P}) -\tfrac{1}{2^{k-1}}(\I - \Z)^{(k-1)} \otimes \left[ (\I\otimes \pred{P}) - \left((\I \otimes \pred{P}) - \tfrac{1}{2}(\I - \Z)\otimes (\pred{P} - \pred{V})\right)\right]\right\}\\
        & \qquad \Rightarrow\ \{\I^{k}\otimes \pred{P} - \tfrac{1}{2^k}(\I - \Z)^{k}\otimes(\pred{P} - \pred{V})\}.
    \end{align*}
\end{proof}

\section{Measurement for additive predicates}
\label{sec:meas-add}

One missing component of additive predicates is the derivation of normal forms. As a consequence, a full formalism for measurement is incomplete. Nonetheless, we can go some distance in characterizing post-measurement conditions.

\subsection{Computing post-measurement states}

When studying Pauli operators, we were able to exploit standard methods from the stabilizer formalism for treating measurements in Pauli bases. However, these techniques no longer apply for general unitary Hermitian operators, even those of trace zero. Hence we need to revisit measurement from the first principles.

First, consider the problem of measuring a single qubit in the $z$-basis. Post measurement, we know the state would satisfy the predicate $\Z \uplus -\Z$ where the factor in this disjunction depends on the measurement outcome. Although it is outside our logical formalism, we see the probability of these outcomes directly in any preconditions of measurement the qubit may satisfy.

\begin{lemma}
    Let $\pred{M} = a\X + b\Y + c\Z$ be an additive predicate and $\pred{M}(\ket\psi)$. Then measurement (which is implicitly in the z-basis) has
    \begin{align*}
        \mathrm{Pr}_{\ket\psi}\{ \text{meas}=+1\} = \frac{1+c}{2} &, \text{ and } \mathrm{Pr}_{\ket\psi}\{ \text{meas}=-1\} = \frac{1-c}{2}.
    \end{align*}
\end{lemma}
\begin{proof}
    We have $\pred{M}(\ket\psi)$ when $\ket\psi$ is the $+1$-eigenvector of the operator $M = a \sigma_x + b \sigma_y + c \sigma_z$. As $\ketbra\psi$ is the projector onto the $+1$-eigenspace of $M$, and $I - \ketbra\psi$ is the projector onto the $-1$-eigenspace, we must have
    $$M = \ketbra\psi - (I - \ketbra\psi) = 2\ketbra\psi - I.$$
    Let $\Pi^Z$ be the projector onto the $+1$-eigenspace of $Z$ (that is $\Pi^Z = \ketbra{0}$). Born's rule states
    \begin{align*}
        \mathrm{Pr}_{\ket\psi}\{ \text{meas}=+1\} &= \tr(\Pi^Z \ketbra\psi) = \tfrac{1}{2}\tr(\Pi^Z(I+M))\\
        &= \frac{1}{2}\left(1 + a\tr(\Pi^Z X) + b\tr(\Pi^Z Y) + c\tr(\Pi^Z Z)\right) = \frac{1 + c}{2}.
    \end{align*}
    Similarly, $\mathrm{Pr}_{\ket\psi}\{ \text{meas}=-1\} = \frac{1 - c}{2}$ follows.
\end{proof}

A similar fact holds in the multi-qubit case; however, it is significantly more challenging to derive. Let us illustrate the key ideas on $2$ qubits. Suppose $(\mathbf{M}_{(1)} \cap \mathbf{M}_{(2)})(\ket\psi)$ where $\mathbf{M}_{(1)}, \mathbf{M}_{(2)}$ are $2$-qubit additive predicates. Let $M_1$ and $M_2$ be the unitary Hermitian operators associated with these. The assertion that $\ket\psi$ is the joint $+1$-eigenvector of $M_1$ and $M_2$ implies $M_1M_2 = M_2M_1$. As above, we have $\ketbra\psi$ as the projector onto this space, and thus
$$\ketbra\psi = \tfrac{1}{4}(I + M_1)(I + M_2) = \tfrac{1}{4}(I + M_1 + M_2 + M_1 M_2).$$
For convenience, let us write $M_0 = I$ and $M_3 = M_1 M_2$. Suppose we measure the first qubit in the z-basis. As in the lemma above, the measurement projector is $\Pi^{Z}\otimes I$ and Born's rule reads
$$\mathrm{Pr}_{\ket\psi}\{ \text{meas}=+1\} = \tr((\Pi^{Z}\otimes I)\ketbra\psi) = \frac{1}{4}\sum_{j=0}^3 \tr((\Pi^{Z}\otimes I)M_j).$$
To compute these traces, we write
$$M_j = I \otimes N_{j0} + X \otimes N_{j1} + Y \otimes N_{j2} + Z \otimes N_{j3},$$
and so
\begin{align*}
    \tr((\Pi^{Z}\otimes I)M_j) &= \tr(\Pi^{Z} \otimes N_{j0}) + \tr( \Pi^{Z}X \otimes N_{j1}) + \tr( \Pi^{Z}Y \otimes N_{j2}) + \tr(\Pi^{Z}Z \otimes N_{j3})\\
    &= \tr(N_{j0}) + \tr(N_{j3}).
\end{align*}
Now, $M_0 = I$, and for $j>0$ we have $M_j$ is trace zero. Thus we may write
$$N_{00} = I \text{ and } N_{01} = N_{02} = N_{03} = 0,$$
and for $j > 0$:
\begin{equation}\label{eqn:other-terms}
  \begin{array}{r@{\:}c@{\:}r@{\:}c@{\:}l}
    N_{j0} &=& &&\tilde{x}_j X + \tilde{y}_j Y + \tilde{z}_j Z\\
    N_{j3} &=& c_j I &+& x_j X + y_j Y + z_j Z.
\end{array}
\end{equation}
So,
\begin{align*}
    \mathrm{Pr}_{\ket\psi}\{ \text{meas}=+1\} &= \tfrac{1}{4}(2 + \tr(N_{13}) + \tr(N_{23}) + \tr(N_{33}))\\
    &= \frac{1 + c_1 + c_2 + c_3}{2}
\end{align*}
where we extract each $c_j$ as:
\begin{equation}\label{eqn:extract-c}
    M_j = c_j Z\otimes I + \text{other terms.}
\end{equation}
For $c_1$ and $c_2$ this is by direct examination of $\mathbf{M}_{(1)}$ and $\mathbf{M}_{(2)}$. However $c_3$ can only be obtained by computing $M_1 M_2$. A similar computation holds for the probability of measuring $-1$, and so we have proven the following result.

\begin{proposition}\label{prop:additive-2-qubit-probability}
    Suppose $(\mathbf{M}_{(1)} \cap \mathbf{M}_{(2)})(\ket\psi)$ where $\mathbf{M}_{(1)}, \mathbf{M}_{(2)}$ are $2$-qubit additive predicates, and suppose we measure the first qubit in the z-basis. As above write $M_1$ and $M_2$ for the operators associated to these predicates and $M_0 = I$ and $M_3 = M_1 M_2$. For $j=0, 1, 2, 3$ define
    $$M_j = I \otimes N_{j0} + X \otimes N_{j1} + Y \otimes N_{j2} + Z \otimes N_{j3}.$$
    Then
    \begin{align*}
        \mathrm{Pr}_{\ket\psi}\{ \text{meas}=+1\} = p_+ = \frac{1 + c_1 + c_2 + c_3}{2}
    \end{align*}
    and
    \begin{align*}
        \mathrm{Pr}_{\ket\psi}\{ \text{meas}=-1\} = p_- = \frac{1 - c_1 - c_2 - c_3}{2}
    \end{align*}
    where the $c_j$ are given in \cref{eqn:extract-c}.
\end{proposition}

From this, we can bootstrap the full post-measurement predicates for a general $2$-qubit state as follows.

\begin{theorem}
On $2$-qubit states, measurement in the z-basis satisfies
$$\hoare{\mathbf{M}_{(1)} \cap \mathbf{M}_{(2)}}{Meas\ 1}{(\Z_1 \cap \pred{M}_+) \uplus ((-\Z)_1 \cap \pred{M}_-)}$$
where
\begin{align}\label{eqn:M-plus}
    \mathbf{M}_+ &= \tfrac{1}{2p_+}\sum_{j=1}^3 ((\tilde{x}_j +x_j) \X + (\tilde{y}_j +y_j) \Y + (\tilde{z}_j +z_j) \Z)\\ \label{eqn:M-minus}
    \mathbf{M}_- &= \tfrac{1}{2p_-}\sum_{j=1}^3 ((\tilde{x}_j - x_j) \X + (\tilde{y}_j - y_j) \Y + (\tilde{z}_j - z_j) \Z),
\end{align}
where $p_\pm$ are given in the proposition above, and the coefficients of $\pred{M}_\pm$ are in \cref{eqn:other-terms}.
\end{theorem}
\begin{proof}
    As above, write $p_+ = \frac{1 + c_1 + c_2 + c_3}{2}$ for seeing outcome $+1$, then the post-measurement state given outcome $+1$ is
    \begin{align*}
        &\tfrac{1}{p_+}(\Pi^{Z}\otimes I) \ket\psi\bra\psi (\Pi^{Z}\otimes I) = \frac{1}{4p_+}\sum_{j=0}^3\left((\Pi^{Z}\otimes I) M_j (\Pi^{Z}\otimes I)\right)\\
        &\quad = \frac{1}{4p_+}\sum_{j=0}^3\left( \Pi^{Z} \otimes N_{j0} + (\Pi^{Z}X\Pi^{Z}) \otimes N_{j1} + (\Pi^{Z}Y\Pi^{Z}) \otimes N_{j2} + \Pi^{Z} \otimes N_{j3}\right)\\
        &\quad= \Pi^{Z} \otimes \frac{1}{4p_+}\left( I + \sum_{j=1}^3 (N_{j0} + N_{j3}) \right)\\
        &\quad= \Pi^{Z} \otimes \left( \tfrac{1}{2} I + \tfrac{1}{4p_+}\sum_{j=1}^3((\tilde{x}_j +x_j) X + (\tilde{y}_j +y_j) Y + (\tilde{z}_j +z_j) Z)\right).
    \end{align*}
    While we wrote this as a density operator, it is a pure state
    $$\tfrac{1}{p_+}(\Pi^{Z}\otimes I) \ket\psi\bra\psi (\Pi^{Z}\otimes I) = \ketbra{0,\psi'}.$$
    As above $\ketbra{\psi'} = \frac{1}{2}(I + M_+)$ so by examination the post-measurement state satisfies $\Z_1 \cap (\mathbf{M}_+)_2$.

    For seeing outcome $-1$, which has probability $p_- = \frac{1 - c_1 - c_2 - c_3}{2}$, the computation is similar:
    \begin{align*}
        &\tfrac{1}{p_-}((I-\Pi^{Z})\otimes I) \ket\psi\bra\psi ((I-\Pi^{Z})\otimes I) \ =\  (I-\Pi^{Z}) \otimes \frac{1}{4p_-}\left( I + \sum_{j=1}^3 (N_{j0} - N_{j3}) \right)\\
        &\qquad =\ \Pi^{Z} \otimes \left( \tfrac{1}{2} I + \tfrac{1}{4p_-}\sum_{j=1}^3((\tilde{x}_j - x_j) X + (\tilde{y}_j - y_j) Y + (\tilde{z}_j - z_j) Z)\right).
    \end{align*}
    So the post-measurement state satisfies $(-\Z)_1\cap (\mathbf{M}_-)_2$
\end{proof}

\begin{example}
Note that in the case that $\pred{M}_{(1)}$ and $\pred{M}_{(2)}$ are derived from Pauli operators, the Proposition above recovers our measurement rules from earlier. Each additive predicate only contains one Pauli term. From \cref{eqn:extract-c} we see that at most one $c_j$ can be nonzero, as otherwise two of $M_1$, $M_2$, and $M_3$ would equal $Z\otimes I$ contradicting independence of $\pred{M}_{(1)}$ and $\pred{M}_{(2)}$. In the case where one $c_j = \pm 1$ the measurement is deterministic (with outcome equal to this $c_j$) and the input state is separable. So suppose this is not the case, and the measurement is uniformly random. One of $M_1$, $M_2$, and $M_3$ must be of the form $\pm X\otimes P$ for some Pauli, as otherwise, one would be $Z\otimes I$ since they are independent and pairwise commuting. Without loss of generality suppose $M_1 = s_1 X\otimes P$, where $s_1 \in \{-1,+1\}$. As $M_3 = M_1M_2$ one of $M_2$ or $M_3$ has of the form
\begin{enumerate}
    \item $\pm I \otimes Q$ where $Q$ commutes with $P$, or
    \item $\pm Z \otimes Q$ where $Q$ anti-commutes with $P$.
\end{enumerate}
Again without loss of generality, we can assume $M_2$ takes one of these forms. Therefore either:
\begin{enumerate}
    \item $M_2 = s_2 I\otimes Q$ and $M_3 = s_1s_2 X\otimes Q$, and so in \cref{eqn:other-terms} the only nonvanishing coefficient is one of $\tilde{x}_2$, $\tilde{y}_2$, or $\tilde{z}_2$ (according to $Q$) and we obtain the postcondition $\pred{M}_+ = \pred{M}_- = s_2\pred{Q}$; or,
    \item $M_2 = s_2 Z \otimes Q$ and $M_3 = -s_1s_2 Y\otimes iPQ$, and so in \cref{eqn:other-terms} the only nonvanishing coefficient is one of $x_j$, $y_j$, or $z_k$ (again according to $Q$) and we have postcondtions $\pred{M}_+ = s_2 \pred{Q}$ and $\pred{M}_- = -s_2 \pred{Q}$.
\end{enumerate}
\end{example}

\begin{example}
If we have a single $T$-gate, what other sort of gates can we synthesize using it, Clifford gates, and measurement in the computational basis? We will focus only on synthesizing another one-qubit gate using a single ancillary qubit that will be measured, and so this example parallels gate injection, which we will study in the next section. By \Cref{prop:prepare-Clifford+oneT}, prior to measurement, we can assume our state $\ket\psi$ satisfies a predicate of the form
$$\tfrac{1}{\sqrt{2}}(\pred{P}_{(0)} + \pred{P}_{(1)}) \cap \pred{P}_{(2)}$$
where the $2$-qubit Pauli operators $P_0$ and $P_1$ anticommute, and $P_2$ commutes with both $P_0$ and $P_1$. Without loss of generality, we can assume the first qubit is measured in the z-basis. Using the notation above $M_1 = \tfrac{1}{\sqrt{2}}(P_0 + P_1)$, $M_2 = P_2$, and $M_3 = \tfrac{1}{\sqrt{2}}(P_0P_2 + P_1P_2)$. As in the previous example, we focus on cases involving $\Z\otimes\I$.

\paragraph{Case 1: $\pred{P}_{(2)} = \pm \Z\otimes\I$} In the notation above, $c_1 = c_3 = 0$ while $c_2 = \pm 1$, and hence the  probability of measuring $Z = +1$ is $0$ or $1$ depending on the sign in $\pred{P}_{(2)}$. Specializing \cref{eqn:other-terms} to this case, we must have
\begin{align*}
    M_1 &= \tfrac{1}{\sqrt{2}} \I \otimes (\tilde{x}\X + \tilde{y}\Y + \tilde{z}\Z) +  \tfrac{1}{\sqrt{2}} \Z \otimes (x\X + y\Y + z\Z)\\
    M_3 &= \pm (\tfrac{1}{\sqrt{2}} \I \otimes (x\X + y\Y + z\Z) + \tfrac{1}{\sqrt{2}} \Z \otimes (\tilde{x}\X + \tilde{y}\Y + \tilde{z}\Z))
\end{align*}
Hence post measurement, our state satisfies
$$\pred{M}'_{\pm} = \pm \tfrac{1}{\sqrt{2}} ( (\tilde{x} \pm x) \X + (\tilde{y} \pm y) \Y + (\tilde{z} \pm z)\Z).$$
As $P_0$ and $P_1$ anticommute, precisely one of $x,y,z$ is nonzero and precisely one of $\tilde{x},\tilde{y},\tilde{z}$ is nonzero, and these cannot both be $x,\tilde{x}$ or $y,\tilde{y}$ or $z,\tilde{z}$. So by \Cref{prop:prepare-Clifford+oneT}, this circuit is equivalent to one that uses Clifford plus one $T$-gate (without measurement).

\paragraph{Case 2: $\pred{P}_{(0)} = \pm \Z\otimes\I$} Note this case also covers when $P_1$, $P_0P_2$, or $P_1P_2$ is $\pm \sigma_z \otimes I$, after relabeling terms as needed. As $P_0$ and $P_1$ anticommute we must have $P_1 = \pm \sigma_x \otimes Q$ or $P_1 = \pm \sigma_y \otimes Q$ for some Pauli operator $Q$ (that may be $I$). Hence $M_1$ does not contribute to a post-measurement predicate. Yet, $P_2$ must commute with both $P_0$ and $P_1$, and hence $P_2 = I \otimes Q'$ where $Q' \in \{\sigma_x, \sigma_y, \sigma_z\}$. But then $M_3$ will not contribute to a post-measurement predicate either, and hence $\pred{M}'_{\pm} = \pred{P}_{(2)}$ and the circuit is equivalent to a Clifford gate.

\paragraph{Case 3: None of $\pred{P}_{(0)}, \pred{P}_{(1)}, \pred{P}_{(2)}$ is $\pm \Z\otimes\I$} This case is somewhat tedious, and so we let the reader verify the details. Regardless, the measurement has probability $\frac{1}{2}$ of obtaining $z = +1$ or $z = -1$. In the subcase where $P_2 = \sigma_x \otimes Q$ or $P_2 = \sigma_y \otimes Q$, then the result is similar to Case 1 above in that between $M_1$ and $M_3$ precisely two of $x,y,z,\tilde{x},\tilde{y},\tilde{z}$ contribute to the output predicate, and so the circuit is equivalent to a Clifford with one $T$-gate circuit (without measurement). In the subcase $P_2 = I \otimes Q$, then the result is similar to Case 2 above in that the state is separable, and hence the post-measurement predicate is $Q$, and the circuit is equivalent to a Clifford gate. Finally in the subcase $P_2 = \sigma_z \otimes Q$, we must have $M_1 = \sigma_x\otimes N_1 + \sigma_y\otimes N_2$ (as otherwise either $M_1$ or $M_3$ would have a $\sigma_z\otimes I$ term); then just as above neither $M_1$ or $M_3$ contribute a post-measurement predicate, which is $Q$, and so the circuit is equivalent to a Clifford gate.
\end{example}

The $n$-qubit analysis follows in a similar way as with two qubits; however, we do not have such a concrete result. Suppose $(\pred{M}_{(1)} \cap \cdots \cap \pred{M}_{(n)})(\ket\psi)$. Continuing our notation from above, let $M_j$ be the unitary Hermitian operator associated with $\pred{M}_{(j)}$. Then
$$\ketbra\psi = \frac{1}{2^n}\prod_{j=1}^n(I + M_j) = \frac{1}{2^n}\sum_{J\subseteq\{1,\cdots,n\}} M_J$$
where the ``multi-index'' $J$ selects a subset of $\{1, \cdots, n\}$ over which $M_J = \prod_{j\in J} M_j$. Here we adopt the convention $M_\emptyset = I$ similar to $M_0 = I$ in the $2$-qubit case. Then Born's rule for measuring the first qubit reads
$$\mathrm{\Pr}_{\ket\psi}\{\text{meas} = +1\} = \frac{1}{2^n}\sum_{J \subseteq \{1,\dots,n\}} \tr\left( (\Pi^Z\otimes I^{(n-1)}) M_J\right).$$

Again we write
$$M_J = I \otimes N_{J0} + X \otimes N_{J1} + Y \otimes N_{J2} + Z \otimes N_{J3}$$
and just as in the $2$-qubit case, have
$$\tr\left( (\Pi^Z\otimes I^{(n-1)}) M_J\right) = \tr(N_{J0} + N_{J3}).$$
Now, $N_{\emptyset 0} = I$ and $N_{\emptyset K} = 0$. For $J \not= \emptyset$, we expand
$$N_{J0} = \sum_{K \not= \vec{0}} q_{JK}P_K \text{ and } N_{J3} = c_J I^{(n-1)} + \sum_{K \not= \vec{0}} r_{JK} P_K,$$
where here $K \in \{0,1,2,3\}^{n-1}$ and for $K = (k_1, \dots, k_{n-1})$ we write $P_K = P_{k_1} \otimes \cdots \otimes P_{k_{n-1}}$. Then for measuring the first qubit to be state $+1$, we have
$$p_+ = \mathrm{Pr}\{\text{meas} = +1\} = \frac{1}{2}\left(1 + \frac{1}{2^{n-1}} \sum_{J\not=\emptyset} c_J\right)$$
and the post-measurement state will be
\begin{align*}
    &\tfrac{1}{p_+}(\Pi^{Z}\otimes I^{(n-1)}) \ket\psi\bra\psi (\Pi^{Z}\otimes I^{(n-1)})\\
    &=\ \Pi^Z \otimes \frac{1}{2^{n-1}}\left( I^{(n-1)} + \frac{1}{2p_+}\sum_{J \not= \emptyset}\sum_{K \not= \vec{0}} (q_{JK} + r_{JK}) P_K\right).
\end{align*}
Now, however, we face a challenge. The post-measurement state is a pure state $\ket{\psi'}$ and
\begin{equation}\label{eqn:post-measurement}
    \ketbra{\psi'} = \frac{1}{2^{n-1}}\left( I^{(n-1)} + \frac{1}{2p_+}\sum_{J \not= \emptyset}\sum_{K \not= \vec{0}} (q_{JK} + r_{JK}) P_K\right).
\end{equation}
But to find the predicates it satisfies, we need to find $(n-1)$-qubit additive predicates $\pred{M}'_1, \dots, \pred{M}'_{n-1}$ such that the associated operators satisfy
$$\prod_{j=1}^{n-1}(I^{(n-1)} + M'_j) = I^{(n-1)} + \frac{1}{2p_+}\sum_{J \not= \emptyset}\sum_{K \not= \vec{0}} (q_{JK} + r_{JK}) P_K.$$

While this does not seem immediately tractable, we can prove a lemma that shows that one feature of measurement from Pauli predicates carries over to general additive predicates: if a term in the intersection involves only $I$ and $Z$ in the measured qubit, then it becomes a term in the post-measurement predicate (possibly with a different sign).

\begin{lemma}
Suppose $M = I \otimes N_0 + Z \otimes N_3$. Then
\begin{itemize}
    \item $(\Pi^Z \otimes I^{(n-1)})M = (\Pi^Z \otimes (N_0 + N_3))(\Pi^Z \otimes I^{(n-1)})$, and
    \item $((I-\Pi^Z) \otimes I^{(n-1)})M = ((I-\Pi^Z) \otimes (N_0 - N_3)) ((I-\Pi^Z) \otimes I^{(n-1)})$.
\end{itemize}
\end{lemma}
\begin{proof}
    Direct computation.
\end{proof}

To apply this lemma, without loss of generality suppose $(\pred{M}_{(1)} \cap \dots \cap \pred{M}_{(n)})(\ket\psi)$ with $M_1 = I \otimes N_{1,0} + Z \otimes N_{1,3}$, and suppose the first qubit is measured (in the z-basis) with outcome $+1$. Then the post-measurement state is
\begin{align*}
    &\tfrac{1}{p_+}(\Pi^{Z}\otimes I^{(n-1)}) \ket\psi\bra\psi (\Pi^{Z}\otimes I^{(n-1)})\\
    &=\ \tfrac{1}{2^n p_+}(\Pi^{Z}\otimes I^{(n-1)})\cdot \prod_{j=1}^n (I^{n} + M_j) \cdot(\Pi^{Z}\otimes I^{(n-1)})\\
    &=\ \tfrac{1}{2}(\Pi^Z \otimes (I^{(n-1)} + N_{1,0} + N_{1,3}))\cdot \tfrac{1}{2^{n-1} p_+}(\Pi^{Z}\otimes I^{(n-1)})\cdot \prod_{j=2}^n (I^{n} + M_j)\cdot (\Pi^{Z}\otimes I^{(n-1)}).
\end{align*}
Hence the output state has $\pred{M}'_{(1)}(\ket{0,\psi'})$ where $M'_1 = N_{1,0} + N_{1,3}$ (up to a normalization term contained in $p_+$). The second conclusion in the lemma handles the case for outcome $-1$, where the post-measurement state has $\pred{M}'_{(1)}(\ket{1,\psi'})$ where $M'_1 = N_{1,0} - N_{1,3}$ (again up to normalization).

\subsection{Application: Gate Injection}\label{sec:magic}

A standard approach to fault-tolerant universal quantum computation is through implementing non-Clifford gates on codes through gate injection using associated ``magic'' states. While we can be explicit about the structure of the unitary gate we wish to inject, let us see what we can prove by simply appealing to Hoare-style judgments. For concreteness, we focus on single-qubit unitaries and assume the two judgments
\begin{equation}\label{eqn:z-type-statements}
\hoare{\X}{U}{\pred{M}} \text{ and } \hoare{\Z}{U}{\Z}.
\end{equation}
The additive predicate $\pred{M}$ cannot be arbitrary. These assumptions imply $U\sigma_zU^\dagger = \sigma_z$ and $U\sigma_xU^\dagger = M$. Since $\sigma_x$ and $\sigma_z$ anti-commute, so must $M$ and $\sigma_z$ and therefore $M = a\sigma_x + b\sigma_y$ where $a^2 + b^2 = 1$. We will parametrize $a = \cos\theta$ and $b = \sin\theta$. Naturally $T$ fits this mold with $\theta = \frac{\pi}{4}$. It is straightforward to deduce
$$\hoare{\Y}{U}{i \cdot (\cos\theta\cdot\X + \sin\theta\cdot\Y)\Z} \leadsto \{-\sin\theta\cdot\X + \cos\theta\cdot\Y\},$$
and so we see $U$ acts as a Bloch sphere rotation in the $\X/\Y$-plane by an angle $\theta$.

\begin{figure}\centering
    \begin{tikzpicture}
        \begin{yquant}
            qubit {$\ket{m}$} m;
            qubit {$\ket{\psi}$} p;
            cnot m | p;
            measure m;
            box {$U^2$} p | m;
            discard m;
            output {$U\ket\psi$} p;
        \end{yquant}
    \end{tikzpicture}
    \caption{Gate injection circuit for $U$.}
    \label{fig:z-type-injection}
\end{figure}

We claim that we can synthesize $U$ using the state $\ket{m}$ that satisfies the predicate $\mathbf{M}$ in the circuit of \Cref{fig:z-type-injection}. That is, we aim to show that this circuit $C$ satisfies the triples
$$\hoare{\pred{M}_1 \cap \Z_2}{C}{(\Z_1 \uplus -\Z_1) \cap \Z_2} \text{ and } \hoare{\pred{M}_1 \cap \X_2}{C}{(\Z_1 \uplus -\Z_1) \cap \pred{M}_2}$$
hence recovering \cref{eqn:z-type-statements} in the separable second factor.

Beginning with $\pred{M}_1 \cap \Z_2 = (\pred{M}\otimes\I) \cap (\I \otimes \Z)$ we evaluate the effect of the circuit on each term of the intersection:
\begin{align*}
    &\hoare{\I \otimes \Z}{NOTC}{I\otimes \Z}, \text{ and}\\
    &\hoare{(\cos\theta\cdot\X + \sin\theta\cdot\Y) \otimes \I}{NOTC}{\cos\theta\cdot\X\otimes\I + \sin\theta\cdot\Y\otimes\Z}.
\end{align*}
Our post-measurement predicate is then
$$((\Z_1 \cap \I_2) \uplus (-\Z_1 \cap \I_2)) \cap ((\Z_1 \cap \Z_2) \uplus (-\Z_1 \cap \Z_2)) \Rightarrow (\Z_1 \cap \Z_2) \uplus (-\Z_1 \cap \Z_2),$$
as the second term on the left side can be obtained using an implication rule on the first. %

Now turning to the case $\pred{M}_1\cap \X_2 = \pred{M}\otimes\I \cap \I \otimes \X$ we again evaluate the effect of the circuit:
\begin{align*}
    &\hoare{\I \otimes \X}{NOTC}{\X\otimes \X}, \text{ and}\\
    &\hoare{(\cos\theta\cdot\X + \sin\theta\cdot\Y) \otimes \I}{NOTC}{\cos\theta\cdot\X\otimes\I + \sin\theta\cdot\Y\otimes\Z}.\\
\end{align*}
Now, however, our precondition to the measurement
$$(\cos\theta\cdot\X\otimes\I + \sin\theta\cdot\Y\otimes\Z) \cap (\X\otimes \X)$$
has too many terms with an $\X$ in the first factor. 
So we multiply the second term into the first (using \Cref{rule-imp_mul}) to obtain
$$(\cos\theta\cdot\I\otimes\X + \sin\theta\cdot\Z\otimes\Y) \cap (\X\otimes \X).$$
Now we apply the discussions from the previous section to write the post-measurement condition as
$$(\Z_1\cap (\cos\theta\cdot\X + \sin\theta\cdot\Y)_2) \uplus ((-\Z)_1 \cap (\cos\theta\cdot\X - \sin\theta\cdot\Y)_2).$$
So we see that upon measuring $0$, the resulting state satisfies $\Z_1\cap \pred{M}_2$ as desired. But upon measuring $1$ we have the resulting predicate $(-\Z)_1\cap (\cos(-\theta)\X + \sin(-\theta)\Y)_2$, and so have accomplished the rotation in the opposite direction. That is, we have implemented $U^\dagger$ and so doing a post-selected correction of $U^2$ as in~\Cref{fig:z-type-injection} produces the output predicate $(-\Z)_1\cap \pred{M}_2$ as desired.

\section{Complexity of program verification}
\label{sec:complexity}

We can now present the algorithm for inferring postconditions and validating triples on quantum circuits. These are noticeably different procedures: validating triples ensures that a program has a given user-specified type, while postcondition inference attempts to derive a valid triple for a program. Given that our logic is rich enough to give infinitely many predicated to any circuit (though many will be equivalent), we will not perform full postcondition inference on a circuit. Instead, we can ask the user to specify the input, i.e., a precondition, and derive the output or postcondition through our inference rules. Alternatively, if the user has a specific postcondition in mind, we can do the same inference procedure and normalize both the pre- and postconditions (applying weakening rules as needed) and check that they are equivalent. Hence, in this section, we will focus on postcondition inference given a variety of programs and preconditions.

\paragraph{Inference on tensor preconditions}

\begin{table}
$$\begin{array}{l|c|c|c}
  & \X  & \Y  & \Z  \\ \hline
X & \X  & -\Y & -\Z \\
Y & -\X & \Y  & -\Z \\
Z & -\X & -\Y & \Z  \\
H & \Z  & -\Y & \X  \\
S & \Y  & -\X & \Z  \\
T         & \X + \Y & \Y - \X & \Z \\
T^\dagger & \X - \Y & \X + \Y & \Z
\end{array}$$
\caption{Axiomatized and derived behavior of common one-qubit gates.}
\label{table:1-qubit-types}
\end{table}

\begin{table}
$$\begin{array}{l|c|c|c|c|c|c}
      & \X ⊗ \I & \I ⊗ \X & \Y ⊗ \I & \I ⊗ \Y & \Z ⊗ \I & \I ⊗ \Z \\ \hline
\CNOT  & \X ⊗ \X & \I ⊗ \X & \Y ⊗ \X & \Z ⊗ \Y & \Z ⊗ \I & \Z ⊗ \Z \\
\mathtt{CZ}    & \X ⊗ \Z & \Z ⊗ \X & \Y ⊗ \Z & \Z ⊗ \Y & \Z ⊗ \I & \I ⊗ \Z
\end{array}$$
\NewDocumentCommand\compressot{}{\mspace{-\medmuskip}⊗\mspace{-\medmuskip}}
$$\begin{array}{l|c|c|c|c|c|c|c|c|c}
     & \X\compressot\X & \X\compressot\Y & \X\compressot\Z & \Y\compressot\X & \Y\compressot\Y & \Y\compressot\Z & \Z\compressot\X & \Z\compressot\Y & \Z\compressot\Z \\ \hline
\CNOT & \X\compressot\I & \Y\compressot\Z & -\Y\compressot\Y & \Y\compressot\I & -\X\compressot\Z & \X\compressot\Y & \Z\compressot\X & \I\compressot\Y & \I\compressot\Z \\
\mathtt{CZ}  & \Y\compressot\Y & -\Y\compressot\X & \X\compressot\I & -\X\compressot\Y & \X\compressot\X & \Y\compressot\I & \I\compressot\X & \I\compressot\Y & \Z\compressot\Z \\
\end{array}$$

\caption{Behavior of common two-qubit gates over all Pauli pairs.}
\label{table:2-qubit-types}
\end{table}

Given a Clifford circuit and a predicate $\pred{P}_1 \otimes \pred{P}_2 \otimes \dots \otimes \pred{P}_n$ (consisting of no intersections or additive terms), we can derive the corresponding postcondition on applying the circuit in $O(m)$ time, where $m$ is the number of gates. This follows from the fact that we can update the postcondition on each gate application in constant time. In practice, this only takes a single lookup since it proves convenient to add a number of derived triples to the system (\Cref{table:1-qubit-types,table:2-qubit-types}). Note that we assume tensors are implemented by arrays, saving us the time of iterating through an $n$ qubit list.

\paragraph{Inference on intersection predicates}

Restricting to just Pauli predicates, we can fully describe the semantics of a Clifford circuit (though we rarely will). Doing so requires determining the postcondition for  $I^{k-1} \otimes \X_k \otimes I^{n-k}$ and $I^{k-1} \otimes \Z_k \otimes I^{n-k}$, for every $1 \leq k \leq n$ where $n$ is the number of qubits. There are precisely $2n$ terms in this intersection, so the time to infer the fully descriptive output predicate is $O(mn)$.

\paragraph{Inference on fully separable preconditions}

When we get to separable preconditions, we have to start doing normalization (\Cref{sec:normal}). The normalization procedure iterates over each tensor in an intersection and then multiplies it by potentially all the remaining tensors. Since there are at most $2n$ elements in the intersection and each tensor is of length $n$, this winds up being an $O(n^3)$ operation. Once the normalization is done, applying the separability {rules} is straightforward. This gives us a complexity of $O(mn+n^3)$, which we can simplify to $O(mn)$ (our previous result) when $m \gg n^2$.

\paragraph{Post-measurement inference on Pauli predicates}

Measurement is where some complexity can start to appear, especially with it being a non-unitary operation. In general, we want to use the \emph{principle of deferred measurement}~\cite[\S 4.4]{Nielsen2010} to push off measurements until the end of the circuit, allowing us to perform normalization only once. A single-qubit post-measurement predicate doubles in size when a random outcome is expected due to the use of disjunctions. Then, performing $m$ measurements could potentially add a $2^m$ factor increase in the number of terms. This is in line with the fact that there could be $2^m$ possible outcomes to track.

In practice, however, it is more common for us to post-select on certain measurement outcomes or perform subsequent operations conditioned on certain outcomes (e.g., error correction). In such cases, it will be possible to simplify the expression or focus only on a pre-determined set of postconditions to understand the measurement behavior. The cost for computing the post-measurement predicate for a single Pauli measurement on an $n$-qubit system with $\ell$ terms in the disjunction is $O(2\ell n^3)$.

\paragraph{The Clifford+$T$ set and exponential blowup}

Naturally, universal quantum computing is the real test case for our logic. We know that our system is capable of fully describing arbitrary quantum computations, so unless quantum computing is efficiently simulable, we cannot efficiently validate the action of arbitrary quantum circuits. This is clear in the case of Toffoli. As we saw in \Cref{sec:toffoli}, despite having seven $T$ gates, checking the Toffoli circuit only involves additive predicates with at most $4$ terms (in the worst case). So in some sense, Toffoli only has an ``effective'' $T$-depth of $2$ (which is essentially the content of \Cref{thm:T-gate-bound}).

Nevertheless, in the worst case, the running time of our validation and inference procedures is $O(2^t)$, where $t$ is the number of $T$-gates, illustrating that our system cannot be efficiently applied to arbitrary circuits.

\paragraph{Post-measurement inference on additive predicates}

At this time, we refrain from providing any asymptotic expressions for the complexity of measuring additive predicates as we only consider very restricted cases of measuring single and 2-qubit systems. These are performed in an ad-hoc way by manipulating the underlying matrix corresponding to the given predicates. Hence, even generalizing the current process may not provide any non-trivial insights into asymptotic complexities beyond taking $O(2^n)$ time for an $n$-qubit system.

\section{Related Work}
\label{sec:related}

A variety of Hoare logics for quantum programs already exist. These logics use a variety of predicates, including quantum observables~\cite{ying2012}, subspaces~\cite{Unruh2019}, and projections~\cite{Zhou2019}, going from most general to least. Notably, each assertion style is more expressive than our stabilizer predicates. This is both an advantage and a limitation. The combination of highly expressive predicates and a fully general consequence rule allows the user to prove arbitrary properties of programs, but this requires significant manual effort. In particular, in the rule for unitary application $\hoare{U^\dag P U}{U}{P}$, some variant of which is present in almost every Hoare logic, $U^\dag P U$ is almost never explicitly computed: Doing so would make using the logic as complex as simulating the program. Instead, such expressions tend to be left symbolic and are simplified by the liberal application of the consequence rule.

By contrast, the logic presented here is designed to be fully automated and efficient at the cost of expressiveness. When a unitary gate is applied to a qubit, the corresponding predicate changes, and this change is efficient, modulo the effect of non-Clifford gates. This allows us to efficiently characterize properties like separability, provided that the $T$ count is low. However, we cannot prove the correctness of Grover's algorithm or Harrow-Hassidim-Lloyd algorithm, as in prior logics~\cite{ying2012,Zhou2019}.

Prior work on lightweight static analysis of quantum programs used the lens of \emph{abstract interpretation}. Abstract interpretation was developed by \textcite{CousotC77} in order to show useful properties of programs at low cost. Abstract interpretation necessarily sacrifices fidelity (being able to perfectly describe a program) in favor of efficiency. \textcite{Perdrix2008} was the first to apply abstract interpretation to quantum programs (expanding on earlier work \cite{Perdrix2007} that used a type system), but his system was quite limited: It could only precisely characterize a qubit as being in the z or x basis, and \emph{conservatively} tracked entanglement, meaning that it would err on the side of saying qubits were entangled if it could not rule that out. Building on Perdix's work, Prose and Zerrari~\cite{prost2009reasoning} developed a Hoare-like logic for conservatively tracking entanglement, applied to Selinger and Valiron's quantum lambda calculus~\cite{Selinger2006}.

More recently, \textcite{YuP21} developed an approach to quantum abstract interpretation based on \emph{reduced density matrices}, specifically $4 \times 4$ partial traces of the full system. The expressiveness of such an approach is unclear and is mostly used to check that qubits are in $\ket{0}$ or $\ket{1}$ states in practice. However, it does demonstrate remarkable performance in assertion checking and admits the possibility of using larger, more informative, reduced density matrices.

\textcite{Honda2015} presented a more powerful system based, like ours, on the stabilizer formalism. It represented states using stabilizer arrays, which can be translated to our logic but are rather less useful than human-readable predicates. It dealt with non-stabilizer states simply by treating them as black boxes (literally represented as $\blacksquare$), which could propagate throughout the program. This could be useful in a few cases, such as where a non-stabilizer state was quickly discarded, but generally meant the system could not meaningfully speak about non-Clifford circuits.

Prior versions of this work~\cite{Rand2021,Sundaram2022} presented the logic as a lightweight type system for quantum programs. While there is a rich literature on semantic subtyping and set-theoretic types~\cite{Frisch2008,castagna2023programming}, in which types can convey rich information about a program,
as our logic became more complex, using types came to feel less natural. In particular, typing derivations, which tend to be expressed in the form of trees, grew unwieldy, and it became clear that we would need a notion of subtyping corresponding to Hoare logic's consequence rule. Our lightweight consequence rule ($\Rightarrow$) proved to be the right middle ground, allowing us to simplify assertions without allowing arbitrary mathematical derivations. However, the switch is not without cost: A type system allows us to fully characterize the behavior of $H$ with the type $(\X \rightarrow \Z) \cap (\Z \rightarrow \X)$, 
while in Hoare logic we need to separately derive \hoare{X}{H}{Z} and \hoare{H}{Z}{X}.

Yuan et al.~\cite{yuan2022}, introduce a type system for conservatively tracking entanglement, with an interesting twist: Qubits can be cast to separable from the broader system (``pure'', in the paper's terminology) at the cost of a runtime check. This is a surprisingly lightweight and readable approach to tracking entanglement. However, the runtime check is potentially costly (whether the program is simulated or executed on a quantum computer) and it is not clear how this type system could be extended to express additional properties.

Concurrently with this work, Wu et al.~\cite{Wu2021} developed QECV, a quantum Hoare logic based on stabilizers for verifying error-correcting codes. This approach is largely complementary to our own (save for our comparatively small steps towards verifying stabilizer codes) but suggests a variety of new possible directions. Their language, as well as their logic, references predicates, allowing for measurement that branches on those predicates and a custom measurement that reflects that. The language also includes loops, which we do not address in this work. Also concurrent with this work, Huang et al.~\cite{huang2025efficient} developed a Hoare-style logic and for reasoning about error-correcting programs, which treats disjunction as the span of two vector spaces, and verifies a range of error correcting codes. Adapting both of these tools to verifying attributes like separability is a promising future direction for this work. %

Finally, there are two approaches to quantum program verification that are adjacent to our own but worth addressing. Quantum assertions~\cite{huang2019, liu2020, li2020} allow one to embed assertions inside programs that will check that a given property holds. While prior work was limited to checking simple assertions, \textcite{li2020} treats arbitrary projections as assertions. However, these systems can fail at runtime (e.g., if the measured state has some probability of being in the desired state) and also require us to check a program's behavior on a quantum device or sufficiently powerful simulator. At the other end of the spectrum are sophisticated logical systems for quantum programs \cite{ying2012,Unruh2019} and powerful tools to formally verify quantum program behavior~\cite{Chareton21,Hietala21}. However, even with an assist from automation, these tools tend to require substantial effort on the part of the programmer. We refer the reader to two recent surveys~\cite{chareton2023formal,lewis2023formal} for an in-depth analysis of the advantages and disadvantages of these approaches.

\section{Future work}
\label{sec:future}

There are still various ways to further enrich our program logic, providing many promising avenues for us to explore. 

\paragraph{Inference on Quantum Channels}

Other than measurement, all the operations whose behavior we infer are unitary circuits. More general quantum operations are given by completely positive trace-preserving maps, i.e., quantum channels. Extending our logic to handle quantum channels could potentially allow us to perform inference on or validate quantum cryptography and communication protocols. A starting point for this would be to use additive predicates and disjunctions to characterize partial traces and post-selection. %

\paragraph{Applications for error-correcting codes}

Implementing a fault-tolerant universal set of gates transversally will reduce the overall cost of error correction. However, as this cannot be achieved using just one code, a common method used switches between two sets of codes, each having a different set of transversal gates~\cite{anderson2014codeswitch}. Extending our logic to either infer the structure of or even validate the code-switching circuit given the predicates describing two codes would prove to be fruitful. Similarly, validating the encoding and decoding circuits for a code given its predicate could also be of value in verifying the implementation of error-correcting codes.

\paragraph{Normalization for additive predicates}

Finding a canonical representation for additive predicates is imperative to effectively validate additive postconditions. A big roadblock to it is that, unlike with Pauli predicates, additive predicates (especially multi-qubit ones) could have terms that neither commute nor anticommute. This makes it hard to find a normalization procedure for them similar to that in~\Cref{sec:normal}. Additionally, this also limits our ability to make multi-qubit separability judgments in the additive case.

\paragraph{Backwards reasoning}

In \Cref{rule:dag} we noted that our logic can be used for backwards, as well as forwards, reasoning about quantum circuits. Right now, this is limited to unitary gates, since measurement isn't reversible in the way unitary operations are. We could flesh out the logic to support backwards reasoning with measurement in the style of \cite{ying2012} and others.

\paragraph{General measurement for additive predicates}

Although we have outlined some cases in~\Cref{sec:meas-add} where we can infer the post-measurement states, this is limited to performing z-basis measurement on single and two-qubit systems. In order to fully exploit the power of additive predicates, it is essential that we have a full characterization for post-measurement states. An immediate consequence of this could be a deeper analysis of predicates for  multi-qubit magic states and applications associated with them.

\paragraph{A logic for quantum programs with classical control} A key component of quantum error correction and many quantum algorithms in practice (whether intermediate or large scale) is that they are interspersed with classical processing. This includes the use of classical control to decide which quantum operations to apply along with any pre- or post-processing. To account for this, we would need to formally extend our logic to explicitly handle classical data types as well as other program elements such as  conditional statements, loops, and recursion. This would involve extending both the language and the program logic itself, in the vein of the classical-quantum states of Feng and Ying \cite{Feng2021} and Ying \cite{Ying2026} and their associated logics.

\section*{Acknowledgments}
This material is based upon work supported by EPiQC, an NSF Expedition in Computing, under
Grant No. 1730449 and the Air Force Office of Scientific Research under Grants No. FA95502110051 and FA95502310406.

\section*{Authors' Contributions}
Aarthi Sundaram, Robert Rand, and Brad Lackey contributed to all aspects of this work, including the logic design, technical details, and writing. Kartik Singhal contributed to the design of the logic and initial drafts of the paper. Youngchan Cho contributed to the correctness proofs, particularly of the soundness of the system, and reviewed the paper's technical details for accuracy.

\printbibliography[heading=bibintoc]

\newpage
\appendix

\renewcommand{\arraystretch}{1.5}

\begin{figure}[H]
    \centering
\section{Full Grammar and Rules}
\small

\begin{enumerate}

    \item Predicates:
    \vspace{-5mm}
    \begin{align*}
        & \pred{G := \I \mid \X \mid \Y \mid \Z} \\
        & \pred{T := G \mid \mathit{c} T \mid T \otimes T \mid T + T} \\
        & \pred{P := T \mid P \cap P \mid P \uplus P \mid P_{S} }
    \end{align*}

\item Core Rules:
$$\begin{array}{cc}
 \hoare{X}{H}{Z} \quad&\quad \hoare{X \otimes I}{\CNOT}{X \otimes X} \\
 \hoare{Z}{H}{X} \quad&\quad \hoare{I \otimes X}{\CNOT}{I \otimes X} \\
 \hoare{X}{S}{Y} \quad&\quad \hoare{Z \otimes I}{\CNOT}{Z \otimes I} \\
 \hoare{Z}{S}{Z} \quad&\quad \hoare{I \otimes Z}{\CNOT}{Z \otimes Z} \\
 \hoare{Z}{T}{Z} \quad&\quad \hoare{X}{T}{\tfrac{1}{\sqrt{2}} (X + Y)}
    \end{array}$$

    \item Tensor Rules:
    \begin{center}
    \begin{tabular}{c @{\quad} c}
    $\inferrule*[right=\ensuremath{\otimes_1}]{\T[i] = \A \and \hoare{A}{U}{B}}{\hoare{T}{U~i}{\T[i \mapsto \B]}}$ &
    $\inferrule*[right=\ensuremath{\otimes_2}]{\T[i] = \A \and \T[j] = \B \and \hoare{A \otimes B}{U}{C\otimes D}}
    {\hoare{T}{U~i~j}{\T [i \mapsto \pred C; j \mapsto \pred D]}}$ \\
    \end{tabular}

    \end{center}
    \item Arithmetic Rules:
    \begin{center}
    \begin{tabular}{c @{\qquad \qquad} c} %
    $\inferrule*[Right=mul]{\hoare{A}{g}{A'} \and \hoare{B}{g}{B'}}{\hoare{AB}{g}{A'B'}}$ &
    $\inferrule*[Right=scale]{\hoare{A}{g}{A'}}{\hoare{\mathit{c}A}{g}{\mathit{c}A'}}$
    \end{tabular}
    \end{center}

    \item Sequence Rule:
        \begin{align*}
            \inferrule*[Right=seq]{\hoare{A}{g_1}{B} \and\hoare{B}{g_2}{C}}{\hoare{A}{g_1 ; g_2}{C}}
        \end{align*}

    \item Consequence Rule:
        \begin{align*}
            \inferrule*[Right=cons]{\pred{A' \Rightarrow A}  \and \hoare{A}{g}{B} \and \pred{B \Rightarrow B'}}{\hoare{A'}{g}{B'}}
        \end{align*}

    \item Conjunction and Disjunction Rules:
    \begin{center}
    \begin{tabular}{c @{\qquad \qquad} c @{\qquad \qquad} c}
    & {$\inferrule*[Right=$\cap$]{\hoare{A}{g}{A'} \and \hoare{B}{g}{B'}}{\hoare{\A \cap \B}{g}{\A' \cap \B'}}$}
    & {$\inferrule*[Right=$\uplus$]{\hoare{A}{g}{A'} \and \hoare{B}{g}{B'}}{\hoare{\A \uplus \B}{g}{\A' \uplus \B'}}$}\\
    \end{tabular}
    \end{center}

    \item Addition Rules: 
    \begin{center}
    \begin{tabular}{c @{\qquad \qquad \qquad} c}
    $\inferrule*[Right=add]{\hoare{A}{g}{C} \and \hoare{B}{g}{D}}{\hoare{A + B}{g}{C + D}}$
    & $\inferrule*[Right=$\otimes$-add]{\hoare{A}{U}{B+C} \and \T[i] = \A}{\hoare{T}{U~i }{T[i \mapsto \B] + T[i \mapsto \Ct]}}$
    \end{tabular}
    \end{center}

\item Separability Rules:
\begin{align*}
    \inferrule*[Right=embed]{\hoare{A}{g}{B}}{\hoare{A_K}{g_K}{B_K}} & \qquad\qquad\qquad\qquad \inferrule*[Right=disjoint]{\mathit{dom}(g) \cap K = \emptyset}{\hoare{A_K}{g}{A_K}}
\end{align*}

    \item Single-Qubit Measurement Rules:
\vspace{-4mm}
\begin{align*}
    \inferrule*{~}{\hoare{X}{\textsf{Meas}}{Z \uplus -Z}} \quad
    \inferrule*{~}{\hoare{Y}{\textsf{Meas}}{Z \uplus -Z}}  \quad
    \inferrule*{~}{\hoare{Z}{\textsf{Meas}}{Z}}  \quad
    \inferrule*{~}{\hoare{I}{\textsf{Meas}}{Z \uplus -Z}}
\end{align*}

\end{enumerate}

\caption{The basic predicates, Hoare triples, and deductive rules for our stabilizer logic. The grammar allows us to describe ill-formed predicates, such as $2\Y$ or $\X \cap (\I \otimes \Z)$, but these are never satisfied.
}
    \label{fig:types}
\end{figure}

\begin{figure}[H]
    \centering
\begin{align*}\label{app-rule-multiqubit_measurement}
    & \inferrule*[Right=Measure]{~}{\hoare{A}{\MEAS\ i}{Z_i \cap \pred{B} \uplus -Z_i \cap \pred{B}}} \\ \text{where} \\
    \nonumber &\qquad \text{(i) } \B = \normal_i{(\A)} \text{ if $\exists j, \A_{(j)}[i] = \Z$ or $\forall j, \A_{(j)}[i] = \I$ };\\
    \nonumber &\qquad \text{(ii) } \B = \normal_i{(\pred{A})} \setminus \A_{(j)} \text{ if $\exists j, \A_{(j)}[i] \in \{\X, \Y\}$};
\end{align*}
\caption{The measurement rule for multiple qubit systems.}
\end{figure}

\begin{figure}[H]
    \centering

\begin{enumerate}

    \item Simplification rules: %
    $$\arraycolsep=15pt \begin{array}{ccc}
    \pred{GG} \leadsto \I  & \pred{IG} \leadsto \pred{G} & \pred{GI} \leadsto \pred{G}\\
    \Z\X \leadsto i\Y &   \X\Y \leadsto i\Z & \Y\Z \leadsto i\X \\
    \X\Z \leadsto -i\Y &  \Y\X \leadsto -i\Z  & \Z\Y \leadsto -i\X \\
    \end{array}$$
    \vspace{-5mm}
    $$\arraycolsep=20pt  
    \begin{array}{cc}
    c\A \otimes \B \leadsto c(\A \otimes \B)
    & \A \otimes c\B \leadsto c(\A \otimes \B) \\
    \end{array}$$

    \item Common implication rules:
    \begin{align*}
    & \bigcap_i \A_i \Rightarrow \bigcap_i \A_{P(i)} & \text{(for any permutation $P$)} \\
    & \A \cap \B \Rightarrow \A \\
    & \A \cap \B \Rightarrow \A \cap \A\B & \text{(in any context)} \\
    & \biguplus_i \A_i \Rightarrow \biguplus_i \A_{P(i)} & \text{(for any permutation $P$)} \\
    & \A \uplus \A \Rightarrow \A \\
    & \sum_i \A_i \Rightarrow \sum_i \A_{P(i)} & \text{(for any permutation $P$)} \\
    & 0\B + \A \Rightarrow \A
    \end{align*}

    \item Single-qubit Separability Rules: 
    \begin{align*}
        & \I^{k-1} \otimes \B \otimes \I^{n-k} \Leftrightarrow \B_k & \\
       & \B_k \cap \T \Leftrightarrow \B_k \cap \T_{[n]\setminus \{k\}} &  \text{where } \T[k] \in \{\B, \I\}
    \end{align*}

    \item Multi-qubit separability rules for Pauli predicates when $S = \{j_1, \dots, j_k\} \subset [n]$:
    \begin{align*}
        & \B \cap \pred{T}_{(1)}\cap \ldots \cap \pred{T}_{(k)}  \Leftrightarrow \B_{\overline{S}} \cap \left(\Ct_{(1)} \cap \ldots \cap \Ct_{(k)}\right)_S, \\
        & \text{where } \forall_{j\in[k]} \ \pred{T}_{(j)}[S] = \Ct_{(j)} \text{, } \forall_{j \in [k]} \ \pred{T}_{(j)}[\overline{S}] = \I^{n-k} \text{ and } \B[S] = \I^k
    \end{align*}

\end{enumerate}
\caption{Simplification and implication rules for our predicates. These cover our applications for normalization and separability judgments. Let $[n] = \{1, \ldots, n\}$ and $S \subset [n]$. The conditions that $\Ct_{(1)}, \ldots, \Ct_{(k)}$ need to satisfy to achieve multi-qubit separability are described in \Cref{sec:multi-sep}.}%
    \label{fig:app-rules}
\end{figure}

\newpage
\section{Transitivity of Clifford groups}

Recall that a group $G$ acting on a set $\Omega$ is \emph{transitive} if for any $x,y\in\Omega$ there exists a $g \in G$ with $g\cdot x = y$. Since Clifford operators act on Pauli operators by conjugation, the Clifford group can never be transitive as $C \cdot I = CIC^\dagger = I$ . However, for nontrivial Paulis, it is.

\begin{proposition}
    Let $P,Q \in \mathcal{P}_n\setminus\{\pm I\}$. Then there exists a $C \in C\ell_n$ such that $CPC^\dagger = Q$.
\end{proposition}

More generally, a group is $m$-transitive if given tuples $(x_1, \dots, x_m), (y_1, \dots, y_m) \in \Omega^m$ with each $x_i \not= x_j$ and $y_i \not= y_j$, then there exists a $g \in G$ with $g\cdot x_i = y_i$ for $i=1, \dots, m$. Again, since the Clifford group acts by conjugation $C \cdot (-P) = -CPC^\dagger = -C\cdot P$ and so the Clifford group cannot be even 2-transitive. However, we modify the definition to require our Pauli elements to be distinct up to sign; then, we do obtain a higher transitivity result in the one-qubit, which follows from simply counting the number of one-qubit Clifford operators.

\begin{lemma}
    Given $P_1,P_2,Q_1,Q_2 \in \mathcal{P}_1\setminus\{\pm I\}$ with $P_1 \not= \pm P_2$ and $Q_1 \not= \pm Q_2$, then there exists a $C \in C\ell_1$ with $CP_1C^\dagger = Q_1$ and $CP_2C^\dagger = Q_2$.
\end{lemma}

Note that from the conditions in the lemma above, we must have $P_1$ and $P_2$ (and respectively $Q_1$ and $Q_2$) anticommute. But for higher qubit Paulis, this is not the case: even if $P_1 \not= \pm P_2$ we could have $P_1$ and $P_2$ commute. Since conjugation preserves commutativity, again, the Clifford group cannot be $2$-transitive. However, it is on pairs of commuting/anticommuting Paulis.

\begin{theorem}\label{theorem:2-transitive}
    Given $P_1,P_2,Q_1,Q_2 \in \mathcal{P}_n\setminus\{\pm I\}$ with $P_1 \not= \pm P_2$ and $Q_1 \not= \pm Q_2$ and either both $P_1,P_2$ and $Q_1,Q_2$ commute or both anticommute. Then then there exists a $C \in C\ell_n$ with $CP_1C^\dagger = Q_1$ and $CP_2C^\dagger = Q_2$.
\end{theorem}

The proof of this theorem follows from the $2$-qubit case (much like building a general Clifford operator out of $\mathit{CNOT}$ and one-qubit Cliffords). For two commuting $2$-qubit Cliffords $P,Q$, using the lemma above (and $\mathit{CNOT}$ if necessary) one can easily produce a $C$ with $C P C^\dagger = \sigma_y\otimes \sigma_y$ and $C Q C^\dagger = \sigma_z \otimes \sigma_z$. Similarly, for two anticommuting $2$-qubit Cliffords $P,Q$, one gets a $C$ with $C P C^\dagger = I\otimes \sigma_y$ and $C Q C^\dagger = I \otimes \sigma_z$. Then the theorem follows from chaining each of $P_1,Q_1$ and $P_2,Q_2$ through the appropriate normal form.

\end{document}